\documentclass[prb,twocolumn,aps,longbibliography, superscriptaddress, nofootinbib]{revtex4-1}

\usepackage[usenames]{xcolor} 
\usepackage{graphicx}
\usepackage{ntheorem}
\usepackage{mathrsfs}

\usepackage{hyperref}

\usepackage{bbm}

\usepackage{amsmath}
\usepackage{amssymb}
\usepackage{longtable}
\usepackage{boldline}

\usepackage{tikz}
\usetikzlibrary{calc,shapes}

\usepackage{placeins}

\newcommand{\defl}{\mathrel{\mathop:}=}
\newcommand{\defr}{\mathrel={\mathop:}}

\usepackage{array}

\definecolor{indigo}{rgb}{0.29, 0.0, 0.51}
\definecolor{RGBgreen}{RGB}{ 18 173 42}

\newcommand{\xcal}{{\scriptstyle{\mathcal{X}}}}
\newcommand{\ycal}{{\scriptstyle{\mathcal{Y}}}}

\begin{document}

\author{Nico Leumer}
\author{Magdalena Marganska}
\affiliation{Institute for Theoretical Physics, University of Regensburg, 93053 Regensburg, Germany}
\author{Bhaskaran Muralidharan}
\affiliation{Department of Electrical Engineering, Indian Institute of Technology Bombay, Mumbai 400076, India}
\author{Milena Grifoni}
\affiliation{Institute for Theoretical Physics, University of Regensburg, 93053 Regensburg, Germany}
 
\title{Exact eigenvectors and eigenvalues of the finite Kitaev chain and its topological properties}
\begin{abstract}
We present  a comprehensive,  analytical treatment of the finite Kitaev chain for arbitrary chemical potential.  We derive the momentum quantization conditions and present exact analytical formulae for the resulting energy spectrum and eigenstate wave functions, encompassing boundary and bulk states. In accordance with an analysis based on the winding number topological invariant, and as expected from the bulk-edge correspondence, the boundary states are topological in nature. They can have zero, exponentially small or even finite energy. A numerical analysis confirms their robustness against disorder.    
\end{abstract}

\maketitle
\date{\today}

\maketitle

\section{Introduction}
The quest for topological quantum computation has drawn a lot of attention to Majorana zero energy modes (MZM), quasi-particles obeying non-Abelian statistics hosted by topological superconductors \cite{Aguado}. 
The archetypal model of a topological superconductor in one dimension was proposed by Kitaev \cite{kitaev:physusp2001}. It consists of a chain of spinless electrons with nearest neighbor superconducting pairing, a prototype for p-wave superconductivity. As shown by Kitaev in the limit of an infinite chain, for a specific choice of parameters, the superconductor enters a topological phase where the chain can host a couple of unpaired zero energy Majorana modes at the end of the chain \cite{kitaev:physusp2001}. 
This model has thus become very popular due to its apparent simplicity and it is often used to introduce topological superconductivity in one dimension \cite{Alicea-2010, Aguado}. Also more sophisticated realizations of  effective p-wave superconductors, based on semiconducting nanowire-superconductor nanostructures \cite{Lutchyn-2010, Oreg-2010, Mourik-2012, Jelena-2012, Deng-2016, Jelena-2017, Zhang-2018,Prada2019},
 ferromagnetic chains on superconductors\cite{Nadj-Perge-2013, Nadj-Perge-2014,Jelena-2013, Zvyagin-2013, Kim-2018} 
 or s-wave proximitized carbon nanotubes \cite{Sau-2013, Marganska-2018, Milz-2018}, all rely on these fundamental predictions of the Kitaev model. While the theoretical models are usually solved in analytic form for  infinite or semi-infinite chains, the experiments are naturally done on finite-length systems. For example, for the iron chain on a superconductor investigated in Ref.~[\onlinecite{Nadj-Perge-2014}] it is expected that the chain length is shorter than the superconducting coherence length \cite{Lee-2014}.   
Spectral properties of a finite-length Kitaev chain have been addressed in more recent papers  \cite{Kao-2014,Hegde-2015,Zvyagin-2015,Zeng-2019}, and have confirmed the presence of bound states of exponentially small  energy in sufficiently long finite Kitaev chains. 

 As noticed by Kao using a chiral decomposition \cite{Kao-2014}, a finite-length Kitaev chain also supports modes with exact zero-energy.
 However, they are only found for discrete values of the chemical potential. As shown by Hegde et al., in Ref.~[\onlinecite{Hegde-2015}], these exact zeros can be associated to a fermionic parity crossing in the open Kitaev chain. 
  Investigations of the finite chain have also been performed by Zvyagin \cite{Zvyagin-2015} using the mapping of   a Kitaev chain  onto an X-Y model for $N$ spin 1/2 particles in transverse magnetic field, for which convenient diagonalization procedures are known \cite{Lieb-1961, Loginov-1997}. Kawabata et al. in Ref.~[\onlinecite{Kawabata-2017}] demonstrated that exact zero modes persist also in a Kitaev chain with twisted boundary conditions.    
  
Despite much being known by now about the  finite length  Kitaev chain, and in particular its low energy properties,  the intricacy of the eigenvector equation still constitutes a challenge.  
In this work we address this longstanding problem.  Exact results for the full energy spectrum  {\it{and}} the associated  bound states  are provided for arbitrary chemical potential by analytical diagonalization in the real space. Our results are not restricted to long chains or to long wavelengths, and thus advance some of the findings in Refs. [\onlinecite{Hegde-2015}], [\onlinecite{Zvyagin-2015}] and respectively [\onlinecite{Zeng-2019}]. They also complete the analysis of the eigenstates of an open Kitaev chain performed by Kawabata et al.\cite{Kawabata-2017} which was restricted to the exact zero modes. This knowledge allows one a deeper understanding of the topological properties of the Kitaev chain and, we believe,  also of other one-dimensional p-wave superconductors. \\
Before summarizing our results, we clarify the notions used further in our paper. 
(i) Since ``phase'' properly applies only to systems in the  thermodynamic limit, we shall use ``topological regime'' to denote the set of parameters for which a topological phase would develop in an infinite system. (ii) We shall call ``topological states'' all boundary states of the finite system  whose existence in the topological regime is enforced by the bulk-boundary correspondence \cite{Aguado}. Hence, the existence of the topological states is associated to a  topological invariant of the bulk system being non trivial. Quite generally,  the topological states can have zero, exponentially small, or even finite energy. (iii) When the gap between the topological states and the higher energy states is of the same order or larger than the superconducting gap $\Delta$, we consider them to be ``robust'' or ``topologically protected''. Their energy may be affected by perturbations, but not sufficiently to make them hybridize with the extended (bulk) states. (iv) All Hamiltonian eigenstates can be written as superpositions of Majorana (self-conjugate) components.  When the energy of the topological state is strictly zero, the whole eigenstate has the Majorana nature and becomes a true ``Majorana zero mode'' (MZM). 
\\  
Using our analytical expressions for the eigenstates of a finite Kitaev chain, we recover in the limit of an infinite chain the region for the existence of the MZM given by the bulk topological phase diagram; the latter can be obtained using the 
Pfaffian\cite{kitaev:physusp2001} or the winding number\cite{Review-Chiu-2016} topological invariant. For a finite-length chain MZM only exist for a set of discrete values of the chain parameters, see Eq. (\ref{equation: eigenvalue zero condition kouachi}) below, in line with Refs. [\onlinecite{Kao-2014}] and [\onlinecite{Kawabata-2017}]. These states come in pairs and, depending on the decay length, they can be localized each at one end of the chain but they can also be fully delocalized over the entire chain. Even in the latter case the states are orthogonal and do not "hybridize" since they live in two distinct Majorana sublattices. 
Similar protection of topological zero energy modes living on different sublattices has recently been observed experimentally in molecular Kagome lattices\cite{Kempkes-2019}. \\

The paper is organized as follows. Section \ref{section: The Kitaev chain and its bulk properties} shortly reviews the model and its bulk properties. Section \ref{section: spectral analysis} covers the finite size effects on the energy spectrum and on the quantization of the wave vectors for some special cases, including the one of zero chemical potential. The spectral properties at zero chemical potential are fully understood in terms of those of two independent Su-Schrieffer-Heeger-like (SSH-like) chains. \cite{wakatsuki:prb2014,Kawabata-2017} The eigenstates at zero chemical potential, the symmetries of the Kitaev chain in real space, as well as the Majorana character of the bound state wave functions are discussed in Section \ref{section: Eigenvectors and symmetries}.  In Section \ref{section: influence of non zero chemical potential}, \ref{section: MZM paths in the topological phase diagram for finite N} and \ref{section:  numerical results and impact of disorder} we turn to the general case of finite chemical potential which couples the two SSH-like chains. While Sec. \ref{section: influence of non zero chemical potential} deals with the energy eigenvalues and eigenvectors of the finite chain, Sec. \ref{section: MZM paths in the topological phase diagram for finite N} provides exact analytical results for the MZM. In Sec. \ref{section:  numerical results and impact of disorder} the influence of disorder on the energy of the lowest lying states is investigated numerically. In Sec. \ref{section: conclusion} conclusions are drawn. Finally, appendices \ref{appendix: comments on Fibonacci and Tetrannaci polynomials}-\ref{appendix: zero energy states} contain details of the factorisation of the characteristic polynomial in real space and the calculation of the associated eigenstates.   

\section{The Kitaev chain and its bulk properties}
\label{section: The Kitaev chain and its bulk properties}

\subsection{Model}
\label{section 2 model}
The Kitaev chain is a one dimensional model based on a lattice of $N$ spinless fermions. It is characterized by three parameters: the chemical potential $\mu$, the hopping amplitude $t$, and the p-wave superconducting pairing constant $\Delta$. The Kitaev Hamiltonian, written in a set of standard fermionic operators $d_j,\,d_j^\dagger$, is \cite{kitaev:physusp2001, Aguado}
\begin{align}\label{equation: Kit. Hamiltonian, fermionic operators, realspace}
	\hat{H}_{\mathrm{KC}}=-\mu\sum\limits_{j=1}^N\,d_j^\dagger d_j+\sum\limits_{j=1}^{N-1}\,\left(\Delta\, d_j^\dagger d_{j+1}^\dagger\,-t\,d^				\dagger_{j+1} d_{j}+h.c.\right),
\end{align}
where the p-wave character allows interactions between particles of the same spin. The spin is thus not explicitly included in the following. We consider $\Delta$ and $t$ to be real parameters from now on.

The Hamiltonian in Eq. (\ref{equation: Kit. Hamiltonian, fermionic operators, realspace}) has drawn particular attention in the context of topological superconductivity, due to the possibility of hosting MZM at its end in a particular parameter range\cite{kitaev:physusp2001}. This can be seen by expressing the Kitaev Hamiltonian in terms of so called Majorana operators $\gamma^{A,B},$ 
\begin{align}\label{equation: Definition Majorana operators}
	\left(\begin{matrix}
	d_j\\
	d_j^\dagger\end{matrix}\right)
	\,\defr\,
	\frac{1}{\sqrt{2}}\,
	\left(\begin{matrix}
	1& i\\
	1 & -i
	\end{matrix}\right)\,
	\left(
	\begin{matrix}
	\gamma_j^A\\
	\gamma_j^B\end{matrix}
	\right), \quad \left(\gamma^{A,B}\right)^\dagger = \gamma^{A,B},
\end{align}
yielding the form
\begin{align}\label{equation: Kitaev chain in Majorana operators}
	\hat{H}_\mathrm{KC}\,&=\,-i\,\mu\,\sum\limits_{j=1}^N\,\gamma_j^A\gamma_j^B\,+\,i\left(\Delta+t\right)\sum\limits_{j=1}^{N-1}				\gamma_j^B\gamma_{j+1}^A \notag\\
	&\qquad +\,i\left(\Delta-t\right)\sum\limits_{j=1}^{N-1} \gamma_{j}^A \gamma_{j+1}^B.
\end{align}
Notice that, in virtue of Eq. \eqref{equation: Definition Majorana operators} it holds $\{\gamma_j^A,\,(\gamma_j^A)^\dagger\}\,=\,2\,(\gamma_j^A)^2\,=\,1$, and similarly for $\gamma_j^B$. For the particular parameter settings $\Delta = \pm t$ and $\mu=0$, which we call the Kitaev points, Eq. (\ref{equation: Kitaev chain in Majorana operators}) leads to a "missing" fermionic quasiparticle $q_\pm$: 
\begin{subequations}
	\begin{align}
		\label{equation: Missing electron a}
		q_+\,&=\,\frac{1}{\sqrt{2}}\left(\gamma_1^A\,+\,i\,\gamma_N^B\right)\qquad \left[\Delta=t\right],\\
		\label{equation: Missing electron b}
		q_-\,&=\,\frac{1}{\sqrt{2}}\left(\gamma_1^B\,+\,i\,\gamma_N^A\right)\qquad \left[\Delta=-t\right].
	\end{align}
\end{subequations}
This quasiparticle has zero energy and is composed of two isolated Majorana states localised at the ends of the chain. In general,
the condition of hosting MZM does not restrict to the Kitaev points ($\mu=0$, $\Delta=\pm t$). Further information on the existence of boundary modes is  evinced from the bulk spectrum and the associated topological phase diagram. 
 If the boundary modes have exactly zero energy, their Majorana nature can be proven by showing that they are eigenstates of the particle-hole operator $\mathcal{P}$. Equivalently, if $\gamma_M^\dagger$ is an operator associated to such a MZM it satisfies $\gamma_M^\dagger  =\gamma_M$ and $\gamma_M^2=1/2$.

The topological phase diagram is shortly reviewed in Sec. \ref{subsection: Topological phase diagram}.
\subsection{Bulk spectrum}
The Hamiltonian from Eq. (\ref{equation: Kit. Hamiltonian, fermionic operators, realspace}) in the limit of $N\rightarrow \infty$ reads in $k$ space
\begin{align}\label{equation: Kitaev Hamilt. in electronic operators}
	\hat{H}_\mathrm{KC}\,=\,\frac{1}{2}\sum_k\,\hat{\psi}^\dagger_k\,\mathcal{H}(k)\hat{\psi}_k,~~
	\hat{\psi}_k\,=\,\left(d_k,\,d_{-k}^\dagger\right)^\mathrm{T},
\end{align}
where we introduced the operators \makebox{$d_k=\frac{1}{\sqrt{N}}\sum\limits_j e^{-i\,j\,kd}\,d_j$} and $k$ lies inside the first Brillouin zone, i.e. \makebox{$k\in\left[-\frac{\pi}{d},\frac{\pi}{d}\right]$} and $d$ is the lattice constant. The $2\times 2$ Bogoliubov- de Gennes (BdG) matrix 
\begin{align}\label{eq:hamiltonian-Kitaev-k}
	\mathcal{H}(k)\,=\,\left[\begin{matrix}
	-\mu-2t\cos(kd) & -2i\Delta\sin(kd)\\
	2i\Delta\sin(kd) & \mu+2t\cos(kd)
	\end{matrix}\right]
\end{align}
is easily diagonalized thus yielding the excitation spectrum
\begin{align}\label{infinite chain: excitation spectrum}
	E_\pm(k)\,=\,\pm\sqrt{4\Delta^2\,\sin^2(kd)\,+\,\left[\mu\,+\,2t\,\cos(kd)\right]^2}.
\end{align}
Note that for $\mu=0$ Eq. (\ref{infinite chain: excitation spectrum}) predicts a gapped spectrum whose width is either $4\,\Delta$ ($\vert \Delta \vert < \vert t \vert$) or $4\,t$ ($\vert t \vert < \vert \Delta \vert$).
\subsection{Topological phase diagram}
\label{subsection: Topological phase diagram}
\begin{figure}[h]
 \includegraphics[width=\columnwidth]{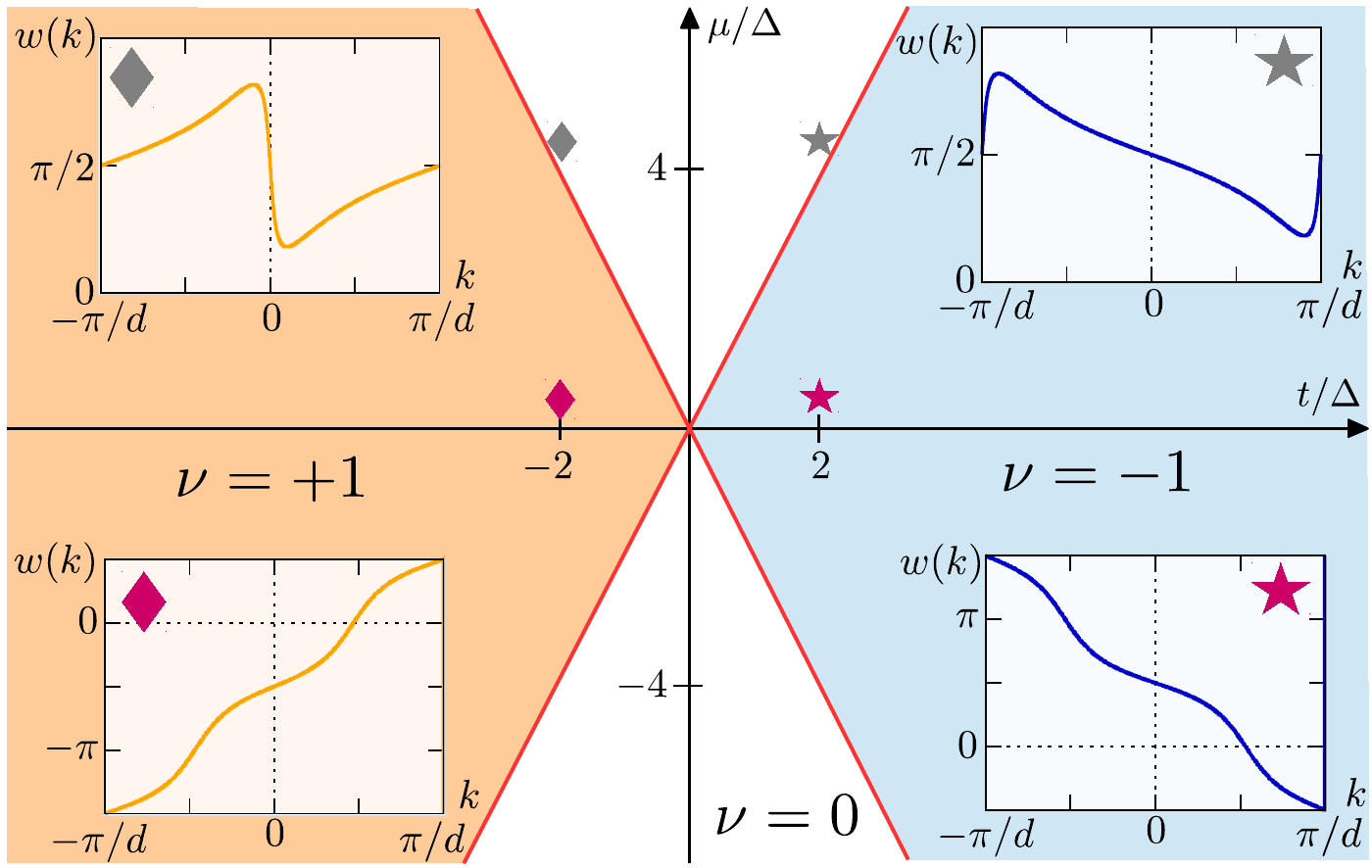}
	\caption{Topological phase diagram of the Kitaev chain for $\Delta>0$, constructed with the winding number invariant Eq. (\ref{equation: winding number invariant}). Distinct topological phases are separated by the phase boundary at $\mu = \pm 2t$, visualised by the red lines. The four insets illustrate the evolution of $w(k)$ along the Brillouin zone at the four marked positions in the phase space, ($t/\Delta =\pm 2$, $\mu/\Delta=0.2$) in the topological and ($t/\Delta =\pm 2$, $\mu/\Delta=4.2$) in the trivial phase. In the phase diagram for $\Delta<0$ the $\nu=+1$ and $\nu=-1$ regions are swapped.}
	\label{fig:topo-phase-diagram}
\end{figure}
The BdG Hamiltonian \eqref{eq:hamiltonian-Kitaev-k} is highly symmetric. By construction it anticommutes with the particle-hole symmetry $\mathcal{P}=\sigma_x\mathcal{K}$, where $\mathcal{K}$ accounts for complex conjugation. The particle-hole symmetry turns an eigenstate in the $k$ space corresponding to an energy $E$ and wavevector $k$ into one associated with $-E$ and $-k$. The time reversal symmetry is also present in the Kitaev chain and is given by $\mathcal{T}=\mathbbm{1}\,\mathcal{K}$. Finally, the product of $\mathcal{TP}=C=\sigma_x$ is the chiral symmetry, whose presence allows us to define the topological invariant in terms of the winding number.~\cite{Review-Chiu-2016} Note that all symmetries square to $+1$, placing the Kitaev chain in the BDI class.~\cite{altland:prb1997} 

The winding number 
is given by \cite{Wen-1989, Review-Chiu-2016}
\begin{align}\label{equation: winding number invariant}
 	\nu =\frac{1}{2\pi} \int\limits_{-\pi/d}^{\pi/d} dk~\partial_k \,w(k),
\end{align}
where $w(k)=\mathrm{arg}\left[2\Delta\,\sin(kd)\,+\,i\,\left(\mu+2t\,\cos(kd)\right)\right]$ and $\partial_k\, w(k)$ is the winding number density. A trivial phase corresponds to $\nu=0$, a non trivial one to finite integer values of $\nu$. The winding number relates bulk properties to the existence of boundary (not necessarily MZM)  states in a finite chain. This property is known as bulk-edge correspondence. Due to their topological nature, their existence is robust against small perturbations, like disorder. This point is further discussed in Sec. \ref{section: numerical results and impact of disorder}. 

The phase diagram constructed using the winding number invariant is shown in Fig.~\ref{fig:topo-phase-diagram}. The meaning of two different values for the winding number is clearer when we recall the Kitaev Hamiltonian in the Majorana basis. In a finite chain the leftmost lattice site consists of the $A$ Majorana  operator \textcolor{blue}{$\gamma_1^A$} connected to the bulk by the $i(\Delta-t)$ hopping and the $B$ Majorana operator \textcolor{orange}{$\gamma_1^B$} connected by the $i(\Delta+t)$ hopping. With $\Delta>0$ and $t<0$ (the $\nu=+1$ phase) the Majorana  state at the left end of the chain will consist mostly of the weakly connected \textcolor{orange}{$\gamma^B_1$}. If $t>0$ (the $\nu=-1$ phase), \textcolor{blue}{$\gamma^A_1$} is connected to the bulk more weakly and contributes most to the left end bound state.

The boundaries between different topological phases can be obtained from the condition of closing the bulk gap, i.e. $E_\pm(k)=0$ (cf. Eq.~\eqref{infinite chain: excitation spectrum}). That is only possible if both terms under the square root vanish. The condition of $\Delta\neq 0$ forces the gap closing to occur at $kd=0$ or $k=\pi d$, and the remaining term vanishes at these momenta if $\mu=\pm2t$. The four insets in Fig. \ref{fig:topo-phase-diagram} show the behavior of $ w (k)$, leading to either a zero (for $w(-\pi)=w(\pi)\,$) or non zero winding number, see Eq. \eqref{equation: winding number invariant}.

Physically speaking, the Kitaev chain is in the topological phase provided that $\Delta\neq 0$ and the chemical potential lies inside the "normal" band ($\vert \mu\vert \le 2 \vert t\vert$). 

\section{Spectral analysis of the finite Kitaev chain}
\label{section: spectral analysis}
One of the characteristics of finite systems is the possibility to host edge states at their ends. To account for the presence and the nature of such edge states, we consider a finite Kitaev chain with $N$ sites and open boundary conditions, yielding $N$ allowed $k$ values. In this section we shall consider the situation in which one of the three parameters $\Delta$, $t$ and $\mu$ is zero. Already for the simple case $\mu=0$ and $\Delta \neq 0$, $t\neq 0$ the quantization of the momentum turns out to be non trivial. The general case in which all parameters are finite is considered in Secs. \ref{section: influence of non zero chemical potential}, \ref{section: MZM paths in the topological phase diagram for finite N} and \ref{section: numerical results and impact of disorder}.

We start with the BdG Hamiltonian of the open Kitaev chain in real space, expressed in the basis of standard fermionic operators  \makebox{$\hat{\psi}\,=\,\left(d_1,\,\ldots,\,d_N,\,d_1^\dagger,\,\ldots,\,d_N^\dagger\right)^\mathrm{T}$}. Then
\begin{align}\label{equation: Kitaev Hamiltonian and the BdG basis}
	\hat{H}_\mathrm{KC}\,=\,\frac{1}{2}\,\hat{\psi}^\dagger\,\mathcal{H}_\mathrm{KC}\,\psi,
\end{align}
where the BdG Hamiltonian $\mathcal{H}_\mathrm{KC}$ is
\begin{align}\label{equation: BdG Hamiltonian in real space in its full glory}
	\mathcal{H}_\mathrm{KC}\,=\,\left[\begin{matrix}
	C & S\\
	S^\dagger & -C
	\end{matrix}\right].
\end{align}
These matrices have the tridiagonal structure
\begin{align}\label{equation: Matrix C}
	C\,=\,\left[\begin{matrix}
	-\mu & -t  		\\
	-t&-\mu & -t 	\\
	& -t &-\mu & -t &\\
	&  & \ddots & \ddots &\ddots &\\
	&&& -t &-\mu & -t\\
	&&&& -t &-\mu & -t\\
	&&&&& -t &-\mu
	\end{matrix}			\right],
	\\\label{equation: Matrix S}
	S\,=\,\left[\begin{matrix}
	0 & \Delta  		\\
	-\Delta&0 & \Delta 	\\
	& -\Delta &0 & \Delta &\\
	&  & \ddots & \ddots &\ddots &\\
	&&& -\Delta &0 & \Delta\\
	&&&& -\Delta &0 & \Delta\\
	&&&&& -\Delta &0
	\end{matrix}\right].
\end{align}
The spectrum can be obtained by diagonalisation of $\mathcal{H}_\mathrm{KC}$ in real space. We consider different situations.
\subsection{$\Delta=0$}
\label{subsection: spectrum delta  = 0}
The BdG Hamiltonian is block diagonal and its characteristic polynomial $P_\lambda(\mathcal{H}_\mathrm{KC})=\mathrm{det}\left[\lambda\,\mathbbm{1}-\mathcal{H}_\mathrm{KC}\right]$ \makebox{factorises} as
\begin{align}
	P_\lambda(\mathcal{H}_\mathrm{KC})\,=\,P_\lambda(C)\,P_\lambda(-C).
\end{align}
The tridiagonal structure of $C$ straightforwardly yields the spectrum of a normal conducting, linear chain \cite{Kouachi}
\begin{align}\label{spectrum: linear chain}
	E^{\Delta=0}_\pm(k_n)\,=\,\pm\left[\mu\,+\,2t\,\cos\left(k_n d\right)\right],\qquad k_n d\,=\,\frac{n\pi}{N+1},
\end{align}
where $n$ runs from $1$ to $N$. Since $k_n\in\mathbbm{R}$, only bulk states exist for $\Delta = 0$.
\subsection{$t=0$}
\label{subsection: spectrum t  = 0}
In the beginning we consider both $t$ and $\mu$ to be zero and include $\mu\,\neq\,0$ in a second step. The parameter setting leads to a vanishing matrix $C$, see Eq. (\ref{equation: Matrix C}), and the characteristic polynomial of the system reads:
\begin{align}
	P_\lambda(\mathcal{H}_\mathrm{KC})\,=\,\mathrm{det}\left[\begin{matrix}
	\lambda\,\mathbbm{1} & -S\\
	S & \lambda\,\mathbbm{1}
	\end{matrix}\right],
\end{align}
where we used the property $S^\dagger\,=\,-S$. Due to the fact that the commutator $[\mathbbm{1},\,S]=0$ vanishes\footnote{Note that $\lambda$ can be zero and it will for odd $N$. Hence, the standard formula to calculate the determinant of a partitioned $2\times 2$ matrix can not be used here, because it requires the inverse of one diagonal block. We use instead Silvester's formula\cite{Silvester}: $\text{det}\left[\begin{matrix}
A & B\\
C & D\\
\end{matrix} \right] =\text{det} \left[AD\,-\,CB\right]$, where $A,\,B,\,C,\,D$ are square matrices of the same size and the only requirement is $%
\left[C,\,D\right]=0$.}, one finds \cite{Silvester}
\begin{align}
	P_\lambda(\mathcal{H}_\mathrm{KC})\,=\,\mathrm{det}\left(\lambda^2\,\mathbbm{1}\,+\,S^2\right).
\end{align}
The characteristic polynomial can still be simplified to the product 
\begin{align}
	P_\lambda(\mathcal{H}_\mathrm{KC})\,=\,P_\lambda(iS)\,P_\lambda(-iS).
\end{align}
The matrix $iS$ is hermitian and describes a {\it{linear}} chain with hopping $i\Delta$. As a consequence, we find the spectrum to be \cite{Kouachi}
\begin{align}\label{equation: spectrum t=mu=0}
	E_\pm(k_n)\,=\,\pm\left[2\Delta\,\cos\left(k_n d\right)\right],\qquad k_n d\,=\,\frac{n\pi}{N+1},
\end{align}
where $n$ runs from $1$ to $N$ and each eigenvalue is twice degenerated. Notice the phase shift by $\pi/2$ compared to the spectrum of an infinite chain Eq. (\ref{infinite chain: excitation spectrum}). We discuss this phase shift in more detail in section \ref{sec:finite-chain_mu0}. 

Furthermore if, and only if, $N$ is odd, we find two zero energy modes, namely for \makebox{$n=(N+1)/2$}. Their existence and the degeneracy is due to the chiral symmetry.

The chemical potential $\mu$ can be included easily. Exploiting the properties of $\mathcal{H}_\mathrm{KC}$, we find the characteristic polynomial to be
\begin{align}
	P_\lambda(\mathcal{H}_\mathrm{KC})\,&=\,\mathrm{det}\left[\begin{matrix}
	\left(\lambda +\mu\right)\,\mathbbm{1} & -S\\
	S & \left(\lambda -\mu\right)\,\mathbbm{1}
	\end{matrix}\right]\notag\\
	&=\,\mathrm{det}\left[\left(\lambda^2-\mu^2\right)\,\mathbbm{1}\,+\,S^2\right]\notag\\
	&=\,\mathrm{det}\left[\Lambda^2\,\mathbbm{1}\,+\,S^2\right],
\end{align}
with $\Lambda^2\,\defl\,\lambda^2-\mu^2$. The same treatment as in the previous $\mu=0$ case yields
$ P_\lambda(\mathcal{H}_\mathrm{KC})\,=\,P_\Lambda(iS)\,P_\Lambda(-iS)$. 
%
%
Consequently the spectrum is
\begin{align}\label{equation: spectrum and quantization for t  is zero}
	E^{t=0}_\pm(k_n)\,=\,\pm\sqrt{\mu^2\,+\,4\Delta^2\,\cos^2(k_n d)},\qquad k_n d\,=\,\frac{n\pi}{N+1},
\end{align}  
where $n$ runs again from $1$ to $N$. Again no boundary modes are found for $t =0$.
\subsection{$\mu=0$}
\label{sec:finite-chain_mu0}
The calculation of the spectrum for $\mu=0$ requires a more technical approach, since the structure of the BdG Hamiltonian Eq. (\ref{equation: BdG Hamiltonian in real space in its full glory}) prohibits standard methods. 
\begin{figure}[ht]\centering
	\includegraphics[width = 0.8\columnwidth]{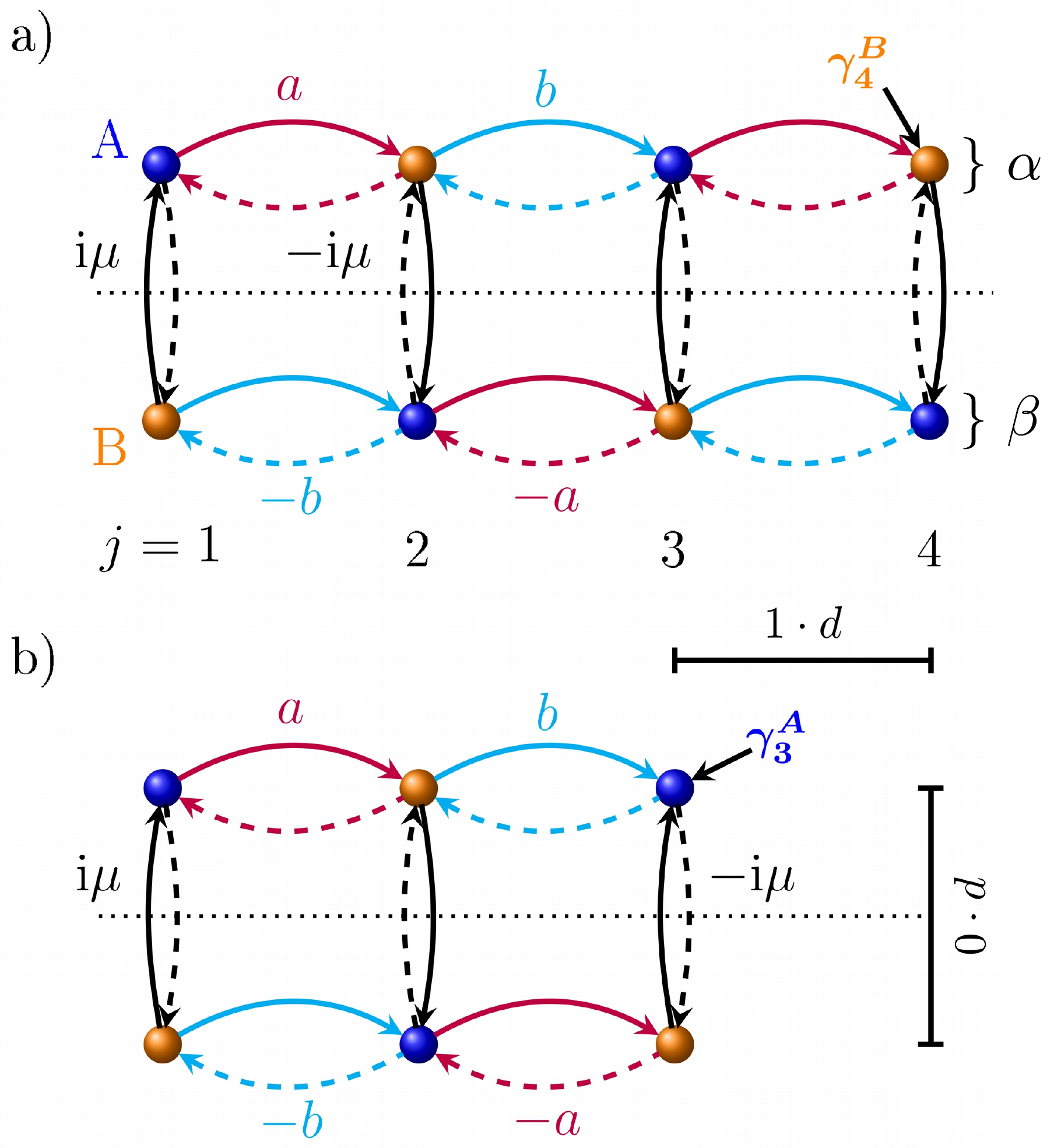}
	\caption{Kitaev chain viewed as two coupled SSH-like chains for a) $N=4$ and b) $N=3$ sites. These two chains $\alpha$ and $\beta$ are coupled by $\pm i \mu$. The hoppings $a=i(\Delta-t)$ in red and $b=i(\Delta+t)$ in blue alternate (dashed lines correspond to $-a$ and $-b$) and connect neighbouring Majorana operators $\gamma_j^A$ (blue spheres) and $\gamma_{j\pm 1}^B$ (orange spheres). The unit cell has size $2d$. }
	\label{figure: Kitaev model as coupled ssh chains.}
\end{figure}

One important  feature of the Kitaev chain can be appreciated inspecting Eq. (\ref{equation: Kitaev chain in Majorana operators}). The entire 
model is equivalent to two coupled SSH-like chains\cite{wakatsuki:prb2014,li:prb2018} containing both the hopping parameters $a\,\defl\,i\left(\Delta-t\right)$ and $b\,\defl\,i\left(\Delta+t\right)$, see Fig. (\ref{figure: Kitaev model as coupled ssh chains.}). Explicitly,  
\begin{align}\label{equation: Kitaev Hamilt. in SSH basis, even N}
	\hat{H}_\mathrm{KC}\,&=\,\left(a\sum_{j=1}^{N_1}\,\gamma_{2j-1}^A\gamma_{2j}^B +b \sum_{j=1}^{N_2}\,\gamma_{2j}^B\gamma_{2j+1}^A 
	\right)+h.c.\notag\\
	&\,+\left(b\sum_{j=1}^{N_1}\,\gamma_{2j-1}^B\gamma_{2j}^A +a \sum_{j=1}^{N_2}\,\gamma_{2j}^A\gamma_{2j+1}^B \right)+h.c.\notag\\
	&\,-i\mu\sum_{j=1}^N\,\gamma_j^A\gamma_j^B,
\end{align}  
where $N_{1,2}$ depend on $N$. If $N$ is even we have $N_1 = N/2$ and $N_2 = N_1-1$, while $N_1=N_2=(N-1)/2$ for odd $N$. Independent of the number of atoms, the first and the second lines in 
Eq. (\ref{equation: Kitaev Hamilt. in SSH basis, even N}) describe two SSH-like chains, coupled by the chemical potential $\mu$. 
We define here the SSH-like basis of the Kitaev chain as:
\begin{align}\label{equation: ssh basis, alternating pattern}
	\hat{\Psi}^\mathrm{even}_\mathrm{SSH}&=\left(\textcolor{blue}{\gamma_1^\mathrm{A}},\,\textcolor{orange}{\gamma_2^\mathrm{B}},\ldots,					\textcolor{blue}{\gamma_{N-1}^\mathrm{A}},\,\textcolor{orange}{\gamma_N^\mathrm{B}} \vert \textcolor{orange}{\gamma_1^\mathrm{B}},\,				\textcolor{blue}{\gamma_2^\mathrm{A}},\ldots,\textcolor{orange}{\gamma_{N-1}^\mathrm{B}},\,\textcolor{blue}{\gamma_N^\mathrm{A}}	\right)^			\mathrm{T},\notag\\
	\notag\\
	\hat{\Psi}^\mathrm{odd}_\mathrm{SSH}&=\left(\textcolor{blue}{\gamma_1^\mathrm{A}},\,\textcolor{orange}{\gamma_2^\mathrm{B}},\ldots,						\textcolor{orange}{\gamma_{N-1}^\mathrm{B}},\,\textcolor{blue}{\gamma_N^\mathrm{A}} \vert \textcolor{orange}{\gamma_1^\mathrm{B}},\,				\textcolor{blue}{\gamma_2^\mathrm{A}},\ldots,\textcolor{blue}{\gamma_{N-1}^\mathrm{A}},\,\textcolor{orange}{\gamma_N^\mathrm{B}}	\right)^			\mathrm{T},
\end{align}
where "$\vert$" marks the boundary between both chains. We call the first one, starting always with $\gamma_1^\mathrm{A}$, the $\alpha$ chain, and the second the $\beta$ chain, such that \makebox{$\Psi^\mathrm{even,\,odd}_\mathrm{SSH}=\left(\vec{\gamma}_\alpha\,\vert \, \vec{\gamma}_\beta\right)^\mathrm{T}$}. The BdG Hamiltonian in the SSH-like basis reads 
\begin{align}\label{equation: Kitaev Hamiltonian/ matrix in SSH basis}
	\mathcal{H}_\mathrm{KC}^\mathrm{SSH}\,=\,\left[\begin{matrix}
	\mathcal{H}_\alpha & \tau \\ 
	\tau^\dagger &   \mathcal{H}_\beta
	\end{matrix}
	\right],
\end{align}
with $\hat{H}_\mathrm{KC}=\frac{1}{2}\hat{\Psi}_\mathrm{SSH}^\dagger\mathcal{H}_\mathrm{KC}^\mathrm{SSH}\hat{\Psi}_\mathrm{SSH}$. The independent SSH-like chains are represented by the square matrices $\mathcal{H}_\alpha$ and $\mathcal{H}_\beta$ of size $N$. Both chains are coupled by the matrices $\tau$ and $\tau^\dagger$, which contain only the chemical potential $\mu$, in a diagonal arrangement specified below.

The pattern of these matrices is slightly different for even and odd number of sites. If $N$ is even we find 
\begin{align}
	\mathcal{H}_\alpha^{\mathrm{even}}\,&=\,\left[\begin{matrix}
	0 & a\\
	-a & 0&b \\
	& -b&0& a\\
	&&\ddots & \ddots & \ddots\\
	&&& -a & 0&b\\
	&&&&-b&0& a\\
	&&&&& -a &0
		\end{matrix}		\right],\\
	\mathcal{H}_\beta^{\mathrm{even}}\,&=\,\left[\begin{matrix}
	0 & b\\
	-b & 0&a \\
	& -a&0& b\\
	&&\ddots & \ddots & \ddots\\
	&&& -b & 0&a\\
	&&&&-a&0& b\\
	&&&&& -b &0
		\end{matrix}\,		\right],
\end{align}
and $\tau^{\mathrm{even}} = -i\mu\, \mathbbm{1}_{N/2}\otimes \tau_z$, where $\tau_z$ denotes the Pauli matrix. The odd $N$ expressions are achieved by removing the last line and column in $\mathcal{H}_\alpha^{\mathrm{even}}$, $\mathcal{H}_\beta^{\mathrm{even}}$ and $\tau^\mathrm{even}$.

As shown in more detail in appendix \ref{appendix: spectrum zero mu}, for $\mu=0$ the characteristic polynomial can be expressed as the product of two polynomials of order $N$
\begin{align}\label{equation: characteristic polynomial at zero mu}
	P_\lambda(\mathcal{H}_\mathrm{KC})_{\mu=0}\,=\,\zeta_N(\lambda,\,a,\,b)\,\,\epsilon_N(\lambda,\,a,\,b),
\end{align}
where the product form reflects the fact that the Kitaev chain is given in terms of two uncoupled SSH-like chains, as illustrated in Fig. (\ref{figure: Kitaev model as coupled ssh chains.}). Even though the polynomials $\zeta_N$ and $\epsilon_N$ belong to different SSH-like chains, both obey a common recursion formula typical of Fibonacci polynomials \cite{Webb, Hoggatt, Oezvatan-2017}
\begin{align}\label{equation: recursion formula fibonacci polynomials}
	\zeta_{j+2}\,=\,\left[\lambda^2\,+\,a^2\,+\,b^2\right]\,\zeta_j\,-\,a^2 b^2\,\zeta_{j-2},
\end{align}
and differ only in their initial values
\begin{align}\label{equation: initial values of epsilon and zeta}
	\left(\begin{matrix}
	\zeta_{-1}		\\
	\zeta_{0}		\\
	\zeta_{1}		\\
	\zeta_{2}		\\
\end{matrix}\right)\,=\,\left(\begin{matrix}
	0					\\
	1					\\
	\lambda				\\
	\lambda^2\,+\,b^2	
\end{matrix}\right),\,\left(\begin{matrix}
	\epsilon_{-1}		\\
	\epsilon_{0}		\\
	\epsilon_{1}		\\
	\epsilon_{2}		\\
\end{matrix}\right)\,=\,\left(\begin{matrix}
	0					\\
	1					\\
	\lambda				\\
	\lambda^2\,+\,a^2	
\end{matrix}\right).
\end{align}
Fundamental properties of Fibonacci polynomials are summarized in appendix \ref{appendix: comments on Fibonacci and Tetrannaci polynomials}. The common sublattice structure of both chains sets the stage for a  relationship between $\zeta_j$ and $\epsilon_j$: The exchange of $a$'s and $b$'s enables us to pass from one to the other
\begin{align}\label{equation: zeta and epsilon, related by abc}
	\zeta_j(\lambda,\,a,\,b)\,=\,\epsilon_j(\lambda,\,b,\,a),\quad \forall j.
\end{align}  
Moreover, Eq.\,(\ref{equation: recursion formula fibonacci polynomials}) implies that a Kitaev chain with even number of sites $N$ is fundamentally different from the one with an odd number of sites. This property is a known feature of SSH chains \cite{Sirker}. The difference emerges since, according to Eqs. (\ref{appendix: closed form zeta_odd}), (\ref{appendix: closed form epsilon_odd}), it holds
\begin{align}\label{equation: odd index polynomials are equal}
	\zeta_\mathrm{odd}(\lambda,\,a,\,b)\,=\,\epsilon_\mathrm{odd}(\lambda,\,a,\,b),
\end{align}
because the number of $a$ and $b$ type bondings in both subchains is the same. This leads to twice degenerate eigenvalues. An equivalent relationship for even $N$ does not exist. The closed form for $\zeta_j$ and $\epsilon_j$, as well as their factorization, is derived in appendix \ref{appendix: spectrum zero mu}.  

The characteristic polynomial can be used to obtain the determinant of the Kitaev chain, here for $\mu=0$, because evaluating it at $\lambda=0$ leads to:
\begin{align*}
	P_{\lambda=0}(\mathcal{H}_\mathrm{KC})_{\mu=0}\,=\,\mathrm{det}\left(\mathcal{H}_\mathrm{KC}\right)_{\mu=0}.
\end{align*}
According to Eq. (\ref{equation: characteristic polynomial at zero mu}) we need only to know $\zeta_N$ and $\epsilon_N$ at $\lambda=0$. The closed form expression for $\zeta_j$ at $\lambda = 0$ reduces to
\begin{align}
	\zeta_j\left\vert_{\lambda=0} \right.\,=\,\left\{\begin{matrix}
	0, &\text{if } j  \text{ is odd}\\
	b^{\,j},&\text{else }
	\end{matrix}\right. ,
\end{align}
while $\epsilon_j \left\vert_{\lambda = 0}\right.$ follows from Eq. (\ref{equation: zeta and epsilon, related by abc}). We find that there are always zero energy eigenvalues for odd $N$, but not in general for even $N$, as it follows from
\begin{align}\label{equation: determinant, pure kitaev chain, N even, mu zero}
	\mathrm{det}\left(\mathcal{H}^{\mu=0}_\mathrm{KC}\right)\,=\,\left\{ \begin{matrix}
	0, & N \text{ odd}\\
	\left[\Delta^2-t^2\right]^N, & N \text{ even}
	\end{matrix}\right. .
\end{align} 
Additional features of the spectrum are discussed in the following.
\subsubsection{Odd $N$  }
The spectrum for odd $N$ is given by two contributions
\begin{align}
		\label{equation: spectrum: mu=0, odd N, zero modes}
	E^{\mu=0}_\pm\,&=\,0,\quad \text{(twofold)}\\
		\label{equation: spectrum: mu=0, odd N, k depending part}
	E^{\mu=0}_\pm(k_n )\,&=\,\pm\sqrt{4\Delta^2\,\sin^2(k_n d)\,+\,4t^2\,\cos^2(k_n d)},
\end{align}
where $k_n d=n\pi/(N+1)$ and $n$ runs from $1$ to $N$, except for $n=(N+1)/2$. This constraint on $n$ is a consequence of the Eqs. \eqref{equation: N odd, entries of the eigenstates}, \eqref{equation: function T odd } below which show that the boundary condition Eq. \eqref{equation: boundary condition for odd N ssh chain} cannot be satisfied for $kd=\pi/2$. Hence, no standing wave can be formed. \\
Each zero eigenvalue belongs to one chain. As discussed below, two decaying states are associated to Eq. (\ref{equation: spectrum: mu=0, odd N, zero modes}), whose wave functions are discussed in Sec. \ref{subsection: N odd eigenvectors}. These states are MZM.

\subsubsection{Even $N$}
\begin{figure}[ht]
	\centering
	\includegraphics[width = 1.0\columnwidth]{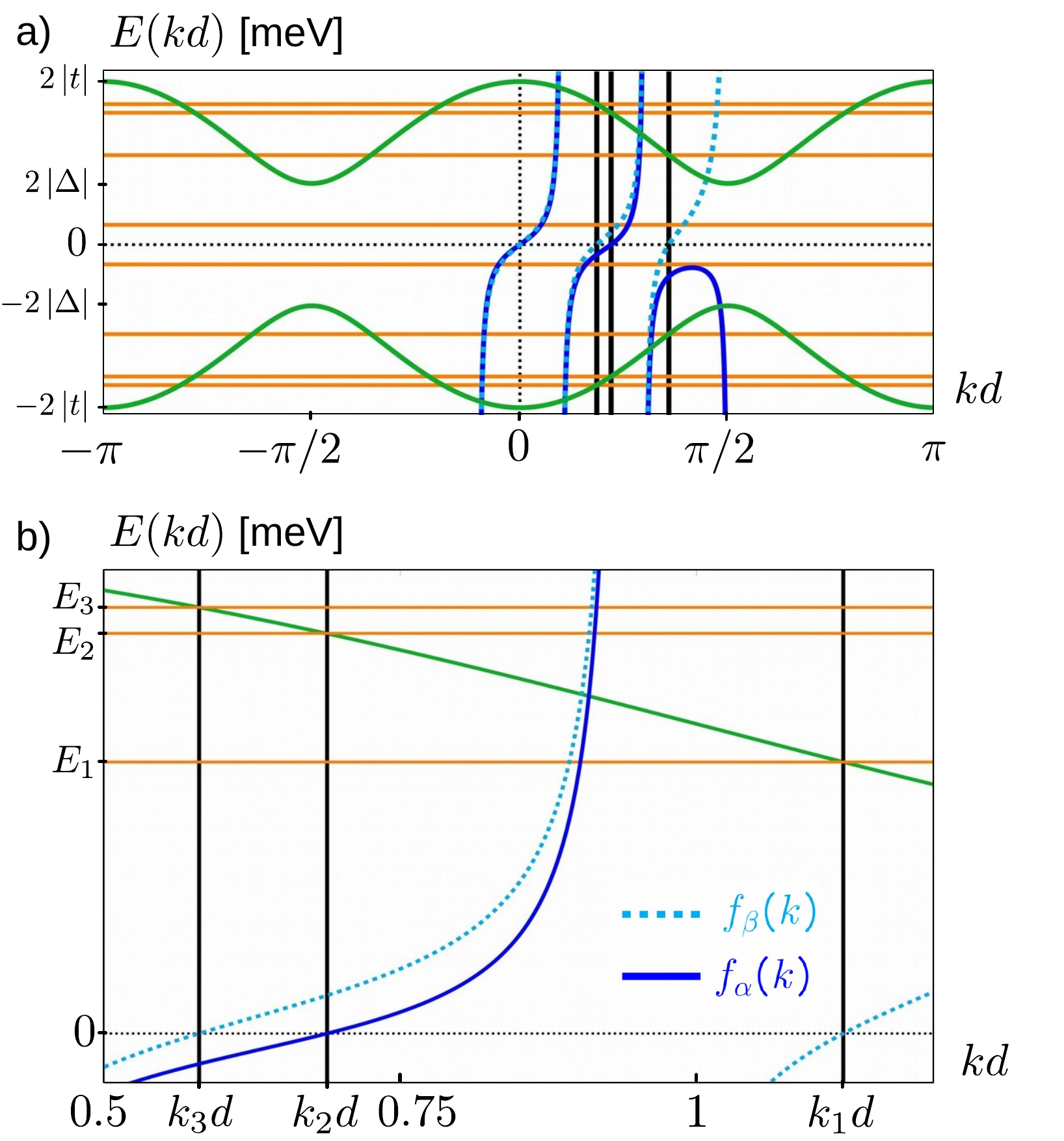}
	\caption{Eigenvalues and the non equidistant quantization of the bulk momentum $k$ for a Kitaev molecule with four sites. a) The horizontal lines mark the numerical eigenvalues $\pm E_{j}$ ($j=0,\,1,\,2,\,3$) and the bulk spectrum of the infinite chain is the green solid
 curve. The tangent-like functions follow $f_\beta(k)$ and $f_\alpha(k)$ from Eq. (\ref{Definition f_pm(q)}). The situation shown in a) is for $k\in\left[-\pi/d, \,\pi/d\right]$. The zeros of $f_{\beta,\alpha}(k)$ define the proper wave vectors $k_{3,\,2,\,1}$ of the finite system and these cut the dispersion relation at the correct positions, such that $E_\pm(k_j)=\pm E_{j}$. b) Zoom of a) for $k\in\left[0.5/d,\,1.2/d\right]$. The chosen parameters $N=4$, $t=4\,$meV, $\Delta = 1.5\,$ meV and $\mu = 0$ meV lead to the bulk eigenvalues $\pm E_{j}\in\left[\, \pm 4.39,\,\pm 6.47,\,\pm 6.89\right]$ (in meV) and to the momenta $k_{3,\,2,\,1}$ approximately $[0.58012/d, 0.68813/d, 1.12386/d]$.}
	 \label{figure: non equidistant bulk quantization for zero mu and N is four}
\end{figure}

In the situation of even $N$ we find for the Kitaev's bulk spectrum at zero $\mu$
\begin{align}\label{equation: spectrum: mu=0, even N}
	E^{\mu=0}_\pm(k)\,&=\,\pm\sqrt{4\Delta^2\,\sin^2(k d)\,+\,4t^2\,\cos^2(k d)},
\end{align}
where the momenta $k$ are in general {\it{not equidistant}} in the first Brillouin zone. Rather, the quantization condition follows from the interplay between $\Delta$ and $t$ and is captured in form of the functions $f_{\beta,\alpha}(k)$ (cf. appendix \ref{appendix: factorisation of the fibonnacis}),    
\begin{align}\label{Definition f_pm(q)}
	f_{\beta,\alpha}(k)\defl\tan\left[k d\left(N+1\right)\right]\,\pm\frac{\Delta}{t}\, \tan\left(k d \right) ,
\end{align}
whose zeros 
\begin{align}\label{equation: mu=0 quantization condition, N even,}
	f_{\beta,\alpha}(k)\,\stackrel{!}{=}\,0, \quad kd\neq0,\pi/2
\end{align}
define the allowed values of $k$. Note that $kd = 0, \pi/2$ are excluded as solutions, due to their trivial character. The functions $f_{\beta,\alpha}(k)$ follow from the factorisation of the polynomials $\epsilon_N$ and $\zeta_N$. 
 The negative sign in Eq. (\ref{Definition f_pm(q)}) belongs to the $\alpha$ subchain, while the positive one to the $\beta$ subchain. The spectrum following from Eq. (\ref{equation: mu=0 quantization condition, N even,}) is illustrated in Fig. \ref{figure: non equidistant bulk quantization for zero mu and N is four}.
%
%
%
%
%
%
%
%
%
%
%
We observe that Eqs. (\ref{equation: spectrum: mu=0, even N}) and (\ref{equation: mu=0 quantization condition, N even,}) hold for all values of $t$ and $\Delta$, independent of whether $\vert \Delta\vert$ is larger or smaller than $\vert t \vert$. The two situations are connected by a phase shift of the momentum $kd\rightarrow kd+\pi/2$, which influences both the spectrum and the quantization condition. In the end all different ratios of $\Delta$ and $t$ are captured by Eqs. (\ref{equation: spectrum: mu=0, even N}) and (\ref{equation: mu=0 quantization condition, N even,}), due to the periodicity of the spectrum.

\begin{figure}[ht]
	\centering
	\includegraphics[width = 1.0\columnwidth]{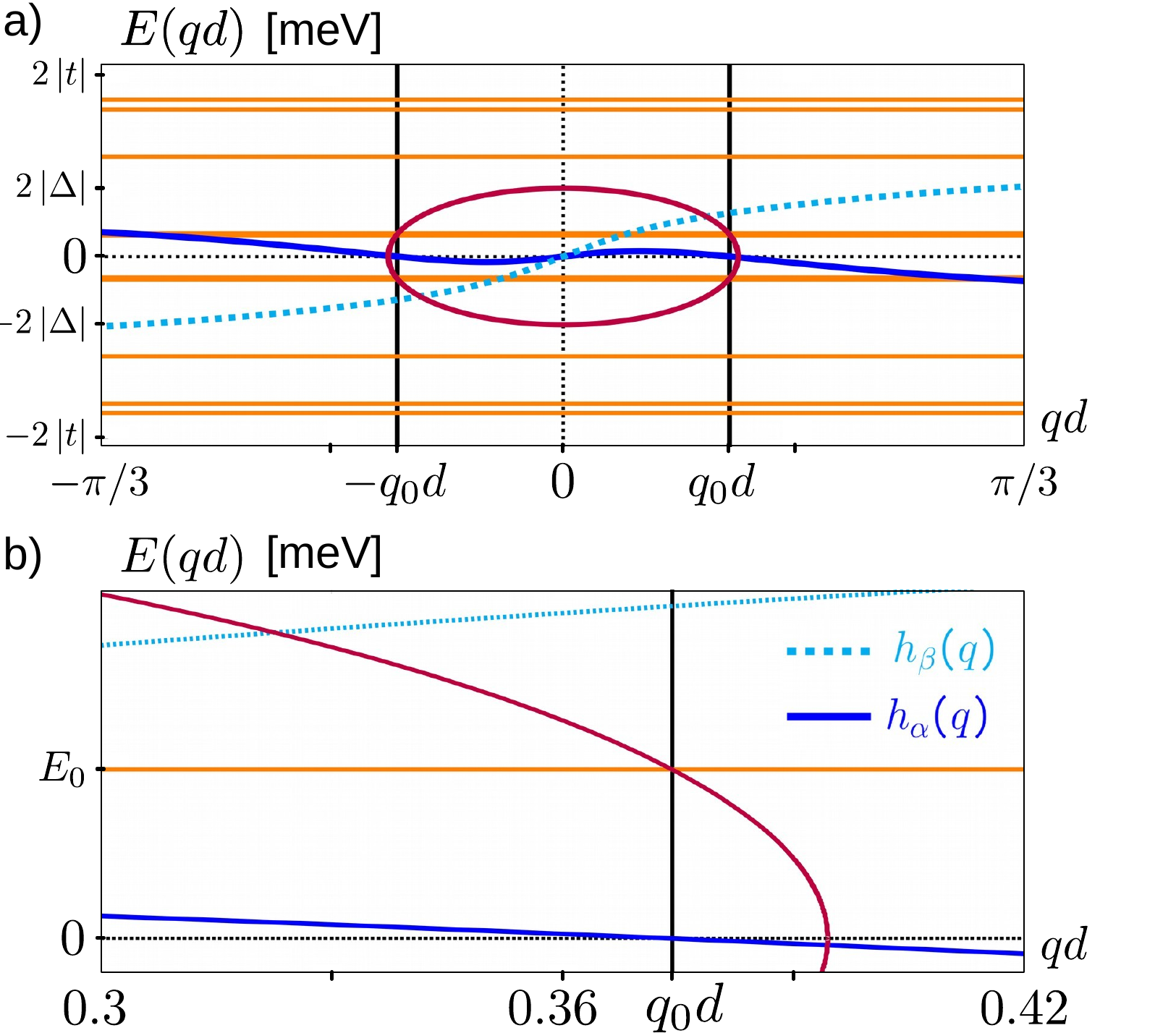}
	\caption{Eigenvalues and the quantised momentum $q_0$ of the gap modes for a Kitaev molecule with four sites. a) The horizontal lines characterise the numerical eigenvalues $\pm E_{j}$ ($j=0,\,1,\,2,\,3$) and the dispersion relation inside the gap is shown as function of continuous $q$ on a finite range. Only one of both hyperbolic tangent-like functions $h_\beta(q)$ or $h_\alpha(q)$ defines a proper $qd\neq 0$. The situation shown in a) is for $q\in\left[-\pi/3d, \,\pi/3d\right]$. b) Zoom of a) for $q\in\left[0.3/d,\,0.42/d\right]$. The momenta $\pm q_0$, the zero of $h_\alpha$, leads to the correct associated energies, such that $E_{\pm}(q_0)=\pm E_{0}$. The chosen parameters $N =4$, $t=4\,$meV, $\Delta = 1.5\,$ meV and $\mu = 0$ meV lead to the eigenvalues $E_{j,\pm}\in\left[\pm 0.97,\, \pm 4.39,\,\pm 6.47,\,\pm 6.89\right]$ (in meV) and to the momentum $q_0\approx 0.37416/d $.}
	 \label{figure: non equidistant decaying states quantization for zero mu and N is four}
\end{figure}

However, when we consider decaying or edge states this periodicity is lost (see Eqs. (\ref{equation: criteria/ number of solutions for edge states even N}) - (\ref{equation: spectrum: mu=0, even N, edges}) below) and $\vert t\vert\lessgtr \vert \Delta\vert$ lead to different quantization rules. 
The hermiticity of the Hamiltonian allows a pure imaginary momentum for $\mu=0$, but a simple exchange of $k$ to $i q$ in Eq. \eqref{Definition f_pm(q)} does not lead to the correct results. We introduce here the functions 
\begin{align}\label{definiton: h_pm(q)}
	h_{\beta,\alpha}(q)\defl\tanh\left[q d\left(N+1\right)\right]\,\pm\,m\,\tanh\left(q d\right),
\end{align}
similar to $f_{\beta,\alpha}(k)$ in Eq. (\ref{Definition f_pm(q)}), where $m$ contains both ratios of $\Delta$ and $t$:
\begin{align}\label{equation: criteria of the prop. factor m}
	m\,\defl \,\left\{	\begin{matrix}
	\frac{\Delta}{t},\quad\,\mathrm{if}\, \vert \Delta\vert \ge \vert t\vert	\\
	\\
	\frac{t}{\Delta},\quad\,\mathrm{if}\, \vert t\vert \ge \vert \Delta\vert	
	\end{matrix}		\right. .
\end{align}
Again, the positive sign in Eq. (\ref{definiton: h_pm(q)}) belongs to the $\beta$ chain and the negative one to the $\alpha$ chain. The exact quantization criterion is provided by the zeros of $h_{\beta,\alpha}(q)$,
\begin{align}\label{equation: criteria/ number of solutions for edge states even N}
	h_{\beta,\alpha}(q)\,\stackrel{!}{=}0, \quad q\neq 0,
\end{align}
as illustrated in Fig. \ref{figure: non equidistant decaying states quantization for zero mu and N is four}. The associated energies follow from the dispersion relation
\begin{align}\label{equation: spectrum: mu=0, even N, edges}
	E(q)=\pm\,\left\{\begin{matrix}
	\sqrt{4t^2\cosh^2(qd)-4\Delta^2\,\sinh^2(qd)}, \,\mathrm{if}\, \vert \Delta\vert \ge \vert t\vert	\\
	\\
	\sqrt{4\Delta^2\cosh^2(qd)-4t^2\,\sinh^2(qd)},\,\mathrm{if}\, \vert t	\vert \ge \vert \Delta \vert	
	\end{matrix}			\right..
\end{align}
%
%
We notice that Eq. (\ref{equation: spectrum: mu=0, even N, edges}) is only well defined for zero or positive arguments  of the square root. Indeed, all solutions of Eq. (\ref{equation: criteria/ number of solutions for edge states even N}), if existent, lie always inside this range, because using $h_{\beta,\alpha}(q) = 0$ in Eq. (\ref{equation: spectrum: mu=0, even N, edges}) yields
\begin{align}\label{equation: spectrum: mu=0, even N, edges, cosh/cosh version}
	E(q)\,=\,\pm\,2\,\frac{\cosh(qd)}{\cosh\left[qd\left(N+1\right)\right]}\,\cdot\,\mathrm{min}\left\{\vert \Delta\vert,~ \vert t \vert\right\}.
\end{align}
Hence, each wavevector from Eq. (\ref{equation: criteria/ number of solutions for edge states even N}) \makebox{corresponds} to two gap modes, since the gap width is $4\,\mathrm{min}\left\{\vert \Delta\vert,~ \vert t \vert\right\}$ and the fraction inside Eq. (\ref{equation: spectrum: mu=0, even N, edges, cosh/cosh version}) is always smaller than one. 
We can restrict ourselves to find only positive solutions $qd$, due to the time reversal symmetry. The number of physically different solutions 
of Eq. (\ref{equation: criteria/ number of solutions for edge states even N}) is zero or two and it follows always from the equation containing the positive factor $m$ or $-m$. Consequently, according to Eq. (\ref{definiton: h_pm(q)}), only zero or two gap modes can form and both belong to the same subchain, $\alpha$ or $\beta$. Moreover a solution exists if, and only if, $\vert m\vert \in\left[1,\,N+1\right]$.
%

In the limiting case when \makebox{$\vert m \vert\rightarrow 1$}, i.e. at the Kitaev points, the solution \makebox{$qd \rightarrow \infty$} and the associated energies $E_\pm$ from Eq. (\ref{equation: spectrum: mu=0, even N, edges, cosh/cosh version}) go to zero. The eigenstate will be a Majorana zero energy mode, see Sec. \ref{subsection: N even eigenvectors}.

In the second special case of $\vert m \vert\rightarrow N+1$ the solution approaches zero. The value $q=0$ is only in this particular scenario a proper momentum, see appendix \ref{appendix: factorisation of the fibonnacis}. 
%
%
%
%
The momentum $q = 0$ yields the energies $E_\pm(0) = \pm 2\,\mathrm{min}\left\{\vert \Delta\vert,~ \vert t \vert\right\}$, which mark exactly the gap boundaries. 

Increasing the value of $\vert m \vert$ beyond $N+1$ entails the absence of imaginary solutions. The number of eigenvalues of a Kitaev chain is still $2N$ for a fixed number of sites and consequently Eq. (\ref{equation: mu=0 quantization condition, N even,}) leads now to $N$ real values for $kd$, instead of $N-1$. In other words, the two former gap modes have moved to two extended states and their energy lies now within the bulk region of the spectrum, even though the system is still fully gaped. This effect holds for the Kitaev chain as well as for SSH chains. Physically this means, that a "boundary" mode with imaginary momentum $q$ and corresponding decay length $\xi\propto 1/q$ reached the highest possible delocalisation in the chain. 

The limit of $N\rightarrow\infty$ yields always two zero energy boundary modes; since the momentum is $qd=\text{arctanh}(\vert 1/ m \vert)$, due to Eqs. (\ref{definiton: h_pm(q)}) (\ref{equation: criteria/ number of solutions for edge states even N}) and according to Eq. (\ref{equation: spectrum: mu=0, even N, edges, cosh/cosh version}) the energy goes to zero. If we consider the odd $N$ situation in the limit of an infinite number of sites, we have there two zero energy boundary modes as well. The results of this section are summarized in table \ref{table: all quantisiation (sub) rules}.
\section{Eigenvectors ($\mu = 0$)}
\label{section: Eigenvectors and symmetries}
We use the SSH-like basis to calculate the eigenvectors of the Hamiltonian Eq. (\ref{equation: Kitaev Hamiltonian/ matrix in SSH basis}) at $\mu=0$. The eigenvectors $\vec{\psi}$ are defined with respect to the SSH-like chains $\alpha$ and $\beta$, see Eq. (\ref{equation: Kitaev Hamiltonian/ matrix in SSH basis}),
\begin{align}
	\vec{\psi}\,=\,\left(\begin{matrix}
	\vec{v}_\alpha\\
	\vec{v}_\beta
\end{matrix}\right),
\end{align}
with the feature that always either $\vec{v}_\beta$ or $\vec{v}_\alpha$ can be chosen to be zero, yielding the solutions $\vec{\psi}^\alpha$ and $\vec{\psi}^\beta$, respectively
\begin{align}\label{equation: eigenvector structure psi_alpha, psi_beta}
\vec{\psi}^\alpha \,=\,\left(\begin{matrix}
	\vec{v}_\alpha\\
	\vec{0}
\end{matrix}\right),\quad\vec{\psi}^{\beta} \,=\,\left(\begin{matrix}
	\vec{0}\\
	\vec{v}_\beta
\end{matrix}\right).
\end{align}
We are left to find the eigenvectors of a single tridiagonal matrix which we did basing on, and extending the results of Ref.~[\onlinecite{Shin-97}]. We focus here on the edge and decaying states, while the rest of our results are in appendix \ref{appendix: eigenvectors}. Remember that in the SSH-like basis Eq. (\ref{equation: ssh basis, alternating pattern}) the Majorana operators $\gamma_j^A$ and $\gamma_j^B$, alternate at each site, thus defining two interpenetrating "A" and "B" type sublattices.

\subsection{Even $N$}
\label{subsection: N even eigenvectors}
We define the vectors $\vec{v}_\alpha$ and $\vec{v}_\beta$ via the entries
\begin{align}
	\vec{v}_\alpha\,&=\,\left(x_1,\,y_1,\,x_2,\,y_2,\,\ldots ,\,x_{N/2},\,y_{N/2}\right)^\mathrm{T},\\
	\vec{v}_\beta\,&=\,\left(\xcal_1,\,\ycal_1,\,\xcal_2,\,\ycal_2,\,\ldots ,\,\xcal_{N/2},\,\ycal_{N/2}\right)^\mathrm{T},
\end{align}
where $x,\,y$ and $\xcal,\,\ycal$ are associated to the A and B sublattices, respectively. The internal structure of $\vec{v}_\alpha$ ($\vec{v}_\beta$) reflects the unit cell of an SSH-like chain and thus simplifies the calculation.
\begin{figure}[ht]
	\includegraphics[width = 1\columnwidth]{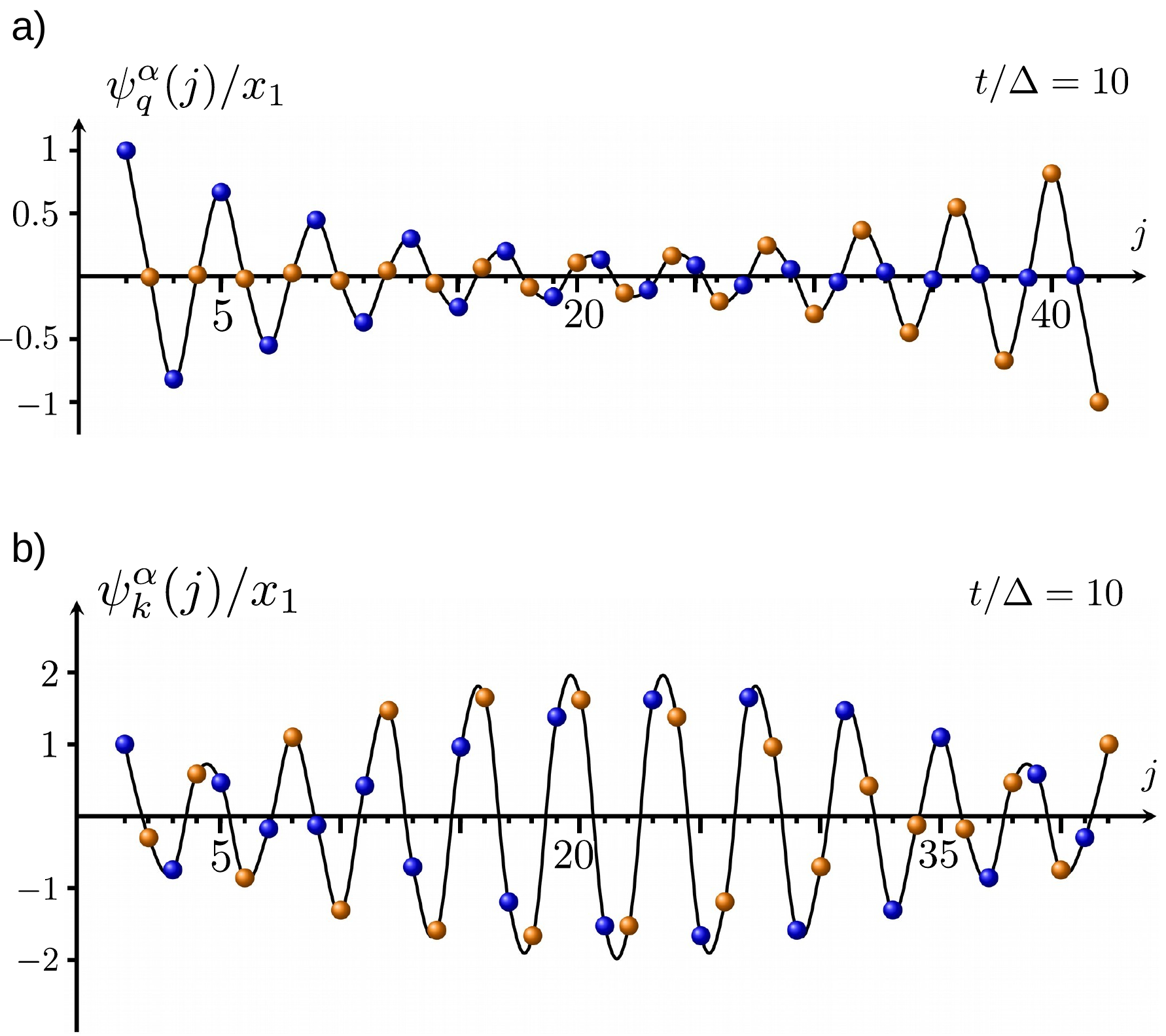}
	\caption{Visualisation of the entries of the eigenstates $\vec{\psi}^\alpha_{q,k} (j)$ of the Kitaev chain with $N =42$ sites and $\mu =0$. Panel a) depicts the gap state $\vec{\psi}^\alpha_q$ and b) the lowest energy bulk state $\vec{\psi}^\alpha_k$. The blue (orange) dots follow $x_l/x_1$ ($i\,y_l/x_1$) at position $j = 2l-1$ ($j= 2l$) for $l=1,\ldots, N/2$, while the black line is only a guide to the eye. The gap state is more localised at the edges. The extended state is largest inside the chain. The chosen parameters are $t=10$ meV and $\Delta = 1$ meV leading to $q = 0.10029/d$ and $E = 0.0539$ meV for the gap state. The shown extended state is associated with $k =1.4806/d$ and $E = 2.6851$ meV. Notice the decaying state in a) as well as the ones depicted in Figs. \ref{figure: wavefunctions N even, t=5} a) and \ref{figure: wavefunctions N even, t=-5, beta chain} are not Majorana states.
	}
	\label{figure: wavefunctions N even, t=10}
\end{figure}
In the real space  $x_l$ ($\xcal_l$) belongs to site $j = 2l-1$ and $y_l$ ($\ycal_l$) to $j=2l$, where $j=1,\,\ldots,\,N$.
\begin{figure}[h]
	\includegraphics[width = 1\columnwidth]{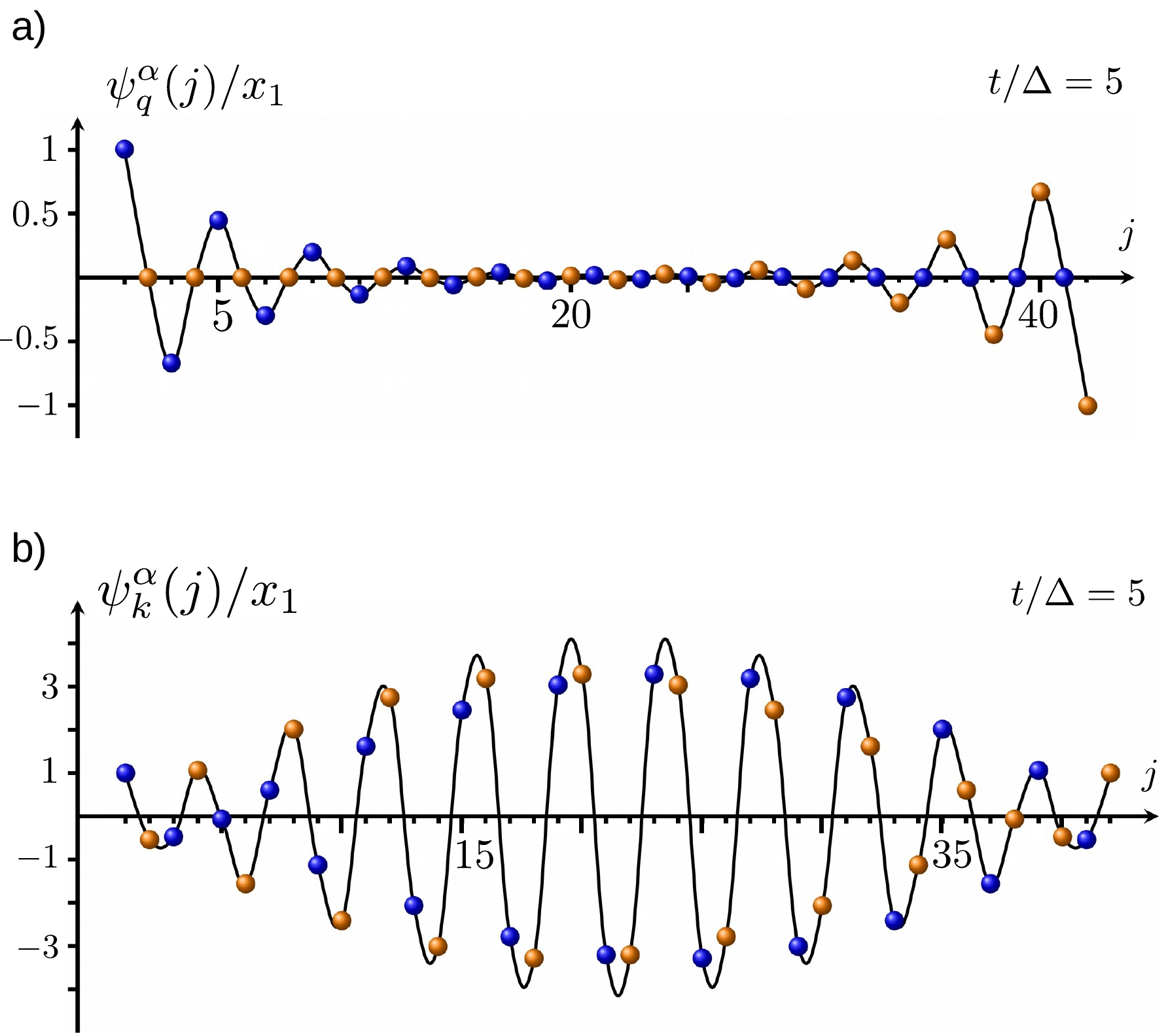}
	\caption{Illustration of the entries of the eigenstates $\vec{\psi}^\alpha_{q,k}(j) $ of a Kitaev chain for $N=42$ sites and $\mu=0$ for $t =5$ meV and $\Delta = 1$ meV. Similar to Fig. \ref{figure: wavefunctions N even, t=10}, but with a modified value for $t/\Delta$. a) shows the gap mode and b) the lowest in energy bulk mode. Notice that for the chosen parameter set the gap state is more localized than the one in Fig. \ref{figure: wavefunctions N even, t=10}. In contrast the extended state has lower weight at the ends of the chain. The gap mode (bulk state) is associated with $q = 0.2027/d$ ($k= 1.4886/d$) and $E =0.6682\cdot 10^{-3}$ meV ($E= 2.1555$ meV). }
	\label{figure: wavefunctions N even, t=5}
\end{figure}

Searching for solutions on the subchain $\alpha$ implies setting \makebox{$\vec{v}_\beta = \vec{0}$} and solving $\left(\mathcal{H}_\alpha^{\mathrm{even}  } -E_\pm \mathbbm{1}_N\right)\vec{v}_\alpha = \vec{0}$. The elements of $\vec{v}_\alpha$ obey
\begin{align}
	 \label{equation: open boundary left side}
	 \,a\,y_1\,&=\,E_\pm\,x_1,\\
	 \label{equation: open boundary right side}
	-a\,x_{N/2}\,&=\,E_\pm\,y_{N/2},
\end{align}
and
\begin{align}
	 \label{equation: recursive formula, entries v_alpha x}
	 b\,x_{l+1}\,-\,a\,x_l\,&=\,E_\pm\,y_l,\\
	 \label{equation: recursive formula, entries v_alpha y}
	 a\,y_{l+1}\,-\,b\,y_l\,&=\,E_\pm\,x_{l+1},
\end{align}
where $l$ runs from $1$ to $N/2 \,-\, 1$. The solution for $\Delta \neq \pm t$ is (in agreement with Ref.~[\onlinecite{Shin-97})] 
\begin{align}
	\label{equation: decaying state, even N, y_j}
	\frac{y_l}{x_1}\,&=\,\frac{E_\pm (\theta)}{a}\,T^\mathrm{e}_l(\theta),\\
	\label{equation: decaying state, even N, x_j}
	\frac{x_l}{x_1}\,&=\,T^\mathrm{e}_l(\theta)-\frac{b}{a} T^\mathrm{e}_{l-1}(\theta),
\end{align}
where $l = 1,\,\ldots,\,N/2$, $\theta$ denotes the momentum $k$ ($q$) for extended (gap) states and $E_\pm$ is the dispersion relation associated to $k$ (Eq. (\ref{equation: spectrum: mu=0, even N})), or $q$ (Eq. (\ref{equation: spectrum: mu=0, even N, edges})). 
\begin{figure}
		\includegraphics[width = 1\columnwidth]{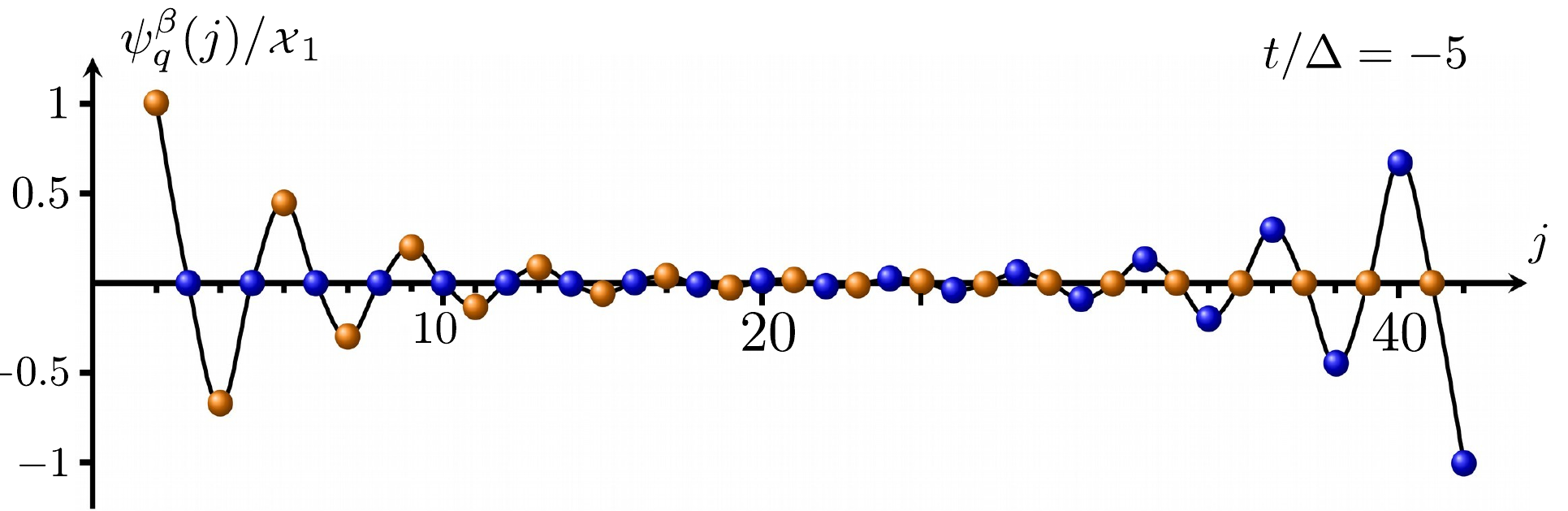}
	\caption{The decaying state $\vec{\psi}^\beta_q$ for $N=42$ sites and $\mu=0$. The black guiding line follows the orange (blue) dots, which correspond to $\xcal_l/\xcal_1$  ($i\,\ycal_l/\xcal_1$) at position $j=2l-1$ ($j= 2l$), $l = 1,\ldots, N/2$. The difference to the edge states on subchain $\alpha$ is the exchanged role of Majorana operators $\gamma_j^A$, $\gamma_j^B$. The chosen parameters are $t=-5$ meV and $\Delta = 1$ meV. The gap mode is associated with $q = 0.2027/d$ and $E =0.6682\cdot 10^{-3}$ meV .  }
	\label{figure: wavefunctions N even, t=-5, beta chain}
\end{figure}
The entries of the eigenvectors are essentially sine functions for the extended states
\begin{align}\label{equation: T even, bulk states}
	T^\mathrm{e}_l (k)\defl\,\frac{\sin(2\,kd\,l)}{\sin(2\,kd)},
\end{align}
and hyperbolic sine functions for the decaying states
\begin{align}\label{equation: T even, gap modes}
	T^\mathrm{e}_l (q)\defl s^{l-1}\,\,\,\frac{\sinh(2\,qd\,l)}{\sinh(2\,qd)},
\end{align}
where the prefactor $s$ depends on the ratio of $\Delta$ and $t$: 
\begin{align*}
	s = \left\{\begin{matrix}
	+1, & \vert \Delta\vert > \vert t\vert\\
	\\
	-1, & \vert t\vert > \vert \Delta\vert
	\end{matrix}\right. .
\end{align*}
An illustration of $\vec{\psi}^\alpha$ is given in Fig. \ref{figure: wavefunctions N even, t=10}. The allowed momenta $k$ or $q$ follow from the open boundary conditions
\begin{align}
	\label{equation: boundary condition y_0, x_N/2+1 has to vanish}
	y_0\,=\,x_{\frac{N}{2}+1}\,=\,0.
\end{align}
The first condition is satisfied due to $T_0(\theta)=0$ for any momentum. The second condition yields the quantization rules $f_\alpha(k)=0$ and $h_\alpha(q)=0$ for the $\alpha$ chain, see Eqs. (\ref{equation: mu=0 quantization condition, N even,}), (\ref{equation: criteria/ number of solutions for edge states even N}).

The eigenvector $\vec{\psi}^\beta$ entails $\vec{v}_\alpha=0$ and the entries of $\vec{v}_\beta$ follow essentially by replacing $a$'s
and $b$'s in the Eqs. (\ref{equation: decaying state, even N, y_j}), (\ref{equation: decaying state, even N, x_j}). We find
\begin{align}\label{equation: decaying state on chain beta, even N, ycal_j}
	\frac{\ycal_l}{\xcal_1}\,&=\,\frac{E_\pm }{b}\,T^\mathrm{e}_l(\theta),\\
	\label{equation: decaying state on chain beta, even N, xcal_j}
	\frac{\xcal_l}{\xcal_1}\,&=\,T^\mathrm{e}_l(\theta)-\frac{a}{b} T^\mathrm{e}_{l-1}(\theta),
\end{align}
where $j = 1,\,\ldots,\,N/2$ and $\Delta \neq \pm t$. The quantisation condition follows from the open boundary condition:
\begin{align*}
	\ycal_0\,&=\,0,\quad \xcal_{\frac{N}{2}+1}\,=\,0,
\end{align*}
and $k$ ($q$) obey $f_\beta(k)=0$ ($h_\beta(q)=0$). Further, from the quantization rules it follows that gap modes belong always to the same subchain $\alpha$ or $\beta$ for even $N$.

As illustrated in Figs. \ref{figure: wavefunctions N even, t=10}, \ref{figure: wavefunctions N even, t=5}, \ref{figure: wavefunctions N even, t=-5, beta chain} our states are symmetric w.r.t. the center of the SSH-like chains. This symmetry is visible in alternative versions of the Eqs. \eqref{equation: decaying state, even N, x_j}, \eqref{equation: decaying state on chain beta, even N, xcal_j} ($l=1,\ldots,N/2$), whereby
%
%
$	\xcal_{\frac{N}{2}+1-l}\,=\,T_l^\mathrm{e}(\theta)\,\xcal_{\frac{N}{2}}$,
%
%
 which holds for all eigenstates. Together with Eqs. \eqref{equation: decaying state, even N, y_j},  we find in general
\begin{align}\label{equation: inversion symmetry of the even N ssh like chain for mu=0}
	x_{\frac{N}{2}+1-l}=y_{l}\,\frac{x_{N/2}}{y_1}
\end{align}
and similarly $\xcal_{\frac{N}{2}+1-l}=\ycal_{l}\,\xcal_{\frac{N}{2}}/\ycal_1$. Recalling the definition of the SSH-like basis, Eq. \eqref{equation: ssh basis, alternating pattern}, and introducing the operators $\psi_{\alpha,\beta}^\dagger$ associated to the states $\vec{\psi}^{\alpha,\beta}$ in Eq. \eqref{equation: eigenvector structure psi_alpha, psi_beta}, we find the expression
\begin{align}\label{equation: operator alpha N even}
	\psi_{\alpha}^\dagger\,=\,\frac{1}{v_\alpha}\left[\sum\limits_{j=1}^{N/2}\,x_j\,\left(\gamma_{2j-1}^A\right)^\dagger\,+\,\sum\limits_{j=1}^{N/2}\,y_j\,\left(\gamma_{2j}^B\right)^\dagger\right],
\end{align}
where $v_\alpha$ is the norm of the vector $\vec{v}_\alpha$. A similar term is found for $\psi_{\beta}^\dagger$. 
We notice that Eq. \eqref{equation: operator alpha N even} and Eq. \eqref{equation: statistics of the even N eigenstates} below are true for all kinds of eigenstates, i.e. extended, decaying states and MZM, of the BdG Hamiltonian in Eq. \eqref{equation: Kitaev Hamiltonian/ matrix in SSH basis} at $\mu=0$. The character (statistics) of the operators depends on whether $(\psi_{\alpha,\beta}^\dagger)^2$ is $0$ or $1/2$. The property $\{\gamma^r_j,\gamma^s_k\}=\delta_{j,k}\delta_{r,s}$ with ($r,s \in \{A,B\}$) yields
\begin{align}\label{equation: statistics of the even N eigenstates}
	\left(\psi_{\alpha}^\dagger\right)^2\,=\,\frac{1}{2\,v_\alpha^2}\sum\limits_{j=1}^{N/2}\,\left(x_j^2\,+\,y_j^2\right).
\end{align}
The symmetry in Eqs. \eqref{equation: inversion symmetry of the even N ssh like chain for mu=0} states that $(\psi_{\alpha}^\dagger)^2$ is essentially determined by $(x_{N/2}/y_1)^2$. For $a\neq0$ and consequently $E\neq 0$, we find from Eqs. \eqref{equation: open boundary left side}, \eqref{equation: open boundary right side} and \eqref{equation: inversion symmetry of the even N ssh like chain for mu=0} that $(x_{N/2}/y_1)^2 = -1$, which yields $(\psi_{\alpha}^\dagger)^2 =0$. Thus the operators associated to the finite energy states $\vec{\psi}^\alpha$, including the ones depicted in Figs. \ref{figure: wavefunctions N even, t=10} a) and \ref{figure: wavefunctions N even, t=5} a), obey fermionic statistics. This result holds also true in the case $a=0$ and $E\neq 0$ as can be seen by using the corresponding eigenstates (appendix \ref{appendix: eigenvectors}). Similar results hold for $(\psi_{\beta}^\dagger)^2$.

We turn now to Majorana zero modes, which at $\mu=0$ only exist at the Kitaev points $\Delta = \pm t$.

When $\Delta = t$ we find two zero energy modes $\vec{\psi}_A^\alpha = \left(\begin{matrix}
\vec{v}_{\alpha, A}\\ \vec{0}\end{matrix}\right)$, $\vec{\psi}_B^\alpha = \left(\begin{matrix}
\vec{v}_{\alpha, B}\\ \vec{0}\end{matrix}\right)$ each localised at one end of the $\alpha$ chain:
\begin{align}\label{equation: Delta = t, N even}
	\vec{v}_{\alpha,\,\textcolor{blue}{A}}\,&=\,\left(1,\,0,\,0,\,\ldots,\,0\right)^\mathrm{T},\\
	\vec{v}_{\alpha,\,\textcolor{orange}{B}}\,&=\,\left(0,\,0,\,\ldots,\,0,\,1\right)^\mathrm{T},
\end{align}
and $(\psi_\alpha^\dagger)^2 = 1/2$ in Eq. \eqref{equation: statistics of the even N eigenstates}. In contrast, both zero energy modes are on the $\beta$ chain for \makebox{$\Delta = -t$}. We find $\vec{\psi}_A^\beta = \left(\begin{matrix} \vec{0}\\
\vec{v}_{\beta, A} \end{matrix}\right)$, $\vec{\psi}_B^\beta = \left(\begin{matrix}\vec{0}\\
\vec{v}_{\beta, B} \end{matrix}\right)$ with
\begin{align}
	\vec{v}_{\beta,\,\textcolor{orange}{B}}\,&=\,\left(1,\,0,\,0,\,\ldots,\,0\right)^\mathrm{T},\\
	\label{equation: Delta = -t, N even}
	\vec{v}_{\beta,\,\textcolor{blue}{A}}\,&=\,\left(0,\,0,\,\ldots,\,0,\,1\right)^\mathrm{T}.
\end{align}
These states are the archetypal Majorana zero modes\cite{kitaev:physusp2001, Aguado}. Due to their degeneracy, these modes can be recombined into fermionic quasiparticles by appropriate linear combination, see in Eq. \eqref{equation: Missing electron a}, \eqref{equation: Missing electron b} from Sec. \ref{section 2 model}. 
\subsection{Odd $N$ }
\label{subsection: N odd eigenvectors}
The composition of the eigenvectors slightly changes for the odd case compared to the even $N$ case
\begin{align}
	\vec{v}_\alpha\,&=\,\left(x_1,\,y_1,\,x_2,\,y_2,\,\ldots ,\,x_{\frac{N-1}{2}},\,y_{\frac{N-1}{2}},\,x_{\frac{N+1}{2}}\right)^\mathrm{T},\\
	\vec{v}_\beta\,&=\,\left(\xcal_1,\,\ycal_1,\,\xcal_2,\,\ycal_2,\,\ldots ,\,\xcal_{\frac{N-1}{2}},\,\ycal_{\frac{N-1}{2}},\,\xcal_{\frac{N+1}{2}}\right)^\mathrm{T}.
\end{align}
Although both odd sized chains share the same spectrum, it is possible to find a linear combination of states which belongs to one chain only. The form of the extended states of the odd chains $(\Delta\,\neq\,\pm t$ \emph{and} $E_\pm \neq 0$) does not differ much from the one of the even chain and the entries of $\vec{v}_\alpha$ are
\begin{align}\label{equation: N odd, entries of the eigenstates}
	\frac{y_l}{x_1}\,&=\,\frac{E_\pm(k_n)}{a}\,T^\mathrm{o}_l(k_n ),\\
	\label{equation: N odd, entries of the eigenstates, x_l}
	\frac{x_l}{x_1}\,&=\,T^\mathrm{o}_l(k_n )\,-\,\frac{b}{a}\,T^\mathrm{o}_{l-1}(k_n ),
\end{align}
where $T^\mathrm{o}_l$ is 
\begin{align}\label{equation: function T odd }
	T^\mathrm{o}_l (k_n )\defl \frac{\sin(2\,k_n d\,l)}{\sin(2\,k_n d)},
\end{align}
with $k_n d\,=\,n\pi/(N+1)$ ($n=1,\,\ldots,\,N$, $n\neq (N+1)/2$). The exchange of $a$'s and $b$'s leads again to the coefficients for the chain $\beta$ (see appendix \ref{appendix: eigenvectors}). 

The significant difference between even and odd $N$ lies in the realization of the open boundary condition. Solving $\left(\mathcal{H}_\alpha^{\mathrm{odd}  } -E_\pm \mathbbm{1}_N\right)\vec{v}_\alpha = \vec{0}$ yields now
\begin{align}\label{equation: boundary condition for odd N ssh chain}
	y_0\,=\,0,\quad 	y_{\frac{N+1}{2}}\,=\,0,
\end{align}
which leads to the momenta $k_n$. 

An SSH-like chain with an odd number of sites hosts only a single zero energy mode, but $\alpha$ and $\beta$ contribute each with one. 
We find on subchain $\alpha$ for $\Delta\neq\pm t$ 
\begin{align}\label{equation: zero mode on alpha N odd}
	y_l\,=\,0,\quad 	x_l\,=\,\left(\frac{\Delta-t}{\Delta+t}\right)^{l-1}\,x_1,
\end{align}
and on subchain $\beta$
\begin{align}\label{equation: zero mode on beta N odd} 
	\ycal_l\,=\,0,\quad 
	\xcal_l\,=\,\left(\frac{\Delta+t}{\Delta-t}\right)^{l-1}\,\xcal_1,
\end{align}
where $l$ runs from $1$ to $(N+1)/2$. 

Regarding the statistics of the operators $\psi_{\alpha}^\dagger,\,\psi_{ \beta}^\dagger$ associated to the states $\vec{\psi}^{\alpha},\, \vec{\psi}^{\beta}$, we proceed like for the even $N$ case. The use of the SSH-like basis from Eq. \eqref{equation: ssh basis, alternating pattern} and the entries of the state $\vec{\psi}^\alpha$ yield now
\begin{align*}
	\psi_{\alpha}^\dagger\,=\,\frac{1}{v_\alpha}\left[\sum\limits_{j=1}^{\frac{N+1}{2}}\,x_j\,\left(\gamma_{2j-1}^A\right)^\dagger\,+\,\sum\limits_{j=1}^{\frac{N-1}{2}}\,y_j\,\left(\gamma_{2j}^B\right)^\dagger\right].
\end{align*}
Again, the Eqs. \eqref{equation: N odd, entries of the eigenstates}, \eqref{equation: N odd, entries of the eigenstates, x_l} and  \eqref{equation: boundary condition for odd N ssh chain} imply a perfect compensation of the $A$ and $B$ sublattice contributions, yielding $(\psi_{\alpha}^\dagger)^2=0$ for $E\ne0$. The zero energy mode, given by its entries in Eq. \eqref{equation: zero mode on alpha N odd}, leads to $(\psi_{\alpha}^\dagger)^2=1/2$.

Further, we find that both zero energy modes $\vec{\psi}^{\alpha, \beta}$ have their maximum at opposite ends of the Kitaev chain and decay into the chain. To better visualize this it is convenient to introduce the decay length
\begin{align}\label{equation: decay length}
	\xi= 2d \left\{ \begin{array}{cc}
	                \left\vert \ln\left(\frac{t-\Delta}{t+\Delta}\right)\right\vert^{-1} & \vert t\vert \geq \vert \Delta \vert, \\[2mm]
	                \left\vert \ln\left(\frac{\Delta-t}{\Delta+t}\right)\right\vert^{-1} & \vert t\vert \leq \vert \Delta \vert,
	                \end{array}\right.
	                \end{align}
and remembering that the atomic site index of $x_l$ is $j=2l-1$ Eq. (\ref{equation: zero mode on alpha N odd}) yields for $\vert t\vert \geq \vert \Delta \vert$
\begin{align}
	x_l  \,&=\, x_1\, (-1)^{l-1}\, e^{-2(l-1)d/\xi},\notag\\
	&=\,x_1\, (-1)^{l-1}\, e^{-(j-1)d/\xi}.
\end{align}
For $\vert t\vert \leq \vert \Delta \vert$ the $x_l$ coefficients are given by the same equation without the $(-1)^{l-1}$ factor. We have moreover $q=\pm1/\xi$, where $q$ is the imaginary momentum yielding $E=0$ in Eq.~\eqref{equation: spectrum: mu=0, even N, edges}.
Thus the localisation of these states is determined only by $t$ and $\Delta$. In the parameter setting of $\Delta =t$ we find:
\begin{align}
	\label{equation: zero mode on alpha, odd N, Kitaev point plus}
	\vec{v}_{\alpha,\,\textcolor{blue}{A}}\,&=\,\left(1,\,0,\,0,\,\ldots,\,0\right)^\mathrm{T},\\
	\vec{v}_{\beta,\,\textcolor{orange}{B}}\,&=\,\left(0,\,0,\,\ldots,\,0,\,1\right)^\mathrm{T},
\end{align}
while both states exchange their position for \makebox{$\Delta = -t$}
\begin{align}
	\vec{v}_{\alpha,\,\textcolor{orange}{B}}\,&=\,\left(1,\,0,\,0,\,\ldots,\,0\right)^\mathrm{T},\\
	\label{equation: zero mode on beta, odd N, Kitaev point minus}
	\vec{v}_{\beta,\,\textcolor{blue}{A}}\,&=\,\left(0,\,0,\,\ldots,\,0,\,1\right)^\mathrm{T}.
\end{align}
\subsection{The particle-hole-operator}
\label{subsection: three symmetries}
In the last section we have shown that some of the zero energy eigenstates of the BdG Hamiltonian of the finite Kitaev chain are Majorana zero modes (MZM) by exploiting the statistics of the corresponding operators $\psi_{\alpha,\beta}$. We further corroborate this statement now by recalling that a MZM is defined as an eigenstate of the Hamiltonian $\mathcal{H}$ and of the particle hole symmetry $\mathcal{P}$. The latter acts on an eigenstate $\vec{\psi}^{\alpha, \beta}$ of energy $E$ by turning it into an eigenstate of $\mathcal{H}$ of energy $-E$. Thus, the energy of such an exotic state has to be zero, since eigenstates associated to different energies are orthogonal.

 The three symmetries, time reversal, chiral and the particle-hole symmetry, discussed in Sec. \ref{subsection: Topological phase diagram}, can be constructed in real space too. Of particular interest is their representation in the SSH-like basis. The antiunitary particle-hole symmetry is  
\begin{align}\label{equation: particle hole symmetry in real space in ssh basis}
	\mathcal{P}\,=\,\mathcal{K}\,\mathbbm{1}.
\end{align}
The time reversal and the chiral symmetry depend on $N$. If $N$ is even we find
\begin{align}
	\label{equation: chiral symmetry even N}
	\mathcal{C}^{\,\mathrm{even}}\,&=\,\left[\begin{matrix}
	\mathbbm{1}_{N/2}\otimes \tau_z & \\
	& -\mathbbm{1}_{N/2}\otimes \tau_z
	\end{matrix}
	\right],\\
	\mathcal{T}^{\,\mathrm{even}}\,&=\,\left[\begin{matrix}
	\mathbbm{1}_{N/2}\otimes \tau_z & \\
	& -\mathbbm{1}_{N/2}\otimes \tau_z
	\end{matrix}
	\right]\,\mathcal{K}.
\end{align}
The expressions for odd $N$ follow by removing the last line and last column in each diagonal block. 

The effect of $\mathcal{P}$ from Eq. (\ref{equation: particle hole symmetry in real space in ssh basis}) can be seen explicitly if one considers $\mathcal{P}\,\vec{\psi}^{\alpha}$. For $N$ even and $\Delta \neq t$
 the elements $x_l$, $y_l$ of $\vec{\psi}^{\alpha}$ are given in Eqs. (\ref{equation: decaying state, even N, y_j}), (\ref{equation: decaying state, even N, x_j}).
Here $y_l/x_1$ is pure imaginary and $x_l/x_1$ is real. Hence, $\vec{\psi}^\alpha$ is not an eigenstate of $\mathcal{P}$ since the prefactor to $y_l$ is finite, i.e., $E_\pm \neq 0$. We conclude that in a finite Kitaev chain with even number of sites Majorana zero modes emerge only at the Kitaev points for $\mu=0$, since the states in Eqs. (\ref{equation: Delta = t, N even})-(\ref{equation: Delta = -t, N even}) are eigenstates of $\mathcal{P}$ as well. In the situation of odd $N$ and $\mu=0$, the eigenstates given by their elements in Eqs. (\ref{equation: zero mode on alpha N odd}), (\ref{equation: zero mode on beta N odd}) are Majorana zero energy modes for an appropriate choice of $x_1$. These states can be delocalised over the entire chain, depending on their decay length $\xi$, while the case of full localisation is only reached at the the Kitaev points, where the MZM turn into the states given by Eqs. (\ref{equation: zero mode on alpha, odd N, Kitaev point plus})-(\ref{equation: zero mode on beta, odd N, Kitaev point minus}). 
\section{Results for the spectrum and eigenstates at finite $\mu$}
\label{section: influence of non zero chemical potential}
\subsection{Spectrum}
The last missing situation is to consider a finite chemical potential $\mu$. 
For this purpose we use  the so called chiral basis $\hat{\Psi}_c\defl\left(\textcolor{blue}{\gamma_1^A},\,\textcolor{blue}{\gamma_2^A},\,\ldots,\,\textcolor{blue}{\gamma_N^A},\,\textcolor{orange}{\gamma_1^B},\,\textcolor{orange}{\gamma_2^B},\,\ldots,\,\textcolor{orange}{\gamma_N^B}\right)^\mathrm{T}$. The Kitaev Hamiltonian transforms via $\hat{H}_\mathrm{KC}\,=\,\frac{1}{2}\,\hat{\Psi}^\dagger_c\,\mathcal{H}_c\,\hat{\Psi}_c,$
%
%
%
%
into a block off-diagonal matrix
\begin{align}\label{equation: Kitaev Hamiltonian in chiral basis}
	\mathcal{H}_c\,=\,\left[\begin{matrix}
	0_{N\times N} & h\\
	h^\dagger & 0_{N\times N}
\end{matrix}			\right],
\end{align}
because there are no $\gamma_j^A\,\gamma_i^A$ ($\gamma_j^B\,\gamma_i^B$) contributions in Eq. (\ref{equation: Kitaev Hamilt. in SSH basis, even N}). The $N \times N$ matrix $h$ is tridiagonal
\begin{align}\label{equation: matrix h in chiral basis}
	h=\left[\begin{matrix}
		-i\mu & a & \\
		-b & -i \mu & a\\
		& -b & -i \mu & a\\
		& & \ddots & \ddots & \ddots \\
		&&& -b & -i \mu & a\\
		&&&& -b & -i \mu 
	\end{matrix}\right],
\end{align}
since the Kitaev Hamiltonian contains only nearest neighbour hoppings. Then the characteristic polynomial is \cite{Silvester}
\begin{align}\label{equation: characteristic polynomial via chiral basis}
	P_\lambda\left(\mathcal{H}_c\right)\,=\,\text{det}\left(\lambda^2\,\mathbbm{1}_N\,-\,h\,h^\dagger\right),
\end{align}
where, however, $h$ and $h^\dagger$ do not commute except for $t=0$ or $\Delta=0$. Thus, such matrices cannot be diagonalised simultaneously. Nevertheless the eigenvalues $\eta_j$ ($\eta_j^*$) of $h$ ($h^\dagger$) are easily derived e.g. following Ref.~[\onlinecite{Kouachi}]. We find
\begin{align}\label{equation: eigenvalues of h,h^dagger}
	\eta_j\,=\,-i\mu\,+\,2\,\sqrt{\Delta^2-t^2}\,\cos\left(\frac{j\pi}{N+1}\right),\quad j\,=\,1,\,\ldots,\,N,
\end{align}
independent of whether $\Delta\ge t$ or $t>\Delta$. 
\subsubsection{Condition for zero energy modes}
Eq. (\ref{equation: eigenvalues of h,h^dagger}) immediately yields the criterion for hosting zero energy modes. According to Eq. (\ref{equation: Kitaev Hamiltonian in chiral basis}), we  have
	\begin{align}\label{equation: determinant of the Kitaev chain in terms of h and h*}
	\mathrm{det}(\mathcal{H}_c)\,=\,\mathrm{det}(h)\,\mathrm{det}(h^\dagger)\,=\, \vert \mathrm{det}(h)\vert^2,
	\end{align}
	and we need only to focus on $\mathrm{det}(h)$. 
	%
	%
	If a single eigenvalue $\eta_j$ of $h$ is zero then  $\mathrm{det}(h)$ vanishes. Thus, for a zero energy mode the chemical potential must satisfy
	\begin{align}\label{equation: eigenvalue zero condition kouachi}
	\mu_n\,\,=\,2\,\sqrt{t^2-\Delta^2}\,\cos\left(\frac{n\pi}{N+1}\right),\quad n = 1,\,\ldots,\,N .
	\end{align}
 Obviously, Eq. (\ref{equation: eigenvalue zero condition kouachi}) cannot be satisfied for generic values of $t^2 -\Delta^2$, because all other quantities are real. The only possibility is $t^2\ge \Delta^2$. There is only one exception for odd $N$, because the value $n = (N+1)/2$ leads to $\mu = 0$ in Eq. (\ref{equation: eigenvalue zero condition kouachi}) for all values of $t$ and $\Delta$, in agreement with our results of section \ref{section: spectral analysis}. This result is exact and confirms findings from Ref.~[\onlinecite{Kao-2014}, \onlinecite{Hegde-2015}]; further it improves a similar but approximate condition on the chemical potential discussed by Zvyagin in Ref.~[\onlinecite{Zvyagin-2015}].
	\begin{figure}[th]
	\centering
		\includegraphics[width=\columnwidth]{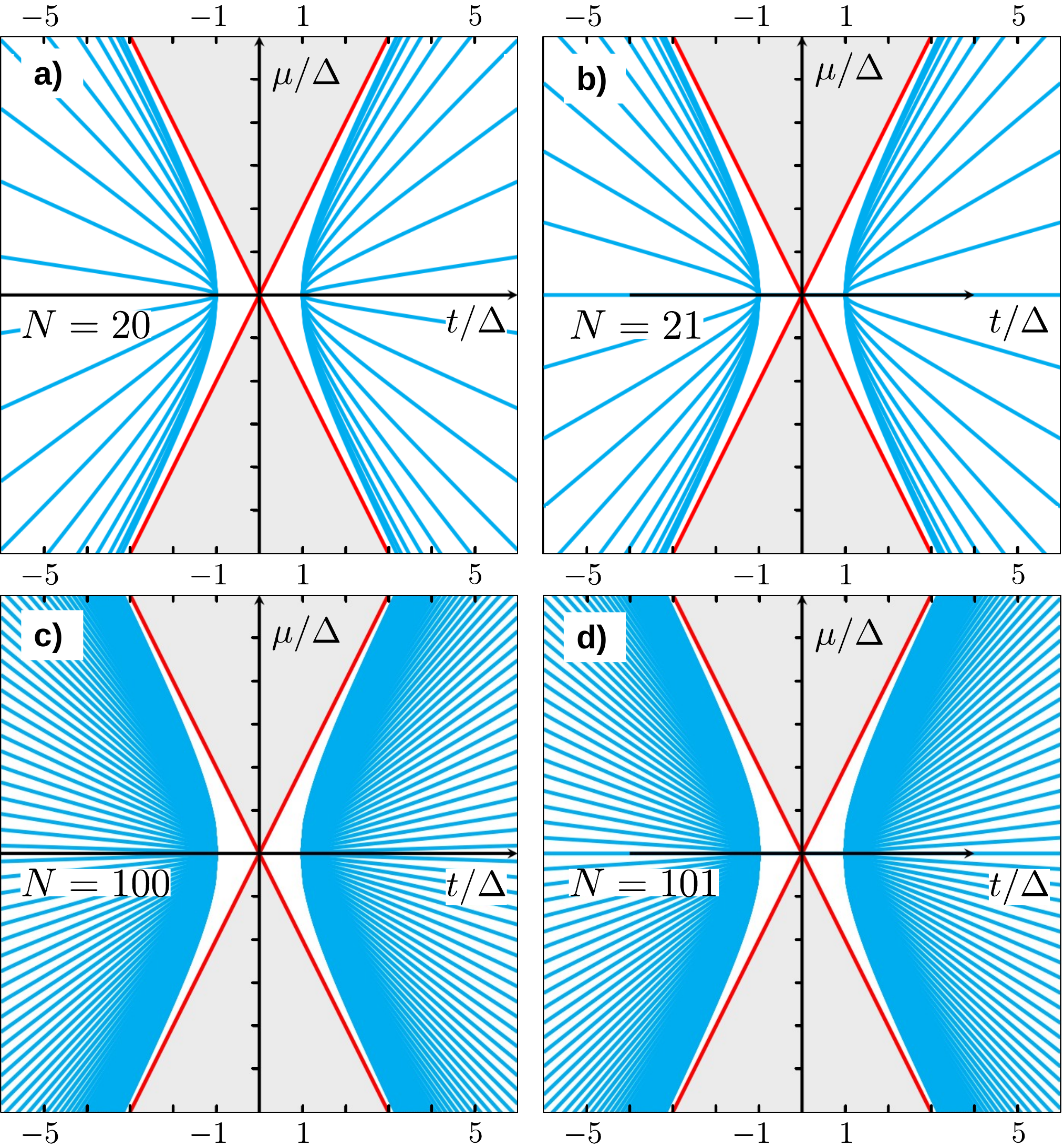}
		\caption{Existence of zero energy solutions for growing system sizes. The red lines mark the boundaries of the topological phase diagram. Zero energy solutions correspond to the blue curves, along which the determinant of the Kitaev Hamiltonian vanishes as a function of $\mu$, $t$ and $\Delta$. a) The shown situation is for a small even $N$, yielding a zero determinant on the horizontal axis only at the Kitaev points $t/\Delta = \pm 1$. Each blue curve departs from one of these two points. b) The situation of small odd $N$ is similar to the even one, but the entire $\mu=0$-axis is now included. c) The solutions $\mu_n$ become dense for larger even $N$, but one sees already the filling of the non trivial phase for $N \rightarrow \infty$. d) Large even $N$ behave similar to large odd $N$, but the latter still include the entire horizontal axis.
	}
		\label{figure: finite size effects and the topological phase diagramm}
	\end{figure}
	An illustration of these discrete solutions $\mu_n$, which we dub "Majorana lines", is shown in Fig. \ref{figure: finite size effects and the topological phase diagramm}. All paths contain the Kitaev points at $\mu = 0$ and $t = \pm \Delta$. Further, their density is larger close to the boundary of the topological phase, as a result of the slow changes of the cosine function around $0$ and $\pi$. 
	
	For growing number of sites $N$, the density of solutions increases. In the limit $N\rightarrow \infty$, $\theta_n= n\pi/N+1$ takes all values in $[0,\,\pi]$ and the entire area between $\mu  = \pm 2\,\sqrt{t^2-\Delta^2}$ for $t^2\ge \Delta^2$ is now occupied with zero energy modes. 
	
	Regarding the remaining part of the topological region, we are going to show in the next section that in that parameter space  only boundary modes with finite energy exist.  Because the energy of these modes is decreasing exponentially with the system size,  in the thermodynamic limit their energy  approaches zero and the full topological region supports zero energy modes. 
\subsubsection{The complete spectrum of the finite Kitaev chain}

To proceed we transform Eq. (\ref{equation: characteristic polynomial via chiral basis}) into an eigenvector problem for $h h^\dagger$, 
\begin{align}\label{equation: eigenvector problem of hhdagger}
	hh^\dagger \vec{v} \,=\,\lambda^2 \vec{v},
\end{align}
where we defined $\vec{v} = \left(\xi_1,\,\xi_2\,\ldots,\,\xi_N\right)^\mathrm{T}$.
Notice that we are not really interested in the eigenvector $\vec{v}$ here; we simply use its entries as dummy variables to release a structure hidden in the product of $h$ and $h^\dagger$. The elements of $h$,  
\begin{align*}
	h_{n,m}\,=\, -i\mu\,\delta_{n,m}\,+\,a\, \delta_{n,m+1}\,-\,b\,\delta_{n+1,m},
\end{align*}
where $n,\,m \,= \,1,\ldots,N$, allow us to calculate the product $hh^\dagger$ entry wise
\begin{align*}
	\left(hh^\dagger\right)_{n,m}\,&=\,\delta_{n,m}\,\left[\mu^2\,-\,a^2\,\left(1-\delta_{n,N}\right)\,-\,b^2\,\left(1-\delta_{n,1}\right) 
		\right]\\
		&\quad +\,i\mu\, \left(a-b\right)\,\left[\,\delta_{n,m+1}\,+\, \delta_{n+1,m}\right]\\
		&\quad +\,ab \,\left(\delta_{n,m+2}\,+ \,\delta_{n+2,m}\right).
\end{align*}
\begin{figure}[t]
	\centering
		\includegraphics[width = 1\columnwidth]{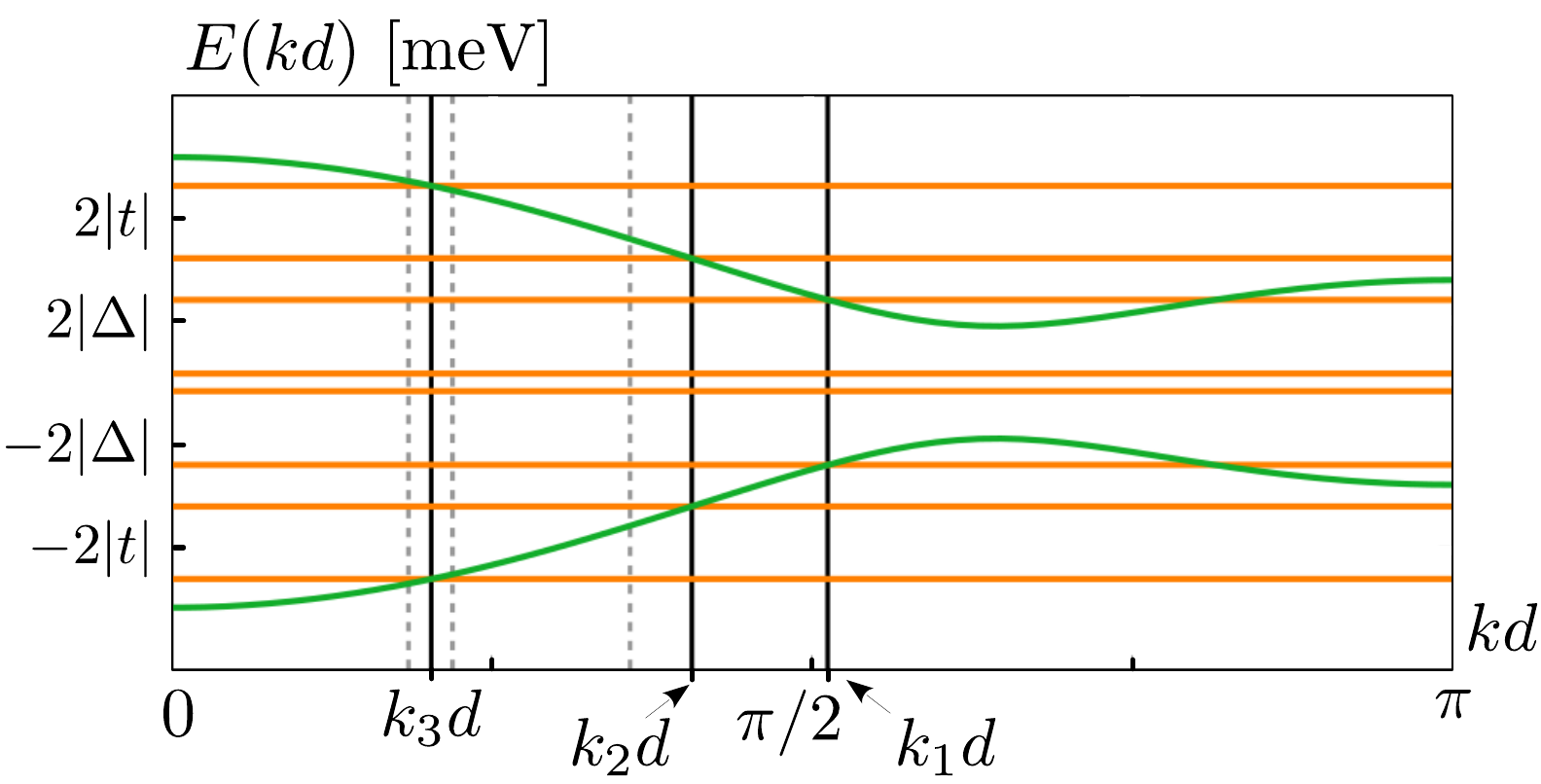}
	\caption{Spectrum of the Kitaev molecule with four sites and $\mu\neq0$. The green line follows the excitation spectrum from Eq. (\ref{infinite chain: excitation spectrum}) and the horizontal lines are the numerical eigenvalues of the Kitaev chain. The momenta $k_{3,\,2,\,1}$ are the proper wavevectors for $\mu \neq 0$ calculated from the full quantization rule, see Eqs.  (\ref{equation: relation k_1,2 via sum of cos}), (\ref{equation: Kitaev chain full quantization rule}), (\ref{equation: definition of F for the quantization rule}). The dashed, light grey lines represent the wavevectors taken from the $\mu=0$ case to highlight the difference. The chemical potential $\mu$ is obviously changing the quantization of a finite chain. The chosen parameters are $t=4\,$meV, $\Delta=1.5\,$meV and $\mu=3\,$meV and lead to the numerical energies $\left[ 0.43,\,  4.034,\,  6.068,\,  9.603\right]$ (in meV). The value of $k_{3,\,2,\,1} d$ is approximately $\left[0.6360,\, 1.2753,\,1.6086\right]$. }
	\label{figure: mu influences the quantization}
\end{figure}		
Thus, importantly, Eq. (\ref{equation: eigenvector problem of hhdagger}) reveals a recursion formula
\begin{align}\label{equation: tetranacci recursion formula}
	\xi_{j+2}\,&=\,\frac{\lambda^2+a^2+b^2-\mu^2}{ab}\,\xi_j-\xi_{j-2}\notag\\
		&\quad -i\mu\left(\frac{a-b}{ab}\right)\,\left(\xi_{j+1} + \xi_{j-1}\right),
\end{align}
for the components of $\vec{v}$. The entries $\xi$ are a generalisation of the Fibonacci polynomials $\zeta_j$ from Eq. (\ref{equation: recursion formula fibonacci polynomials}), to which they reduce for $\mu= 0$, and may be called Tetranacci polynomials\cite{McLaughlin-1979, Waddill-1992}. 
Further,  we find the open boundary conditions from Eq. (\ref{equation: eigenvector problem of hhdagger}) to be
\begin{align}\label{equation: open boundary condition in general}
	\xi_0 \, =\, \xi_{N+1}\,=\,b\,\xi_{N+2}\,-\,a\, \xi_N\,=\,b\,\xi_1\,-\,a \,\xi_{-1}\,=\,0,
\end{align}
where we used Eq. (\ref{equation: tetranacci recursion formula}) for simplifications.

Appendix \ref{appendix: spectrum finite u} contains the description of how to deal with those polynomials, the boundary conditions and further the connection of Eq. (\ref{equation: tetranacci recursion formula}) to Kitaev's bulk spectrum $\lambda = E_\pm(k)$ in Eq. (\ref{infinite chain: excitation spectrum}). Essentially one has to use similar techniques as it was  done for the Fibonacci polynomials, where now the power law ansatz $\xi_j\propto r^j$ leads to a characteristic equation for $r$ of order four. Thus, we find in total four linearly independent fundamental solutions $r_{\pm 1\,, \pm 2}$, which can be expressed in terms of two complex wavevectors denoted by $\kappa_{1,2}$ through the equality
\begin{align}
	r_{\pm j}\,=\,e^{\pm i \kappa_j },\quad j= 1,2.
\end{align}
These wavevectors are not independent, but coupled via
\begin{align}\label{equation: relation k_1,2 via sum of cos}
	\cos(\kappa_1)+ \cos(\kappa_2)\,=\,-\frac{\mu t}{t^2-\Delta^2},\quad \forall \,t,\,\Delta,\, \mu\in\mathbbm{R}.
\end{align}
\begin{widetext}

\begin{center}
	\centering
	\begin{table}
	\caption{Overview of the quantization rule for the wave vectors of the finite Kitaev chain in different scenarios. The wavevectors $k$, $k_n$, $\kappa_{1,2}$ used together with Eq. (\ref{infinite chain: excitation spectrum}) and $q$ with Eq. \eqref{equation: spectrum: mu=0, even N, edges} yield the correct finite system energies and $k,\,k_n,\,q\in\mathbbm{R}$, $\kappa_{1,2}\in\mathbbm{C}$. Notice: $n = 1,\,\ldots,\,N$ and $m = \frac{t}{\Delta}$ ($m = \frac{\Delta}{t}$) for $\vert t \vert > \vert \Delta\vert$ ($\vert \Delta \vert > \vert t\vert$).} 
	\begin{ruledtabular}
	
	\begin{longtable}{p{2.4 cm} p{7 cm} p{4.2 cm} p{2.2 cm} p{1.4 cm} }

	\hline\hline \\
	
	Requirements & Quantisation rule &  Zero modes & Equation for \newline eigenstate \newline elements  & Majorana\newline character\\
	\\
	\hline\\
	$\Delta = 0:$		 & $k_n d = \frac{n\pi}{N+1}$						& Yes, if for some $n$: \newline $\mu=\mu_n = 2t \cos\left(\frac{n\pi}{N+1}\right) $ &		& No \\
	\\
	$t = 0:$ 			 & $k_n d = \frac{\pi}{2}+\frac{n\pi}{N+1}$						& No & 						&	No	\\
	\\ \\ \\
	$\mu = 0,$ $N$ odd:  & $k_n d = \frac{n\pi}{N+1},\quad n\neq \frac{N+1}{2}$ \newline\newline
						   $qd =\mathrm{arctanh}\left(1/\vert m \vert\right)$	 		& No \newline\newline Yes & (\ref{equation: N odd, entries of the eigenstates})\,-\,(\ref{equation: function T odd })\newline\newline 
						   (\ref{equation: zero mode on alpha N odd}), (\ref{equation: zero mode on beta N odd})\newline (\ref{equation: zero mode on alpha, odd N, Kitaev point plus})\,-\,(\ref{equation: zero mode on beta, odd N, Kitaev point minus})	&No	\newline\newline Yes\newline Yes	\\
	\\ \\ \\
	$\mu = 0,$ $N$ even: & $\tan\left[kd \left(N+1\right)\right]\,=\,\mp\frac{\Delta}{t}\tan\left(kd\right)$\newline\newline
						  $\tanh\left[qd \left(N+1\right)\right]\,=\,\mp m\tanh\left(qd\right)$ & No \newline\newline only if $\Delta = \pm t$\newline otherwise	& (\ref{equation: decaying state, even N, y_j}), (\ref{equation: decaying state, even N, x_j})\newline\newline (\ref{equation: Delta = t, N even})\,-\,(\ref{equation: Delta = -t, N even})\newline (\ref{equation: decaying state, even N, y_j}), (\ref{equation: decaying state, even N, x_j}) & No	\newline\newline Yes \newline No	\\
	\\ \\ \\
	$t,\,\Delta,\,\mu\in\mathbbm{R}$ 				&  $\displaystyle\frac{\sin^2\left[\frac{\kappa_1+\kappa_2}{2}  \left(N+1\right)\right]}{\sin^2\left[\frac{\kappa_1-\kappa_2}{2}  \left(N+1\right)\right]} = \frac{1+\left(\frac{\Delta}{t}\right)^2\,\cot^2\left(\frac{\kappa_1-\kappa_2}{2}\right) }{1+\left(\frac{\Delta}{t}\right)^2\,\cot^2\left(\frac{\kappa_1+\kappa_2}{2}\right)}$ &  only for $\mu=\mu_n\in \mathbbm{R}$ and \newline \makebox{$\mu_n = 2\sqrt{t^2-\Delta^2} \cos\left(\frac{n\pi}{N+1}\right) $}	&\eqref{equation: closed formula for all Tetranacci xi} \eqref{equation: entries of u}\newline(\ref{equation: MZM x_l}), (\ref{equation: MZM ycal_l}) \newline (\ref{equation: MZM xcal_l}), (\ref{equation: MZM y_l})  & No \newline Yes\newline Yes\\
	\\
	\hline\hline
	\end{longtable}
	
	\end{ruledtabular}
	\label{table: all quantisiation (sub) rules}
	\end{table}
\end{center}

\end{widetext}

For $\mu=0$ we can recover from Eq. (\ref{equation: relation k_1,2 via sum of cos}) our previous results, whereby one has only pure real $(k)$ or pure imaginary $(iq)$ wavevectors\footnote{In fact $\mu=0$ supports complex wavevectors too, but their real part has to be zero or $\pi/2$, i.e. one has to use $iq$ or $\frac{\pi}{2}+iq$.}. 
Further, Eqs. (\ref{infinite chain: excitation spectrum}) and (\ref{equation: relation k_1,2 via sum of cos}) yield
\begin{align*}
	E_\pm (\kappa_1) = E_\pm(\kappa_2).
\end{align*}
The linearity of the recursion formula Eq. (\ref{equation: tetranacci recursion formula}) states that the superposition of all four fundamental solutions is the general form of $\xi_j$. Since the boundary conditions translate into a homogeneous system of four coupled equations and a trivial solution for $\xi_j$ has to be avoided, we find that the determinant of the matrix describing these equations has to be zero. After some algebraic manipulations, this procedure leads finally to the full quantization rule of the Kitaev chain
\begin{align}\label{equation: Kitaev chain full quantization rule}
	F(\kappa_1, \kappa_2) = F(\kappa_1, -\kappa_2),
\end{align} 
where we introduced the function $F(\kappa_1, \kappa_2)$ as
\begin{align}\label{equation: definition of F for the quantization rule}
	F(\kappa_1, \kappa_2)&\,=\,\sin^2\left[\frac{\kappa_1+\kappa_2}{2}\,\left(N+1\right)\right]\times\notag\\
	&\quad \times \left[1+\left(\frac{\Delta}{t}\right)^2\,\cot^2\left(\frac{\kappa_1+\kappa_2}{2}\right)\right].
\end{align}

Similar quantization conditions are known for an open X-Y spin chain in transverse field\cite{Loginov-1997}. 
Notice that the quantization rule is symmetric with respect to $\kappa_{1,2}$. \mbox{Table \ref{table: all quantisiation (sub) rules}} gives an overview of the quantization rules for different parameter settings $\left(\Delta,\,t,\,\mu\right)$. The bulk eigenvalues of a finite Kitaev chain with four sites and $\mu\neq 0 $ are shown in Fig. \ref{figure: mu influences the quantization}. 

The previous relations open another route to finding the condition   leading to modes with exact zero energy. A convenient form of Eq. (\ref{equation: relation k_1,2 via sum of cos}) is
\begin{align}\label{equation: connection of kappa1pmkappa2}
	\cos\left(\frac{\kappa_1+\kappa_2}{2}\right)\, \cos\left(\frac{\kappa_1-\kappa_2}{2}\right)\,=\,-\frac{1}{2}\,\frac{\mu t}{t^2-\Delta^2},
\end{align}
and the dispersion relation can be transformed into
\begin{align}\label{equation: dispersion relation with kappas}
	E^2&=\frac{1}{\cos^2\left(\frac{\kappa_1\pm\kappa_2}{2}\right)}
	\left[4(t^2-\Delta^2)\cos^2\left(\frac{\kappa_1\pm\kappa_2}{2}\right)-\mu^2  \right]\times \notag	\\
	&\quad \times\left[\frac{t^2}{t^2-\Delta^2}-\cos^2\left(\frac{\kappa_1\pm\kappa_2}{2}\right)\right].
\end{align}
Both combinations $\kappa_1\pm\kappa_2$ yield the same energy, due to Eq. (\ref{equation: connection of kappa1pmkappa2}). If one of the brackets in Eq. (\ref{equation: dispersion relation with kappas}) vanishes for $\kappa_1+\kappa_2$ ($\kappa_1-\kappa_2$), the second one does so for $\kappa_1-\kappa_2$ ($\kappa_1+\kappa_2$ ) too. Hence, zero energy is achieved exactly if 
\begin{align}\label{equation: cot part in the quantization rule is zero}
	1+\left(\frac{\Delta}{t}\right)^2\,\cot^2\left(\frac{\kappa_1\pm\kappa_2}{2}\right)\,=\,0.
\end{align}
%
%
This  puts restrictions on $\Delta,\,t,\,\mu$. Together with  the quantization rule in Eq. (\ref{equation: Kitaev chain full quantization rule}), this ultimately leads to (\ref{equation: eigenvalue zero condition kouachi}) and  to the condition $2\sqrt{t^2-\Delta^2} < \vert \mu \vert < 2 \vert t \vert$, defining the region where exact zero modes can form. 


 Regarding the remaining part of the topological phase diagram, we find that in the limit $N\rightarrow \infty$ the difference between even and odd $N$ vanishes and the part of the $\mu=0$ axis between $t=-\Delta$ and $t=\Delta$ for even $N$ leads to zero energy states too, in virtue of Eq. (\ref{equation: spectrum: mu=0, even N, edges, cosh/cosh version}). Analogously, the area around the origin in Fig. (\ref{figure: finite size effects and the topological phase diagramm}), defined by $2\sqrt{t^2-\Delta^2}>\vert \mu \vert $ and  $\mu < 2 \vert t \vert$ with $\mu \neq 0$ does not  support zero energy modes for all finite $N$. 
 Instead, this area contains solutions with exponentially small energies, see Eq. (\ref{equation: dispersion relation with kappas}), which become zero exclusively in the limit $N\rightarrow \infty$. The wavevectors obey $\kappa_1 = i q_1$, $\kappa_2 = \pi + i q_2$ with real $q_{1,2}$ for $\Delta^2>t^2$, and  $\kappa_1 = i q_1$, $\kappa_2 =  i q_2$ otherwise, which follows from Eq. (\ref{equation: Kitaev chain full quantization rule}) after some manipulations; see also Ref.~[\onlinecite{Loginov-1997}]. Thus, the entire non trivial phase hosts zero energy solutions for $N\rightarrow\infty$.
\subsection{Eigenstates}
The calculation of an arbitrary wave function of the Kitaev Hamiltonian without any restriction on $t$, $\Delta$, $\mu$ is performed at best in the chiral basis yielding the block off-diagonal structure in Eq. \eqref{equation: Kitaev Hamiltonian in chiral basis}. A suitable starting point is to consider a vector $\vec{w}$ in the following form
\begin{align*}
	\vec{w}\,=\,\left(
	\begin{matrix}
	\vec{v}\\
	\vec{u}
	\end{matrix}	
	\right)
\end{align*}
with $\vec{v}=\left(\xi_1\,\ldots,\,\xi_N\right)$, $\vec{u}=\left(\sigma_1\,\ldots,\,\sigma_N\right)$. Solving for an eigenstates with eigenvalue $\lambda$ demands on $\vec{v},\,\vec{u}$
\begin{align}
    \label{eq:hu-is-lambdav}
	h\,\vec{u}\,&=\,\lambda\,\vec{v},\\
	\label{equation: u follows from v}
	h^\dagger\,\vec{v}\,&=\,\lambda\,\vec{u},
\end{align}
with $h$ from Eq. \eqref{equation: matrix h in chiral basis}. Thus, 
%
%
$	h\,h^\dagger \vec{v}\,=\,\lambda^2\,\vec{v}$,
%
%
and we recover Eq. \eqref{equation: eigenvector problem of hhdagger} and the entries of $\vec{v}$ obey Eq. \eqref{equation: tetranacci recursion formula} again. In appendix \ref{appendix: section: The closed formula of Tetranacci polynomials} we derived the closed formula for $\xi_j$, namely
\begin{align}\label{equation: closed formula for all Tetranacci xi}
	\xi_j\,=\,\sum\limits_{i=-2}^1\,\xi_i \,X_i(j), \quad j \in \mathbbm{Z}
\end{align}
where $\xi_{-2},\,\ldots \xi_1$ are the initial values of the polynomial sequence dependent on the boundary conditions, and $X_i(j)$ inherit the selective property
\begin{align}
	X_i(j)\,=\,\delta_{i,j}\qquad \mathrm{for~only\,} i,\,j=-2,\,\ldots,\,1.
\end{align}
That these functions $X_i(j)$ exist and that they indeed satisfy Eq. \eqref{equation: closed formula for all Tetranacci xi} for arbitrary values of $j$ is discussed in Appendix \ref{appendix: section: The closed formula of Tetranacci polynomials}.

The remaining task is to obtain the initial values $\xi_{-2},\,\ldots \xi_1$,
 which follow from the open boundary conditions  Eq. \ref{equation: open boundary condition in general}. Further, one has one free degree of freedom, which we  to choose to be the  entry $\xi_1$ of $\vec{v}$. In total our initial values are $\xi_1,\xi_0=0,\xi_{-1}=b \,\xi_1/a$ and $\xi_{-2}$ follows from $\xi_{N+1}=0$ and Eq. \eqref{equation: closed formula for all Tetranacci xi}, 
\begin{align*}
	\xi_{-2}\,=\,-\xi_1\,\frac{a\,X_1(N+1)+b \,X_{-1}(N+1)}{a\,X_{-2}(N+1)}.
\end{align*}
%
%
Demanding further $b\,\xi_1\,-\,a \,\xi_{-1}\,=\,0$ quantizes  the momenta $\kappa_{1,2}$ and in turn $\lambda=E_\pm(\kappa_{1,2})$, according to Eq. \eqref{equation: Kitaev chain full quantization rule} (the relation between $X_j$ and $\kappa_{1,2}$ is discussed in the Appendix \ref{appendix: section: The closed formula of Tetranacci polynomials}). Notice that the form given by  Eqs. \eqref{equation: closed formula for all Tetranacci xi}, \eqref{equation: entries of u} (below) hold for all eigenstates of the Kitaev BdG Hamiltonian;  the distinction between extended/decaying states and MZM is made by the values of $k_{1,2}$, or equivalently $\kappa_{1,2}$. 

The second part of the eigenstate $\vec{w}$, i.e. $\vec{u}$,  follows in principle from Eq. \eqref{equation: u follows from v}. A simpler and faster way is to consider 
%
%
 $	h^\dagger\,h\,\vec{u}\,=\,\lambda^2\,\vec{u}$.
%
%
A comparison of $h$ and $h^\dagger$ reveals that they transform into each other by exchanging $a$ and $b$ and switching $\mu$ into $-\mu$. Consequently, the structure of the entries of $\vec{u}$ follow essentially from the ones of $\vec{v}$. Thus, we find that the $\sigma_i$ obey
Eq. \eqref{equation: tetranacci recursion formula} too, yielding 
\begin{align}\label{equation: entries of u}
	\sigma_j\,=\,\sum\limits_{i=-2}^1\,\sigma_i \,X_i(j), \quad j \in \mathbbm{Z}
\end{align}
with the same functions $X_i(j)$. The boundary condition on $\vec{u}$ reads
\begin{align*}
	\sigma_0 \, =\, \sigma_{N+1}\,=\,a\,\sigma_{N+2}\,-\,b\, \sigma_N\,=\,a\,\sigma_1\,-\,b \,\sigma_{-1}\,=\,0.
\end{align*}
Proceeding as we did for $\vec{v}$ yields $\sigma_0\,=\,0$, $\sigma_{-1}=a \sigma_1/b$ and 
\begin{align}
	\sigma_{-2}\,=\,-\sigma_{1}\,\frac{b\,X_1 (N+1)-\,a\,X_{-1}(N+1)}{b\,X_{-2}(N+1)},
\end{align}
where $\sigma_1$ is fixed by the first line of Eq. \eqref{equation: u follows from v}
\begin{align}
\label{eq:sigma1-with-lambda}
	\sigma_1\,=\,\frac{i\mu\,\xi_1\,+\,b\,\xi_2}{\lambda}.
\end{align}
The last open boundary condition $a\,\sigma_{N+2}\,-\,b \,\sigma_{N}\,=\,0$ is satisfied for $\lambda=E_\pm(\kappa_{1,2})$ and thus already by the construction of $\vec{v}$. We notice that we assumed $\lambda\neq 0$ in order to obtain $\sigma_1$; the  case $\lambda=0$ is discussed in Sec. \ref{section: MZM paths in the topological phase diagram for finite N} below. 

In the limit of $\mu=0$ on $\vec{w}$ it holds that
\begin{align*}
	X_{-2}(2l+1)\vert_{\mu=0} \,&=\,0,\\
	X_{0}(2l+1)\vert_{\mu=0} \,&=\,0,\\
	X_{-1}(2l)\vert_{\mu=0} \,&=\,0,\\
	X_{1}(2l)\vert_{\mu=0} \,&=\,0,
\end{align*}
for all values of $l$; thus only two initial values for $\xi_j$ and two for $\sigma_j$ are necessary to fix the sequences.

Finally, one can prove easily that the functions $X_i(j)$ are always real and in consequence all $\xi_j$ ($\sigma_j$) are real (pure imaginary) if $\xi_1$ is chosen to be real. Thus, the corresponding operators $\psi_{\vec{w}}$  $(\psi^\dagger_{\vec{w}})$ can never square to $1/2$.

\section{MZM eigenvectors at finite $\mu$}	
\label{section: MZM paths in the topological phase diagram for finite N}
The technique outlined above (in particular, Eq.~\eqref{eq:sigma1-with-lambda}) cannot be used directly for exact zero energy modes, because then $\lambda = 0$ in Eqs.~\eqref{eq:hu-is-lambdav},~\eqref{equation: u follows from v}.
In this section we demonstrate the Majorana nature of the zero energy solutions satisfying Eq. (\ref{equation: eigenvalue zero condition kouachi}), and we give the explicit form of the associated MZM using a different technique, which (similar to the Chebyshev polynomials method in Ref.~[\onlinecite{Kawabata-2017}]) requires only the use of Fibonacci, not Tetranacci polynomials. This simplification is caused by the fact that setting $\lambda=0$ decouples the two Majorana sublattices, while setting $\mu=0$ decouples the two SSH-like chains.
%
%
%
%
%
%
\begin{figure}[htb]\centering
	\includegraphics[width=\columnwidth]{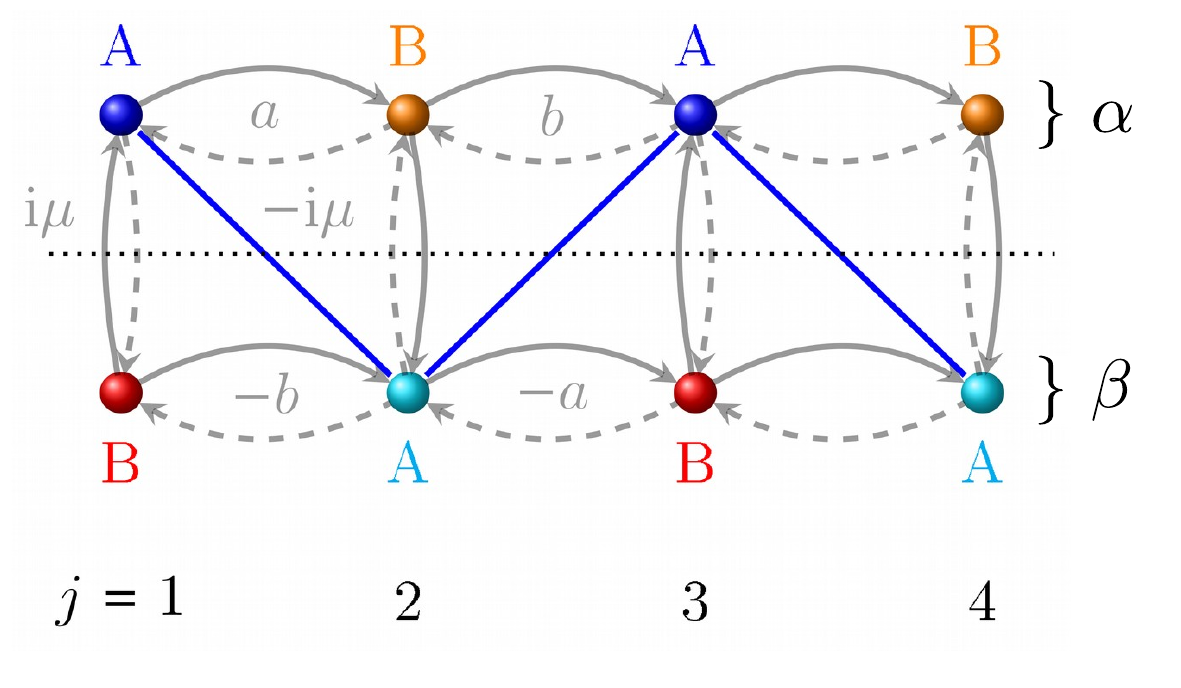}
	\caption{Illustration of the sawtooth pattern of $\vec{\psi}_A$. The real space position of the entries $x_l$ ($\ycal_l$) of $\vec{\psi}_A$ is 
	at $j=2l-1$ ($j=2l$) on chain $\alpha $ ($\beta$) and marked by the blue (light blue) spheres. The blue line connects these entries as guide to the eye. 
		}
	\label{Figure: sawtooth pattern of  psi_A}
\end{figure}

We use the SSH-like description Eq. (\ref{equation: Kitaev Hamiltonian/ matrix in SSH basis}) of the Kitaev chain where $\mu\neq 0 $ couples both chains together. Consequently an eigenstate $\vec{\psi}\,=\,\left(\vec{v}_\alpha,\,\vec{v}_\beta\right)^\mathrm{T}$ has in general no zero entries and we use the same notation for the components of $\vec{v}_\alpha$, $\vec{v}_\beta$ as in the sections \ref{subsection: N even eigenvectors} and \ref{subsection: N odd eigenvectors}. 
 
The zero energy values are twice degenerated, as one can see from Eq. (\ref{equation: determinant of the Kitaev chain in terms of h and h*}), and the associated zero modes are connected by the chiral symmetry $\mathcal{C}$. Thus, we get zero energy states by superposition $\vec{\psi}_{A,B} \,\defl \left(\vec{\psi}\pm \mathcal{C}\vec{\psi}\right)/2$ too. The chiral symmetry Eq. (\ref{equation: chiral symmetry even N}), contains an alternating pattern of $\pm 1$, such that $\vec{\psi}_A$ $(\vec{\psi}_B)$ includes only non zero entries on the Majorana sublattice $A$ ($B$). Hence, $\vec{\psi}_A$ 
($\vec{\psi}_B$) contains only $x_l$ ($\xcal_l$) and $\ycal_j$ ($y_j$) terms and the last component depends on whether $N$ is odd or even. In the latter case we have
\begin{align*}
	\vec{\psi}_A\,&=\,\left(x_1,\,0,\,x_2,\,0,\ldots,x_{\frac{N}{2}},0\,\left\vert\,0,\ycal_1,\,0,\ycal_2,\ldots,0,\,\ycal_{\frac{N}{2}}\right.\right)^\mathrm{T},\\
	\vec{\psi}_B\,&=\,\left(0,\,y_1,\,0,\,y_2,\ldots,0,y_{\frac{N}{2}}\left\vert\,\xcal_1,\,0,\xcal_2,\,0\ldots,\,\xcal_{\frac{N}{2}},\,0\right.\right)^\mathrm{T}.
\end{align*}
The form of the odd $N$ eigenvectors is quite similar, see Eqs. (\ref{equation: appendix: vec_+ for odd N}), (\ref{equation: appendix: vec_- for odd N}).
 
The composition of $\vec{\psi}_A$ is illustrated in Fig. \ref{Figure: sawtooth pattern of  psi_A}, where its entries are shown to form a sawtooth like pattern, following the action of $\gamma_j^A$ on both SSH-like chains.

The full calculation is given in appendix \ref{appendix: zero energy states}. We focus here on $\vec{\psi}_A$ exclusively, because the $\vec{\psi}_B$ components follow essentially from $\vec{\psi}_A$ by exchanging $a$ and $b$ and replacing $i\mu$ by $-i\mu$. The chemical potential has still to obey Eq. (\ref{equation: eigenvalue zero condition kouachi}).

The components of the zero mode $\vec{\psi}_A$ have to satisfy $(l = 1,\,\ldots,\,N/2-1)$ 
\begin{align}\label{equation: recursion formula x_l}
	b\,x_{l+1}\,-\,a\,x_l\,+\,i\mu\,\ycal_l\,&=\,0,\\
	b\,\ycal_{l+1}\,-\,a\,\ycal_l\,+i\mu\,x_{l+1}\,&=\,0,
\end{align}
for even $N$, and the open boundary conditions are
\begin{align*}
	\ycal_0\,=\,y_0\,=\,\xcal_{\frac{N}{2}+1} \,=\,x_{\frac{N}{2}+1} \,=\,0.
\end{align*}
The situation for the entries of $\vec{\psi}_A$ for odd $N$ is similar
\begin{align}
	b\,x_{j+1}\,-\,a\,x_j\,+\,i\mu\,\ycal_j\,=\,0,\\
	\label{equation: recursion formula ycal_l}
	b\,\ycal_{i+1}\,-\,a\,\ycal_i\,+i\mu\,x_{i+1}\,=\,0,
\end{align}
where $j = 1,\ldots,\,(N-1)/2$, $i = 1,\ldots,\,(N-3)/2$. The open boundary condition changes to
\begin{align*}
	\ycal_0\,=\,y_0\,=\,\ycal_{\frac{N+1}{2}} \,=\,y_{\frac{N+1}{2}} \,=\,0.
\end{align*}
Solving these recursive formulas leads in both cases to
\begin{align}\label{equation: MZM x_l}
	x_l\,&=\,x_1\,\frac{\sin[\theta_n\,(2l-1)]}{\sin(\theta_n)}\,\left(-\frac{a}{b}\right)^{l-1},
\end{align}
with $\theta_n = n\pi/(N+1)$, $n =1,\,\ldots,\,N$ and 
\begin{align}\label{equation: MZM ycal_l}
	\ycal_l\,&=\,-x_1\,\mathrm{sgn}(t+\Delta)\frac{\sin(2\,\theta_n\,l)}{\sin(\theta_n)}\,\left(-\frac{a}{b}\right)^{\frac{2l-1}{2}},
\end{align}
where $x_1$ is a free parameter and $-a/b\ge 0$ due to $t^2\ge \Delta^2$. Recalling that $a = i \left(\Delta -t\right)$,  $b = i \left(\Delta +t\right)$, Eqs. (\ref{equation: MZM x_l}), (\ref{equation: MZM ycal_l}) predict an oscillatory exponential decay of the 
coefficients $x_l$, $\ycal_l$. For example
\begin{align}
	x_l\,&=\,x_1\,\frac{\sin[\theta_n\,(2l-1)]}{\sin(\theta_n)}\,e^{-\left(l-1\right) d/\xi},
\end{align}
where the decay length is defined by $\xi= d/ \left\vert\ln \left(\frac{t-\Delta}{t+\Delta}\right)\right\vert$, for $t>\Delta>0$. Summarizing: the zero energy modes $\vec{\psi}_{A,B}$ look like small or strong suppressed standing waves with $n-1$ nodes for $n=1,...,N_\mathrm{max}$ and $N_\mathrm{max}  = N/2$  ($N_\mathrm{max} = N-1/2$) for even (odd) $N$. The expressions for $\xcal_l$ and $y_l$ are obtained in a similar way
\begin{align}\label{equation: MZM xcal_l}
	\xcal_l\,&=\,\xcal_1\,\frac{\sin[\theta_n\,(2l-1)]}{\sin(\theta_n)}\,\left(-\frac{b}{a}\right)^{l-1},\\
	\label{equation: MZM y_l}
	y_l\,&=\,-\xcal_1\,\mathrm{sgn}(t-\Delta)\frac{\sin(2\,\theta_n\,l)}{\sin(\theta_n)}\,\left(-\frac{b}{a}\right)^{\frac{2l-1}{2}},
\end{align}
and $\xcal_1$ can be freely chosen. The open boundary conditions for $l=0$ are satisfied by construction of $\ycal_l$ ($y_l$), while the remaining ones follow due to the structure of $\theta_n$. 
\begin{figure}[ht]
	\centering
	\includegraphics[width=\columnwidth]{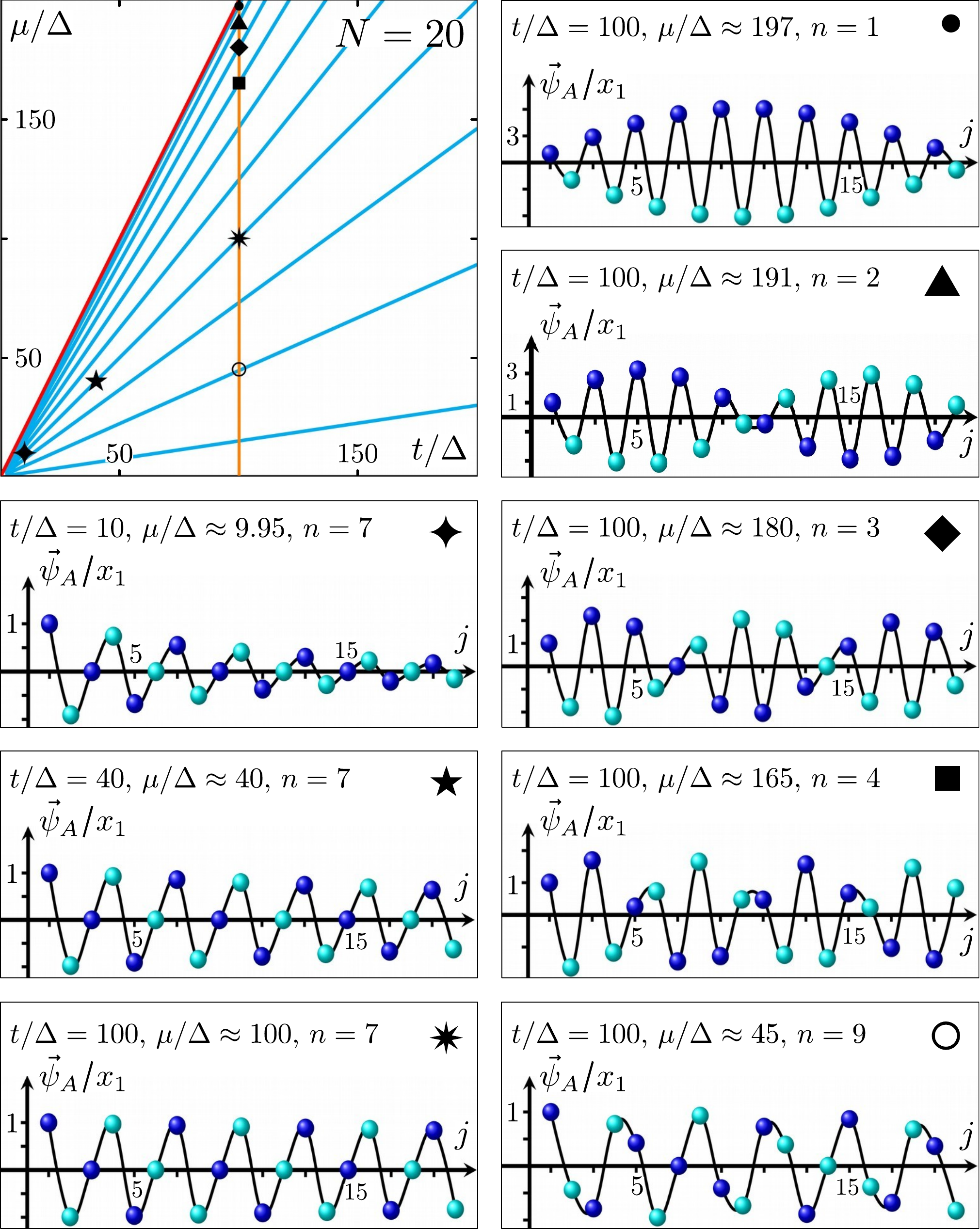}
	\caption{Majorana zero mode $\vec{\psi}_A$ for various parameter sets. The considered parameters are denoted by different symbols on the topological phase diagram. The dark (light) blue spheres follow $x_l/x_1$ ($\ycal_l/x_1$) at position $j=2l-1$ ($j=2l$) from Eq. (\ref{equation: MZM x_l}) (\ref{equation: MZM ycal_l}). The decay length of the MZM increases for larger ratios of $t/\Delta$ for a fixed value of $\theta_n$, until the state is delocalised over the entire system. Lowering the chemical potential, e.g. following the vertical orange line, but keeping $t/\Delta$ fixed, changes the shape of the MZM's. Small decaying length and large enough chemical potentials leads to Majorana modes which have highest weight in the center of the chain. 
}
\label{figure: Majorana zero modes}
\end{figure}

The zero mode $\vec{\psi}_A$ is shown in Fig. \ref{figure: Majorana zero modes} for a various range of parameters. For not too large ratios $ t/\Delta  >1$, the zero mode $\vec{\psi}_A$ is mostly localised at one end of the Kitaev chain and decays away from it in an oscillatory way. The eigenstate $\vec{\psi}_B$ is concentrated on the opposite end. While the oscillation depends on the chemical potential $\mu_n$ associated to the zero mode, according to Eq. (\ref{equation: eigenvalue zero condition kouachi}), the decay length is only set by the parameters $\Delta$ and $t$. Thus, as the ratio of $t/\Delta$ is increased, the zero energy mode gets more and more delocalized.

The zero energy states $\vec{\psi}_{A,B}$ are MZM's, since they are eigenstates of the particle hole operator $\mathcal{P}$ Eq. (\ref{equation: particle hole symmetry in real space in ssh basis}) for real or pure imaginary values of $x_1$, $\xcal_1$. Further, the states $\vec{\psi}=\vec{\psi}_A + \vec{\psi}_B$ and $\mathcal{C} \vec{\psi}=\vec{\psi}_A - \vec{\psi}_B$ are MZM's too. On the other hand a fermionic state is constructed with $\vec{\psi}_\pm = \vec{\psi}_A \pm i\, \vec{\psi}_B$, similar to what was found in Sec. \ref{section: Eigenvectors and symmetries} for the $\mu=0$ case, or at the Kitaev points in Eqs. \eqref{equation: Missing electron a} and \eqref{equation: Missing electron b} in Sec. \ref{section 2 model}.

There are three limiting situations we would like to discuss: $t\rightarrow \pm \infty$, $N\rightarrow \infty$, and how the eigenstate changes if the sign of the chemical potential is reverted. For the first situation we notice that larger hopping amplitudes affect the decay length $\xi$. Because $-a/b \rightarrow 1$ for $t\rightarrow \pm\infty$, this implies also that $\xi\rightarrow \infty$ in that limit. Hence oscillations are less suppressed for large values of $t$, as illustrated in Fig. \ref{figure: Majorana zero modes}. Already a ratio of $t/\Delta \approx 100$ is enough to avoid a visible decay for $N\approx 20$. This effect can be found as long as $N$ is finite, but one has to consider larger values of the ratio $t/\Delta$. 

What happens instead for larger system sizes? Regardless of how close $-a/b$ is to $1$, for a finite $t$, at some point the exponent $j$ $(l)$ in $x_l$ $\ycal_l$, $\xcal_l$ and $y_l$ leads to significantly large or small values. Thus, the state $\vec{\psi}_A$ ($\vec{\psi}_B$) becomes more localised on the left (right) end for $t> 0$, and on the right (left) one for $t<0$.

If we change the chemical potential to its negative value we find that $y_l$, $\ycal_l$ only change their sign. For odd $N$ and for $\theta_n = \pi/2$, i.e. $\mu=0$, one recovers the result in the Eqs. (\ref{equation: zero mode on alpha N odd}) -  (\ref{equation: zero mode on beta N odd}).

\section{Numerical results and impact of disorder}
\label{section: numerical results and impact of disorder}
In this section we  discuss the impact of disorder on the topological boundary states. To this extent we investigate numerically the lowest energy eigenvalues of the finite Kitaev chain.
\subsection{The clean Kitaev chain} 
\begin{figure}[ht]
	\centering
	\includegraphics[width=\columnwidth]{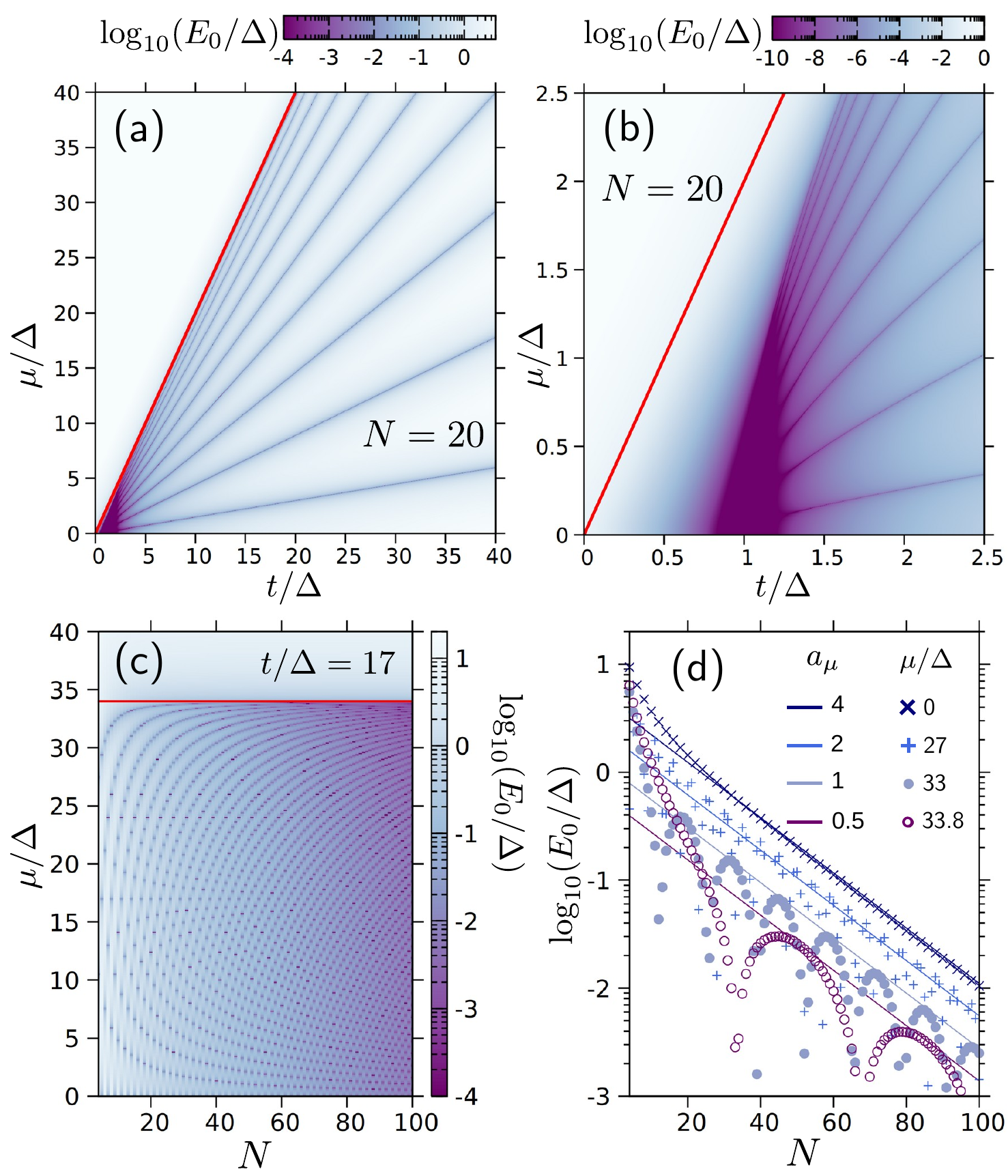}
	\caption{Numerical results for the energy $E_0$ of the lowest positive energy state as a function of $t/\Delta$, $\mu/\Delta$ and system size $N$. (a) $E_0$ as a function of $t$ and $\mu$ for $N=20$. The red line marks the boundary of the bulk topological phase. The $N/2$ dark lines coincide with the Majorana lines given by Eq. (\ref{equation: eigenvalue zero condition kouachi}). (b) Zoom into the neighbourhood of the Kitaev point at $\mu=0$ and $t=\Delta$, showing the absence of zero energy solutions for $\Delta<t$ in the nominally non-trivial phase. (c) $E_0$ as a function of $\mu$ and the system size $N$ for $t/\Delta = 17$. The red line marks $\mu=2t$, the boundary of the bulk topological phase. As $N$ increases, the number of $\mu$ yielding zero energy solutions also increases according to Eq. (\ref{equation: eigenvalue zero condition kouachi}), and the maximum energies of the bound states decrease. (d) The values of $E_0$ for the same set of parameters as in (c), projected onto the $N$-$E_0$ plane. The ground state energy $E_0$ shows a $\mu$ dependent, oscillatory behavior in the system size $N$, where the maximum energies follow with very good accuracy an $a_\mu \exp(-Nd/\xi)$ rule for sufficiently large values of $N$, with $\xi$ from Eq. (\ref{equation: decay length}) and the numerical prefactor $a_\mu$.}
	\label{figure: numerical data groundstate energy}
\end{figure}

The features predicted analytically above are also clearly visible in the numerical calculations. The lowest positive energy eigenvalues $E_0$ of a finite Kitaev chain, with the Hamiltonian given by Eq. \eqref{equation: Kit. Hamiltonian, fermionic operators, realspace} and varying parameters, are shown in Fig.~\ref{figure: numerical data groundstate energy}. The phase diagram in Fig.~\ref{figure: numerical data groundstate energy} (a) is the numerical equivalent of that shown in Fig.~\ref{figure: Majorana zero modes}, but for a smaller range of $t$ and $\mu$. Because of the necessarily discrete sampling of the parameter space, the zero energy lines are never met exactly, hence along the Majorana lines we see only a suppression of $E_0$.  Along the border of the topological regime, $\mu \lesssim 2t$, all the boundary states in a finite system have finite energy, as shown in Fig.~\ref{figure: numerical data groundstate energy} (b). Figure ~\ref{figure: numerical data groundstate energy} (c) displays $E_0$ for fixed $t/\Delta = 17$, as a function of $\mu$ and $N$. The number of near-zero energy solutions increases linearly with $N$, according to Eq.~\eqref{equation: eigenvalue zero condition kouachi}. 
It is worth noting that a spatial overlap between Majorana components of the end states in a short system does not need to lead to finite energy (cf. the right column of Fig.~\ref{figure: Majorana zero modes}). The decay length $\xi$ of the in-gap eigenstates, defined in Eq.~\eqref{equation: decay length}, is determined by the ratio $t/\Delta$ and is the same both for the near-zero energy states along the Majorana lines and for the finite energy states between them. It is the {\it maximum} energy of the boundary states that decreases as $E_{0,max} \propto \exp(-Nd/\xi)$ as $N$ is increased, as illustrated in Fig.~\ref{figure: numerical data groundstate energy} (d), in agreement with Eq. \eqref{equation: spectrum: mu=0, even N, edges, cosh/cosh version} and Ref. [\onlinecite{kitaev:physusp2001}, \onlinecite{Zvyagin-2015}, \onlinecite{Zeng-2019}]. The {\it minimum} energy of zero can be reached for any chain length, provided that the chemical potential is appropriately tuned. 

\subsection{Topological protection against Anderson disorder}
\label{subsection: disorder}
\begin{figure}[b]
	\begin{center}
		\includegraphics[width=\columnwidth]{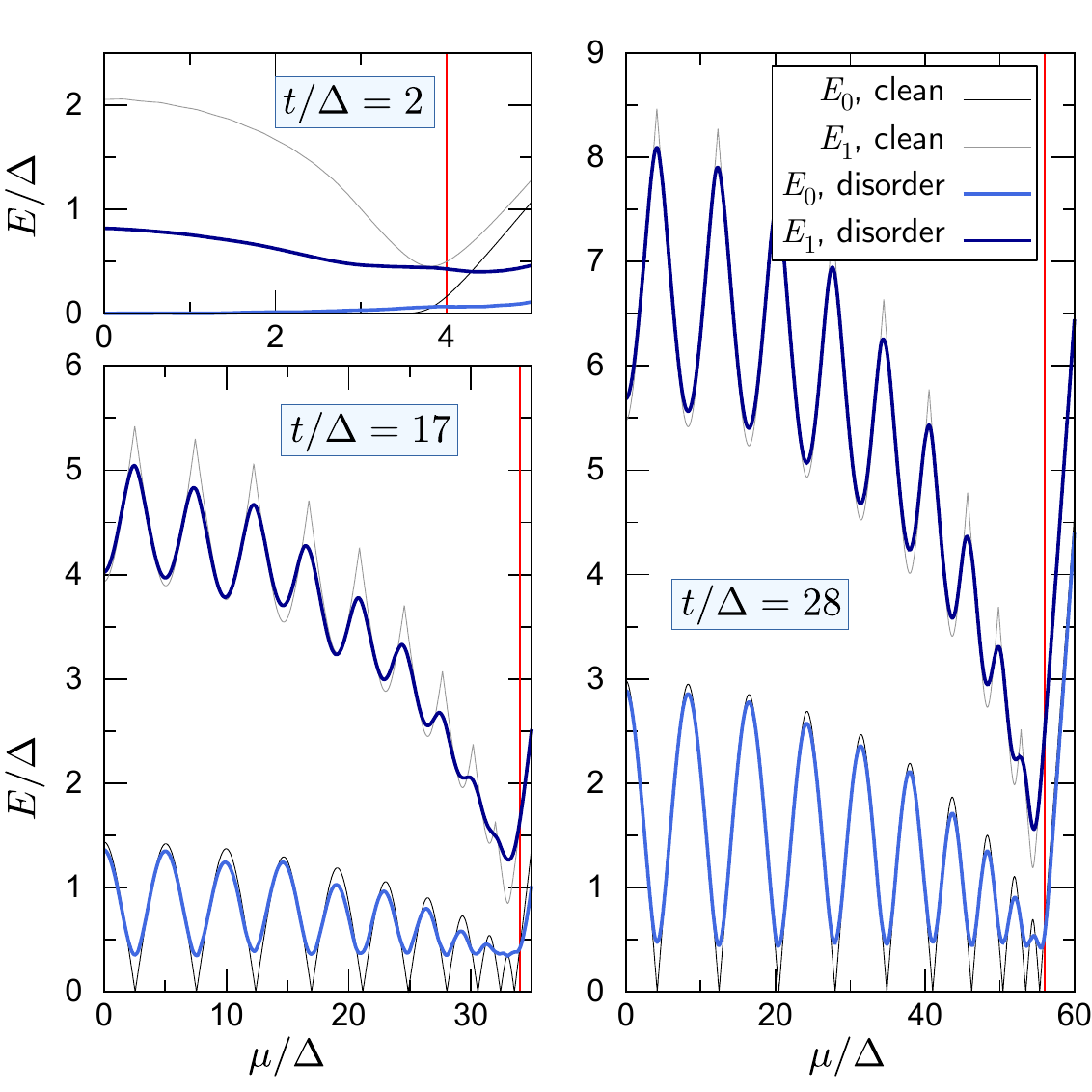}
	\end{center}
	\caption{\label{figure: disorder}
		The lowest ($E_0$) and second lowest ($E_1$) energy eigenvalue of a Kitaev chain with $N=20$ sites as a function of $\mu$ for various ratios of $t/\Delta$. Thin lines correspond to the eigenvalues of a clean chain, thick lines to those of a chain with random on-site disorder $\varepsilon_i \in [-4\Delta,4\Delta]$. The red line marks the boundary of the topological phase.  
	}
\end{figure}
%
%
 One of the most sought after properties of topological states is their stability under perturbations which do not change the symmetry of the Hamiltonian. In order to see whether the non-Majorana topological modes enjoy greater or lesser topological protection than the true Majorana zero modes, we have calculated numerically the spectrum of a Kitaev chain with Anderson-type disorder as a function of $\mu$ for $N=20$ and three different values of $t/\Delta$. The disorder was modeled as an on-site energy term $\varepsilon_i$ whose value was taken randomly from the interval $[-W,W]$. The energies $E_0$, $E_1$ of the two lowest lying states are plotted in Fig.~\ref{figure: disorder}. In all plots $W=4\Delta$, i.e. twice larger than the $\pm2\Delta$ gap at $\mu=0$. Each curve is an average over 100 disorder configurations.
\begin{figure}[hbt]
	\begin{center}
		\includegraphics[width=\columnwidth]{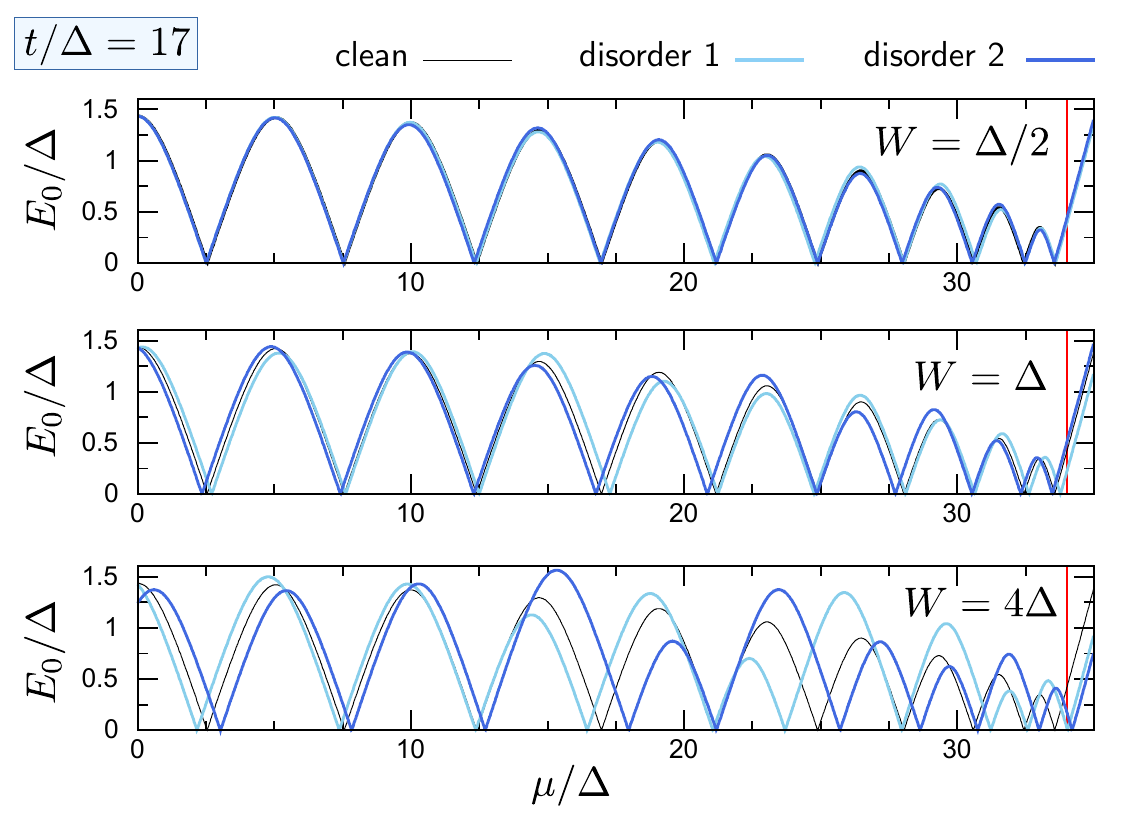}
	\end{center}
	\caption{\label{figure: disorder-noaverage}
		The lowest energy eigenvalue $E_0$ of a Kitaev chain with $N=20$ sites as a function of $\mu$ for $t/\Delta = 17$. Thin lines correspond to the eigenvalues of a clean chain, thick lines to those of a chain with random on-site disorder $\varepsilon_i \in [-W,W]$, for two disorder realizations. The red line marks the boundary of the topological phase.  
	}
\end{figure}
Even with this high value of the disorder it is clear that the energy of the in-gap states is rather robust under this perturbation. For $t/\Delta=2$, i.e. close to the Kitaev point where the boundary states are most localized, they are nearly immune to disorder - its influence is visible only at high $\mu$ and in the energy of the first extended state. For higher ratios of $t/\Delta$, closer to the value of $N+1$ (cf. Eq.~\eqref{equation: criteria/ number of solutions for edge states even N} and the discussion under Eq.~\eqref{equation: spectrum: mu=0, even N, edges, cosh/cosh version}), 
the lowest energy states seem to be strongly perturbed and the Majorana zero modes entirely lost. This is however an artifact of the averaging - the energy $E_0$ plotted in Fig.~\ref{figure: disorder-noaverage} for several disorder strengths $W$ shows that for any particular realization of the disorder the zero modes are always present, but their positions shift to different values of $\mu$. The existence of these crossings is in fact protected against local perturbations and they correspond to a switch of the fermionic parity \cite{Hegde-2015}.

\section{Conclusion}
\label{section: conclusion}
Due to its apparent simplicity, the Kitaev chain is often used as the archetypal example for topological superconductivity in one dimension. Indeed, its bulk spectrum and the associated topological phase diagram are straightforward to calculate, and the presence of Majorana zero modes (MZM) at special points of the topological phase diagram, known as Kitaev points ($\mu=0$, $t/\Delta = \pm 1$ in the notation of this paper), is easy to demonstrate. However, matters become soon complicated when generic values of the three parameters $\mu$, $\Delta$ and $t$ are considered. 

In this work we have provided exact analytical results for the eigenvalues and eigenvectors of a finite Kitaev chain valid for any system size. Such knowledge has enabled us to gain novel insight into the properties of these eigenstates, e.g. their precise composition in terms of Majorana operators and their spatial profile.

Our analysis confirms the prediction of Kao [\onlinecite{Kao-2014}], whereby for finite chemical potential ($\mu\neq0$) zero energy states only exists for discrete sets of $\mu (\Delta,\,t)$ which we dubbed "Majorana lines". We calculated the associated eigenvectors and demonstrated that such states are indeed MZM. Importantly, such MZM come in pairs, and because they are made up of Majorana operators of different types, they are orthogonal. In other words the energy of these modes is exactly zero, even when the two MZM are delocalized along the whole chain (which depends on the state's decay length $\xi$). 

Beside of the Majorana lines, but still inside the topologial region, finite energy boundary states exist. We studied the behavior of the energy $E_0$ of the lowest state numerically as a function of $t/\Delta$ , $\mu/\Delta$ and of the system size $L=Nd$. We found that with good accuracy the maximum energy $E_{0,max} \propto \exp(L/\xi)$. This energy, and hence the energy of all the boundary states, tends to zero in the thermodynamic limit $N\to \infty$. For fixed $N$  the ratio $t/\Delta$ can be varied until the decay length $\xi$ becomes of the order of the system size $L$ and hence the associated  $E_{0,max} $ is not exponentially close to zero energy.

All the boundary states in the topological region, whether of exact zero energy or not, are of topological nature, as predicted by the bulk-edge correspondence. 
This fact is important in the context of topological quantum computation. In fact, whether a state has exact zero energy or not is not relevant for computation purposes, as long as this state is topologically protected. 


Although our treatment using Tetranacci polynomials for $\mu\neq 0$ is general (Secs. \ref{section: influence of non zero chemical potential}, \ref{section: MZM paths in the topological phase diagram for finite N}), we have dedicated special attention to two parameter choices in which the Tetranacci polynomials reduce to the generalized Fibonacci polynomials. The first case if that of zero chemical potential discussed in Sec. \ref{section: spectral analysis}, \ref{section: Eigenvectors and symmetries} of the paper, where the Kitaev chain turns out to be composed of two independent SSH-like chains. This knowledge allows one a better understanding of the difference between an even and an odd number $N$ of sites of the chain. This ranges from different quantization conditions for the allowed momenta of the bulk states, to the presence of MZM. While MZM are always present for $odd$ chains, they only occur at the Kitaev points for even chains. When $\mu$ is allowed to be finite, the Kitaev points develop into Majorana lines hosting MZM for both even and odd chains. In the thermodynamic limit the distinction between even and odd number of sites disappears. At the Majorana lines the fact that $E=0$ decouples the two Majorana sublattices, and again allows us to use the simpler Fibonacci polynomials.

\section{Acknowledgments}

The authors thank the Elite Netzwerk Bayern for financial support via the IGK "Topological Insulators" and the Deutsche Forschungsgemeinschaft via SFB 1277 Project B04. We acknowledge useful discussions with A. Donarini, C. de Morais Smith and M. Wimmer.
\appendix

\section{A note on Fibonacci and Tetranacci Polynomials}
\label{appendix: comments on Fibonacci and Tetrannaci polynomials}

An object of mathematical studies are the Fibonacci numbers $F_n$ $(n\in\mathbbm{N}_0)$ defined by
\begin{align}
	F_{n+2}\,=\,F_{n+1}\,+\,F_n,\quad F_0\,=\,0,\,F_1\,=\,1,
\end{align}
which frequently appear in nature. A more advanced sequence is the one of the Fibonacci polynomials\cite{Webb} $F_n (x)$, where
\begin{align}\label{appendix equation: Webbs Fibonacci Polynomials}
	F_{n+2} (x) \,&=\, x \,F_{n+1}(x)\, +\, F_n(x)\\\notag
	 F_0(x)\,&=\,0,\,F_1(x)\,=\,1,
\end{align}
with an arbitrary complex number $x$ which gives different weight to both terms. The polynomial character becomes obvious after a look at the first few terms 
\begin{align}\label{appendix equation: Webbs Fibonacci Polynomials first terms}
	0,\,1,\,x,\,x^2+1,\,x^3+2x.
\end{align}
The so called generalized Fibonacci polynomials\cite{Hoggatt} are defined by
\begin{align}\label{appendix equation: Hoggatts generalized Fibonacci Polynomials}
	F_{n+2} (x,y) \,&=\, x \,F_{n+1}(x,y)\, +\, y\,F_n(x,y)\\\notag F_0(x,y)\,&=\,0,\,F_1(x,y)\,=\,1,
\end{align}
where $x$, $y$ are two complex numbers. The second weight changes the first elements of the sequence in Eq. \eqref{appendix equation: Webbs Fibonacci Polynomials} to
\begin{align}\label{appendix equation: Hoggatts generalized Fibonacci Polynomials, first terms}
	0,\,1,\,x,\,x^2+y,\,x^3+2x\,y.
\end{align} 
There is a general mapping between the sequences $F_n(x)$ in Eq. \eqref{appendix equation: Webbs Fibonacci Polynomials} and $F_n(x,y)$ in Eq. \eqref{appendix equation: Hoggatts generalized Fibonacci Polynomials}, namely
\begin{align}\label{appendix equation: Fibonacci recursion formula, transformation between unit and unitless}
	F_n(x,y)\,=\,y^{(n-1)/2}\,F_n(x/\sqrt{y}),	
\end{align}
where $F_n(x/\sqrt{y})$ obeys Eq. \eqref{appendix equation: Webbs Fibonacci Polynomials} with $x/\sqrt{y}$ instead of $x$. 

A last generalization is to consider arbitrary initial values $F_{i}(x,y)=f_i$, $i=0,1$ and keeping\cite{Oezvatan-2017}
\begin{align}
	F_{n+2} (x,y) \,&=\, x \,F_{n+1}(x,y)\, +\, y\,F_n(x,y).
\end{align}
This changes the first terms into
\begin{align*}
	f_0,\,f_1,\,xf_1\,+\,yf_0,\,x^2f_1\,+\,y\,(x\,f_0+f_1),\\ x^3 f_1\,+\,y^2f_0\,+\,xy\, (f_0+2f_1).
\end{align*}
The Fibonacci polynomials $\zeta_{2n-1},\,\zeta_{2n}$, $\epsilon_{2n-1},\,\epsilon_{2n}$ we consider in the spectral analysis are of the last kind with $x = \lambda^2+a^2+b^2$, $y = -a^2\,b^2$ ($a = i(\Delta-t),\,b = i(\Delta+t)$) and with different initial values for odd and even index as well as different ones for $\zeta$ and $\epsilon$. The first terms for $\zeta_{2n-1}$ are
\begin{align*}
	\zeta_1 &= \lambda,\\
	\zeta_3 &= \lambda \,(\lambda^2+a^2+b^2),\\
	\zeta_5 &= \lambda \left[(\lambda^2+a^2+b^2)-a^2b^2\right],\\
	\zeta_7 &= \lambda \,(\lambda^2+a^2+b^2) \left[(\lambda^2+a^2+b^2)-2 a^2b^2\right],
\end{align*}
while one finds for $\zeta_{2n}$
\begin{align*}
	\zeta_0 & = 1,\\
	\zeta_2 & = \lambda^2+b^2,\\
	\zeta_4 & = (\lambda^2+a^2)^2+a^2\lambda^2,\\
	\zeta_6 & = (\lambda^2+a^2)^3 +b^2 (\lambda^2+a^2)^2+ a^2\lambda^2  (\lambda^2+a^2)+a^2b^4.
\end{align*}
The expressions for $\epsilon_{2n}$ ($\epsilon_{2n-1}$) follow from the ones of $\zeta_{2n}$ ($\zeta_{2n-1}$) by exchanging $a$ and $b$.

A closed form for Fibonacci numbers/ polynomials is called a Binet form, see for example Ref.~[\onlinecite{Oezvatan-2017}]. In the case of $\zeta_n$ this form is given in Eqs. \eqref{appendix: closed form zeta_odd}, \eqref{appendix: closed form zeta_even}.

In order to obtain the general quantization condition for the wavevectors of the Kitaev chain we face further generalizations of Fibonacci polynomials, so called Tetranacci polynomials $\tau_n$, defined by
\begin{align}\label{appendix equation: general Tetranacci polynomials}
	\tau_{n+4}\,=\,x_3\,\tau_{n+3}\,+\,x_2\,\tau_{n+2}\,+\,x_1\,\tau_{n+1}\,+\,x_0\,\tau_n
\end{align}  
with four complex variables $x_0,\,\ldots,\,x_3$ and four starting values $\tau_0,\,\ldots,\,\tau_3$. These polynomials are a generalisation of Tetranacci numbers \cite{McLaughlin-1979, Waddill-1992} and their name originates from the four terms on the r.h.s. of Eq. \eqref{appendix equation: general Tetranacci polynomials}. The form of Tetranacci polynomials $\xi_j$ we deal with in this work, is provided by Eq. \eqref{appendix equation: tetranacci recursion formula}.

\section{Spectrum for $\mu =0$}
\label{appendix: spectrum zero mu}

\subsection{Characteristic polynomial in closed form}
\label{appendix: characteristic polynomial}

The full analytic calculation of the spectrum is at best performed in the basis of Majorana operators $\gamma_j^{A(B)}$, ordered according to the chain index
\begin{align}
	\hat{\Psi}_{M,\,co}\,\defl\,\left(\gamma_1^A,\,\gamma_1^B,\,\gamma_2^A,\,\gamma_2^B,\,\ldots,\,\gamma_N^A,\,\gamma_N^B\right)^\mathrm{T}.
\end{align}
Then the BdG Hamiltonian becomes block tridiagonal
\begin{align}
	\mathcal{H}_{M,\,co}^{\mathrm{KC}}\,=\,\left[\begin{matrix}
	A & B\\
	B^\dagger &A&B\\
	& B^\dagger &A&B\\
	&&\ddots &\ddots&\ddots\\
	&&& B^\dagger &A&B\\
	&&&&B^\dagger &A
	\end{matrix}
	\right]_{2N\times 2N},
\end{align}
where $A$ and $B$ are $2\times 2$ matrices
\begin{align}
	A=\left[\begin{matrix}
	0 & -i\mu\\
	i\mu &0
	\end{matrix}\right],\quad B=\left[\begin{matrix}
	0 & a\\
	b &0
	\end{matrix}\right].
\end{align}
Since we are interested in the spectrum, we have essentially only to calculate (and factorise) the characteristic polynomial $P_\lambda\left(\mathcal{H}^{\mathrm{KC}}\right)=\mathrm{det}\left(\lambda\mathbbm{1}-\mathcal{H}_\mathrm{KC}\right)$ which reads simply 
\begin{align}
	P_\lambda\,=\,\mathrm{det}\left[\begin{matrix}
	\lambda \mathbbm{1} & -B\\
	-B^\dagger &\lambda \mathbbm{1}&-B\\
	& -B^\dagger &\lambda \mathbbm{1}&-B\\
	&&\ddots &\ddots&\ddots\\
	&&& -B^\dagger &\lambda \mathbbm{1}&-B\\
	&&&&-B^\dagger &\lambda \mathbbm{1}
	\end{matrix}
	\right]_{2N\times 2N},
\end{align}
at zero $\mu$. In the following we will consider $\lambda$ to be just a "parameter", which is not necessarily real in the beginning. Further, we shall impose (only in the beginning) the restriction $\lambda\neq 0$. However, our results will even hold without them. The validity of our argument follows from the fact that the determinant $P_\lambda$ is a smooth function
\begin{align}\label{smoothnes of the char. polynomial}
	P_\lambda\in C^\infty\left(\mathbbm{P}\right),
\end{align}
in the entire parameter space $\mathbbm{P}\defl\mathbbm{C}^3$, which contains $a$, $b$ and $\lambda$. 

The technique we want to use to evaluate $P_\lambda$ is essentially given by the recursion formula of the $2\times 2$ matrices $\Lambda_j$ \cite{Salkuyeh, Molinari}:
\begin{align}\label{equation: recursion formula salkuyeh/molinari}
	\Lambda_j\,=\,\lambda \mathbbm{1}_2- B^\dagger \Lambda_{j-1}^{-1} B,\quad \Lambda_1\defl \lambda \mathbbm{1}_2,
\end{align}
where $j = 1,...,N$ and $P_\lambda = \prod\limits_{j=1}^N \mathrm{det}\left(\Lambda_j \right)$.

The matrices $B$ and $B^\dagger$ are pure off-diagonal matrices and since $\lambda \mathbbm{1}_2$ is diagonal, one can prove that $\Lambda_j$ has the general diagonal form of \makebox{$\Lambda_j\defl\left[\begin{matrix} x_j&0\\ 0&y_j\end{matrix}\right]$} (for all $j$). 
The application of Eq. (\ref{equation: recursion formula salkuyeh/molinari}) leads to a recursion formula for both sequences of entries
\begin{align*}
	x_{j+1}\,&=\,\lambda\,+\,\frac{b^2}{y_j},\\
	y_{j+1}\,&=\,\lambda\,+\,\frac{a^2}{x_j},
\end{align*}
and the initial values are $x_1 = y_1 = \lambda$. We find $x_j$ and $y_j$ to be fractions in general, and define $\zeta_j$, $\epsilon_j$, $\beta_j$ and $\delta_j$ by
\begin{align*}
	x_j&\,\defr \,\frac{\zeta_j}{\beta_j},\\
	y_j&\,\defr\,\frac{\epsilon_j}{\delta_j},
\end{align*}
to take this into account. The initial values can be set as
\begin{align}
	\label{appendix equation: inital values, zeta epsilon 1}
	\zeta_1&=\epsilon_1=\lambda,\\
	\label{appendix equation: inital values, beta delta 1}
	\beta_1&=\delta_1=1,
\end{align}
and after a little bit of algebra we find their growing rules to be 
\begin{align}
	\label{equation: recursion formula, zeta}
	\zeta_{j+1}\,&=\,\lambda\, \epsilon_j\,+\,b^2\,\delta_j,\\
	\label{equation: recursion formula, epsilon}
	\epsilon_{j+1}\,&=\,\lambda\, \zeta_j\,+\,a^2\,\beta_j,\\
	\label{equation: beta and epsilon}
	\beta_{j+1}\,&=\,\epsilon_j,\\
	\label{equation: delta and zeta}
	\delta_{j+1}\,&=\,\zeta_j,
\end{align}
where $j$ starts from $1$. The definitions $\zeta_0\defl\delta_1=1$ and $\epsilon_0\defl\beta_1=1$, enable us to get rid of the $\delta_j$ and $\beta_j$ terms inside Eqs. (\ref{equation: recursion formula, zeta}), (\ref{equation: recursion formula, epsilon}). Hence
\begin{align}
	\label{equation: recursion formula, tridiagonal kind, zeta}
	\zeta_{j+1}\,&=\,\lambda\, \epsilon_j\,+\,b^2\,\zeta_{j-1},\\
	\label{equation: recursion formula, tridiagonal kind, epsilon}
	\epsilon_{j+1}\,&=\,\lambda\, \zeta_j\,+\,a^2\,\epsilon_{j-1}.
\end{align}
which leads to the relations 
\begin{align}
	\label{appendix: equation: zeta_2}
	\zeta_2&=\lambda^2+b^2,\\
	\label{appendix: equation: epsilon_2}
	\epsilon_2\,&=\,\lambda^2+a^2.
\end{align}

We already extended the sequences of $\zeta_j$ and $\epsilon_j$ artificially backwards and we continue to do so, using the Eqs. (\ref{equation: recursion formula, tridiagonal kind, zeta}) and (\ref{equation: recursion formula, tridiagonal kind, epsilon}), starting from $j=-1$ with $\zeta_{-1} = \epsilon_{-1} =0$. Please note there are no corresponding $x_0$, $y_0$ or even $x_{-1}$, $y_{-1}$ expressions, since they would involve division by 0.

The last duty of $\beta_j$ and $\delta_j$ is to simplify the determinant $P_\lambda$ by using the Eqs. (\ref{appendix equation: inital values, beta delta 1}), (\ref{equation: beta and epsilon}) and (\ref{equation: delta and zeta})
\begin{align}\label{appendix: equation: closed form determinant}
	P_\lambda \,=\, \prod\limits_{j=1}^N \mathrm{det}\left(\Lambda_j \right)\,=\,\prod\limits_{j=1}^N x_j\,y_j\,=\,\zeta_N\,\epsilon_N,
\end{align}
which reduces the problem to finding only $\zeta_N$ and $\epsilon_N$. 

Please note that the determinant is in fact independent of the choice of the initial values for $\zeta_1$, $\epsilon_1$, $\beta_1$ and $\delta_1$
in the Eqs. (\ref{appendix equation: inital values, zeta epsilon 1}) and (\ref{appendix equation: inital values, beta delta 1}). Further, Eqs. (\ref{equation: recursion formula, tridiagonal kind, zeta}), (\ref{equation: recursion formula, tridiagonal kind, epsilon}) and (\ref{appendix: equation: closed form determinant}) together show the predicted smoothness of $	P_\lambda$ in $\mathbbm{P}$ and all earlier restrictions are not important anymore. Finally we consider $\lambda$ to be real again.

Even though it seems that we are left with the calculation of two polynomials, we need in fact only one, because both are linked via the exchange of $a$ and $b$. Note that $\lambda$ is considered here as a number and thus does not depend on $a$ and $b$. Further, the dispersion relation is invariant under this exchange.

The connection of $\zeta_j$ and $\epsilon_j$ for all $j\ge -1$ is
\begin{align*}
	\zeta_j\equiv\zeta_j(a,b)\,&=\,\epsilon_j(b,a),\\
	\epsilon_j\equiv\epsilon_j(a,b)\,&=\,\zeta_j(b,a),
\end{align*}
and can be proven via induction using Eqs. (\ref{equation: recursion formula, tridiagonal kind, zeta}), (\ref{equation: recursion formula, tridiagonal kind, epsilon}). Decoupling $\zeta_j$ and $\epsilon_j$ yields
\begin{align}\label{appendix: equation: fibonacci recursion formula}
	\zeta_{j+2}\,=\,\left[\lambda^2+a^2+b^2\right]\,\zeta_{j}\,-\,a^2 b^2\,\zeta_{j-2},
\end{align}
where one identifies them as (generalized) Fibonacci polynomials \cite{Hoggatt, Webb}. The qualitative difference between even and odd number of sites is a consequence of Eq. (\ref{appendix: equation: fibonacci recursion formula}) and the initial values for $\zeta_j$.

The next step is to obtain the closed form expression of $\zeta_j$ ($\epsilon_j$), the so called Binet form. We focus exclusively on $\zeta_j$. 

One way to keep the notation easier is to introduce $x\defl\lambda^2+a^2+b^2$, $y\defl a^2b^2$, $v_j\defl\zeta_{2j}$ and $u_j\defl \zeta_{2j-1}$, such that $u_j$ ($v_j$) obey
\begin{align*}
	u_{j+1}\,=\,x\,u_j\,-\,y\,u_{j-1}.
\end{align*}
The Binet form can be obtained by using a power law ansatz $u_j\,\propto r^j$, leading to two fundamental solutions 
\begin{align}
	r_{1,2}\,=\,\frac{x\,\pm\,\sqrt{x^2-4\,y}}{2}.
\end{align}
Please note that this square root is always well defined, which can be seen 
in the simplest way by setting $\lambda$ to zero.  Consequently, the difference between $r_1$ and $r_2$ is never zero.

A general solution of $u_j$ ($v_j$) can be achieved with a superposition of $r_{1,2}$ with some coefficients $c_{1,2}$, 
\begin{align*}
	u_j\,=\,c_1\,r_1\,+\,c_2\,r_2,
\end{align*}
due to the linearity of their recursion formula. Both constants $c_1$ and $c_2$ are fixed by the initial values for $\zeta_j$, for example $u_0\,=\,\zeta_{-1}\,=\,0$ and $u_1\,=\,\zeta_1\,=\,\lambda$ and similar for $v_j$. After some simplifications, we finally arrive at
\begin{align}
	\label{appendix: closed form zeta_odd}
	\zeta_{2j-1}\,&=\,\lambda\,\frac{r_1^j-r_2^j}{r_1-r_2},	\\
	\label{appendix: closed form zeta_even}
	\zeta_{2j}\,&=\,\frac{\left[\lambda^2+b^2\right]\,\left(r_1^j-r_2^j\right)\,-\,r_1\,r_2\left(r_1^{j-1}-r_2^{j-1}\right)}{r_1-r_2},
\end{align}
in agreement with Ref.~[\onlinecite{Hoggatt, Webb, Oezvatan-2017}]. The validity of the solutions is guaranteed by a proof via induction, where one needs mostly the properties of $r_{1,2}$ to be the fundamental solutions. The exchange of $a$ and $b$ leads to the expressions
\begin{align}
	\label{appendix: closed form epsilon_odd}
	\epsilon_{2j-1}\,&=\,\lambda\,\frac{r_1^j-r_2^j}{r_1-r_2},	\\
	\label{appendix: closed form epsilon_even}
	\epsilon_{2j}\,&=\,\frac{\left[\lambda^2+a^2\right]\,\left(r_1^j-r_2^j\right)\,-\,r_1\,r_2\left(r_1^{j-1}-r_2^{j-1}\right)}{r_1-r_2},
\end{align}
where we used that $r_{1,2}$ is symmetric in $a$ and $b$
At this stage we have the characteristic polynomial in closed form for all $\Delta$, $t$ and more importantly for all sizes $N$ at zero $\mu$.

We can already anticipate the twice degenerated eigenvalues of the odd sized Kitaev chain, because from the closed forms of $\epsilon_j$ and $\zeta_j$ it follows immediately
\begin{align}\label{equation: alpha odd is equal to gamma odd}
	\zeta_{\mathrm{odd}}\,=\,\epsilon_\mathrm{odd}.
\end{align}
Notice that Eq. (\ref{equation: alpha odd is equal to gamma odd}) is important to derive the characteristic polynomial via the SSH description of the Kitaev BdG Hamiltonian at $\mu=0$ and to show the equivalence to the approach used here. It is recommended to use the determinant formula in Ref.~[\onlinecite{Usmani}] together with Eqs. (\ref{equation: recursion formula, zeta}), (\ref{equation: recursion formula, epsilon}) for the proof.  

The main steps of the factorisation are mentioned in the next section.
\subsection{Factorisation of generalized Fibonacci polynomials}
\label{appendix: factorisation of the fibonnacis}
The trick to factorise our Fibonacci polynomials\cite{Webb, Hoggatt} bases on the special form of $r_{1,2}$. The ansatz is to look for the eigenvalues $\lambda$ in the following form
\begin{align}\label{equation: factorisation ansatz for bulk, fibonacci}
	x=2\sqrt{y}\,\cos(\theta),
\end{align}
which is actually the definition of $\theta$. The hermiticity of the Hamiltonian enforces real eigenvalues and consequently $\theta$ can be chosen either real, describing extended solutions, or pure imaginary, which is connected to decaying states. The ansatz leads to an exponential form of the fundamental solutions
\begin{align*}
	r_1\,&=\,\sqrt{y}\,e^{i\,\theta},\\
	r_2\,&=\,\sqrt{y}\,e^{-i\,\theta},
\end{align*}
and we consider $\theta\in\mathbbm{R}$ first. Thus, we find the eigenvalues for odd $N$
\begin{align*}
	\epsilon_{N}\,=\,\zeta_{N}\,=\,\lambda\,\frac{\sin\left(\frac{N+1}{2}\,\theta\right)}{\sin\left(\theta\right)}\,\sqrt{y}^{\frac{N-1}{2}}\,				=\,0.
\end{align*}
One obvious solution is $\lambda = 0$. The introduction of $2\,kd=\theta$, where $d$ is the lattice constant of the Kitaev chain, leads to:
\begin{align}\label{equation: quanisation condition odd cases mu zero}
	\frac{\sin\left[\left(N+1\right)\,kd\right]}{\sin\left(2\,kd\right)}=\,0,
\end{align} 
and solutions inside the first Brillouin-zone are given by
\begin{align*}
	k_n d\,=\,\frac{n\pi}{N+1}\,
\end{align*}
where $n$ runs from $1,\,...,\,N$ without $(N+1)/2$. Please note that Eq. (\ref{equation: quanisation condition odd cases mu zero}) cannot be satisfied for $N=1$. 

The even $N$ case requires more manipulations. We first rearrange Eq. (\ref{appendix: closed form zeta_even}) as
\begin{align*}
	\zeta_{2j}\,=\,\frac{\left(\lambda^2+b^2-r_2\right)\,r_1^j\,-\,\left(\lambda^2+b^2-r_1\right)\,r_2^j}{r_1-r_2}.
\end{align*}
The expressions $\lambda^2+b^2-r_{1,2}$ are simplified to
\begin{align*}
	\lambda^2+b^2-r_{1}\,&=\,x-a^2-r_{1}\,=\,\sqrt{y}\,e^{-i\theta}-a^2,\\
	\lambda^2+b^2-r_{2}\,&=\,\sqrt{y}\,e^{i\theta}-a^2.
\end{align*}
In the end $\zeta_{2j}$ becomes
\begin{align}\label{equation: zeta 2j in theta form}
	\zeta_{2j}\,=\,\left(\sqrt{y}\right)^{j+1}\frac{\sqrt{y}\,\sin\left[\theta\left(j+1\right)\right]\,-\,a^2\,\sin\left(\theta j\right)}{\sin\left(\theta\right)}.
\end{align}
Note that the competition of $\Delta$ and $t$ is hidden inside the square root
\begin{align*}
	\sqrt{y}\,=\,\left\{\begin{matrix}
	\Delta^2-t^2,& \mathrm{if~}\vert\Delta \vert>\vert t\vert\\
	t^2-\Delta^2, & \mathrm{else}
	\end{matrix}\right.,
\end{align*}
affecting both the quantization condition and the dispersion relation $E_\pm(k)\,=\,\lambda(\theta)$, which follow from Eq. (\ref{equation: factorisation ansatz for bulk, fibonacci}). However, both situations lead to the same result, because the momenta and the spectrum are shifted by $\pi/2$  (with respect to $kd$). From $\zeta_N$ it follows:
\begin{align*}
	&\Delta\,\frac{\sin\left[kd\,\left(N+1\right)\right]\,\cos(kd)}{\sin(2\,kd)}\\
	&\quad +\,t\,\frac{\cos\left[kd\,\left(N+1\right)\right]\,\sin(kd)}{\sin(2\,kd)}\,=\,0,
\end{align*}
or in shorter form
\begin{align}
	\tan\left[kd\,\left(N+1\right)\right]\,=\,-\frac{\Delta}{t}\tan(kd),\quad kd\neq 0, \frac{\pi}{2}
\end{align}
for even $N$. The polynomial $\epsilon_N$ can be treated in the same way leading to
\begin{align}
	\tan\left[kd\,\left(N+1\right)\right]\,=\,\frac{\Delta}{t}\tan(kd),\quad kd\neq 0, \frac{\pi}{2}.
\end{align}
From Eq. (\ref{equation: factorisation ansatz for bulk, fibonacci}) follows the bulk spectrum for all $N$,
\begin{align*}
	\lambda(\theta)=E_\pm(kd)\,=\,\pm\sqrt{4\Delta^2\sin^2(kd)\,+\,4t^2\cos^2(kd)},
\end{align*}
in agreement with Eqs. (\ref{equation: spectrum: mu=0, odd N, k depending part}) and (\ref{equation: spectrum: mu=0, even N}).

The case of decaying states is similar, but not just done by replacing $k$ by $iq$. The following case is only valid for even $N$, since we have already all $2N$ eigenvalues of the odd $N$ case.

Our ansatz is modified to
\begin{align*}
	x=-2\sqrt{y}\,\cosh(\theta),
\end{align*}
by an additional minus sign, which is important to find the decaying state solutions. After some manipulations $\zeta_N=0$ yields the quantization conditions
\begin{align}
	\label{appendix: equation: decaying states, quantistaion condition 1}
	\frac{t\,\tanh\left[qd\,\left(N+1\right)\right]\,+\,\Delta\,\tanh(qd)}{\sinh(2\,qd)}\,&=\,0,\quad \vert\Delta\vert\ge\vert t\vert,\\					\label{appendix: equation: decaying states, quantistaion condition 2}
	\frac{\Delta\,\tanh\left[qd\,\left(N+1\right)\right]\,+\,t\,\tanh(qd)}{\sinh(2\,qd)}\,&=\,0,\quad \vert t\vert\ge\vert \Delta\vert,
\end{align}
where $qd = \theta/2$. The conditions for $qd=0$ as solution, corresponding to infinite decay length $\xi$, turn out to be $\pm t/\Delta=N+1$ (if $\vert t\vert\ge\vert \Delta\vert$) or $\pm\Delta /t=N+1$ (else) and follow by applying the limit $qd\rightarrow 0$ on Eqs. (\ref{appendix: equation: decaying states, quantistaion condition 1}) and (\ref{appendix: equation: decaying states, quantistaion condition 2}).

A last simplification can be done for $qd\neq 0$
\begin{align*}
	\tanh\left[qd\,\left(N+1\right)\right]\,&=\,-m\tanh(qd),
\end{align*}
where we introduced
\begin{align*}
	m = \left\{\begin{matrix}
	\frac{\Delta}{t} & \mathrm{if} \vert \Delta \vert > \vert t \vert\\
	\\
	\frac{t}{\Delta} & \mathrm{if} \vert t \vert > \vert \Delta \vert
	\end{matrix}\right. .
\end{align*}

The criterion to find a wave vector is that $(-m)\ge1$, but not larger than $N+1$, which leads then to exactly two solutions $\pm q$ and otherwise to none. The corresponding eigenvalues can be obtained from 
\begin{align*}
	E_\pm (qd)\,=\,\pm\sqrt{4\,t^2\cosh^2(qd)-4\Delta^2\sinh^2(qd)},\quad \vert\Delta\vert\ge\vert t\vert,\\
	E_\pm (qd)\,=\,\pm\sqrt{4\,\Delta^2\cosh^2(qd)-4t^2\sinh^2(qd)},\quad \vert t\vert\ge\vert \Delta\vert,
\end{align*}
which can be zero. The results for $\epsilon_N$ can be obtained by replacing $t$ with $-t$ everywhere.
\section{Eigenvectors for zero $\mu$}
\label{appendix: eigenvectors}
The simplest way to calculate the eigenstates of the Kitaev Hamiltonian is the use of the SSH-like basis for $\mu = 0$ from Eq. (\ref{equation: Kitaev Hamiltonian/ matrix in SSH basis}). We define the eigenvector $\vec{\psi}$ as
\begin{align*}
	\vec{\psi}\,=\,\left(\vec{v}_\alpha,\,\vec{v}_\beta\right)^\mathrm{T},
\end{align*}
for all $N$ to respect the structure of the Hamiltonian. Moreover, one can search for solutions belonging only to one block $(\vec{v}_\alpha,\,\vec{0}_\beta)^\mathrm{T}$ or $(\vec{0}_\alpha,\,\vec{v}_\beta)^\mathrm{T}$, without any restriction. In other words either is $\vec{v}_\alpha$ zero or $\vec{v}_\beta$ and we will mention only non zero entries from now on. We report here only about the calculation of $\vec{v}_\alpha$, because the one for $\vec{v}_\beta$ can be performed analogously.

The general idea behind the eigenvector calculation of tridiagonal matrices is given in Ref.~[\onlinecite{Shin-97}], but we consider here all possible configurations of parameters.
\subsection{$N$ even}
\label{appendix: eigenvectors even N}
The sublattice vectors are defined via the $N$ entries
\begin{align*}
	\vec{v}_\alpha\,&=\,\left(x_1,\,y_1,\,x_2,\,y_2,\,\ldots ,\,x_{N/2},\,y_{N/2}\right)^\mathrm{T},\\
	\vec{v}_\beta\,&=\,\left(\xcal_1,\,\ycal_1,\,\xcal_2,\,\ycal_2,\,\ldots ,\,\xcal_{N/2},\,\ycal_{N/2}\right)^\mathrm{T}.
\end{align*}
Solving \makebox{$\left(\mathcal{H}_\alpha^\mathrm{even}-\lambda\,\mathbbm{1}\right)\,\vec{v}_\alpha=0$} leads to
\begin{align}
	\label{equation: eigenvector system for even N and for gamma, start A}
	 \,a\,y_1\,&=\,\lambda\,x_1,\\
	 \label{equation: eigenvector system for even N and for gamma, start B}
	-a\,x_{N/2}\,&=\,\lambda\,y_{N/2},
\end{align}
and
\begin{align}
	\label{equation: eigenvector system for even N and for gamma, bulk A}
	 b\,x_{l+1}\,-\,a\,x_l\,&=\,\lambda\,y_l,\\
	 \label{equation: eigenvector system for even N and for gamma, bulk B}
	 a\,y_{l+1}\,-\,b\,y_l\,&=\,\lambda\,x_{l+1},
\end{align}
where $l$ runs from 1 to $(N/2)-1$. The coupled equations (\ref{equation: eigenvector system for even N and for gamma, start A}) - (\ref{equation: eigenvector system for even N and for gamma, bulk B}) for the entries of the eigenvector are continuous in all parameters. However, resolving to the $x_l$'s and $y_l$'s may lead to problems for certain values of $\Delta$, $t$ and $\lambda$. 

\textbf{Case 1}. $\vert\Delta\vert \neq \vert t\vert$. The parameter setting excludes $\lambda = 0$, as we found from our spectral analysis in Sec. \ref{section: spectral analysis}. Eqs. (\ref{equation: eigenvector system for even N and for gamma, bulk A}), (\ref{equation: eigenvector system for even N and for gamma, bulk B}) are used to disentangle $x$'s and $y$'s. Both sequences obey
\begin{align}\label{equation: eigenvector system, N even, Fibonacci}
	y_{l+1}\,=\,\frac{\lambda^2+a^2+b^2}{ab}\,y_l\,-\,y_{l-1},
\end{align}
where $l=1,\,\ldots\,N/2$. Thus the $y$'s and $x's$ are Fibonacci polynomials \cite{Hoggatt, Webb}. The difference to the previous ones found for the spectrum is that the new version can be dimensionless in physical units, depending on the initial values. The transformation formula\cite{Hoggatt} to pass from the unitless recursion formula to the other one is given by Eq. \eqref{appendix equation: Fibonacci recursion formula, transformation between unit and unitless}.

The Binet form of the dimensionless sequences is obtained with same treatment as  for the spectrum. The power ansatz $y_l\propto f^l$ yields the fundamental solutions $f_{1,2}$, obeying
\begin{align}
	\label{appendix: equation: fundamental solution f_1}
	f_1\,+\,f_2\,&=\,\frac{\lambda^2+a^2+b^2}{ab},\\
	\label{appendix: equation: fundamental solution f_2}
	f_1\,\cdot\,f_2\,&=\,1,\\
	f_1\,&\neq\, f_2.
\end{align}
Due to the linearity of the recursion formula, the most generic ansatz for $y_l$ is
\begin{align*}
	y_l\,=\,c_1\,f_1^l\,+\,c_2\,f_2^l,
\end{align*}
where the constants $c_{1,\,2}$ follow from the initial values of $y_{1,\,2}$. The calculation of both constants leads to
\begin{align*}
	y_l\,=\,y_2\,T_{l-1}\,-\,y_1\,T_{l-2},
\end{align*}
where $T_l$ is simply\cite{Shin-97}
\begin{align*}
	T_l\defl\frac{f_1^l-f_2^l}{f_1-f_2}.
\end{align*}
Analogously we find
\begin{align*}
	x_l\,=\,x_2\,T_{l-1}\,-\,x_1\,T_{l-2}.
\end{align*}
A short comment on the initial values $y_{1,\,2}$. A hermitian matrix is always diagonalisable, regardless of degenerations in its spectrum and an eigenvector is well defined only up to the prefactor. Consequently we have the freedom to choose one component of $\vec{v}_\alpha$. This choice will in turn define all remaining initial values.

Consider for example $x_1$ to be a fixed value of our choice. We find $y_1$, $x_2$ and $y_2$ to be
\begin{align*}
	y_1\,&=\,\frac{\lambda}{a}\,x_1,\\
	y_2\,&=\,\frac{\lambda}{a}\,\frac{\lambda^2+a^2+b^2}{ab}\, x_1,\\	
	x_2\,&=\,\frac{\lambda^2+a^2}{ab}\,x_1.
\end{align*}
The $y_2$ can be rewritten as $y_2 = y_1\left[f_1 +f_2\right]$ which leads to a simpler form of all $y_l$'s \cite{Shin-97}
\begin{align}
	y_l\,=\,\frac{\lambda}{a}\,x_1\,T_l.
\end{align}
After a bit of algebra, one finds $x_l$ to be
\begin{align}\label{appendix equation: half the boundary condition N even, alpha chain}
	x_l\,=\,x_1\,\left[T_l-\frac{b}{a}T_{l-1}\right].
\end{align}
So far we found the general solutions of the recursion formulas Eqs. (\ref{equation: eigenvector system for even N and for gamma, start A}) - (\ref{equation: eigenvector system for even N and for gamma, bulk B}). The comparison of Eq. (\ref{equation: eigenvector system for even N and for gamma, start B}) and Eq. (\ref{equation: eigenvector system for even N and for gamma, bulk A}) leads to 
\begin{align}\label{eigenvectors: quantization condition on level of the entries}
	x_{\frac{N}{2}+1}\,=\,0,
\end{align}
because the recursion formulas themselves do not care about any index limitation. The last equation means only that the wave function of a finite system has to vanish outside, at the boundary, yielding the quantization rule. 

The extended states can be obtained with
\begin{align*}
	f_1\,&=\,e^{2i\,kd},\\
	f_2\,&=\,e^{-2i\,kd},
\end{align*}
where Eqs. (\ref{appendix: equation: fundamental solution f_1}), (\ref{appendix: equation: fundamental solution f_2}) relate $kd$ and $\lambda$, and   $T_l$ is recast as
\begin{align}\label{appendix equation: T_l even, extended state}
	T_l\,=\,\frac{\sin(2\, kd\, l)}{\sin(2\,kd)}.
\end{align} 
The last equation for $T_l$ yields via Eqs. (\ref{appendix equation: half the boundary condition N even, alpha chain}), (\ref{eigenvectors: quantization condition on level of the entries}) the quantization condition. Thus, the momenta $k$ obey
\begin{align*}
	\tan\left[kd\,\left(N+1\right)\right]\,=\,\frac{\Delta}{t}\,\tan\left(kd\right),
\end{align*}
where each solution defines two states with the energy $E_\pm$ from Eq. (\ref{equation: spectrum: mu=0, even N}).

The decaying states depend strongly on the interplay of $\Delta$ and $t$. The ansatz is
\begin{align*}
	f_1\,&=\,s\,e^{2\,qd},\\
	f_2\,&=\,s\,e^{-2\,qd},
\end{align*}
where $s$ is defined as
\begin{align*}
	s = \left\{\begin{matrix}
	+1, & \vert \Delta\vert >\vert t\vert\\
	\\
	-1, & \vert t\vert > \vert \Delta\vert
	\end{matrix}\right. .
\end{align*}
Finally the coefficient $T_l$ becomes 
\begin{align*}
	T_l (qd)\defl s^{l-1}\,\,\,\frac{\sinh(2\,qd\,l)}{\sinh(2\,qd)}.
\end{align*}
The proper $q$, if existent, leads to two states and satisfies
\begin{align*}
	\tanh\left[qd\,\left(N+1\right)\right]\,=\,m\,\tanh\left(qd\right),
\end{align*}
where $m$ is
\begin{align}
	m\,\defl \,\left\{	\begin{matrix}
	\frac{\Delta}{t},\quad\,\mathrm{if}\, \vert \Delta\vert \ge \vert t\vert	\\
	\\
	\frac{t}{\Delta},\quad\,\mathrm{if}\, \vert t\vert \ge \vert \Delta\vert	
	\end{matrix}		\right. .
\end{align}
In total we have already all $N$ non normalized states with respect to the chain $\alpha$ and this approach holds as long as $\vert\Delta\vert \neq \vert t \vert$. \\
The remaining cases start again from the Eqs. (\ref{equation: eigenvector system for even N and for gamma, start A}) - (\ref{equation: eigenvector system for even N and for gamma, bulk B}).

\textbf{Case 2}. Eigenvectors at the Kitaev point. We consider now $\Delta\,=\,-t$, or $b=0$, and we have to solve
\begin{align*}
	a\, y_1\,&=\,\lambda\,x_1,\\
	-a\, x_{N/2}\,&=\,\lambda\,y_{N/2},\\
	-a\, x_l\,&=\,\lambda\,y_l,\\
	a\, y_{l+1}\,&=\,\lambda\,x_{l+1},
\end{align*}
where $l$ runs from $1$ to $(N-2)/2$. A zero energy mode is obviously not existing on the $\alpha$ subchain, because $\lambda\,=\,0$ would lead to $\vec{v}_\alpha\,=\,0$ which is not an eigenvector by {definition}. These zero modes belong to the subchain $\beta$ for $\Delta = -t$. The only possible eigenvalues for the extended modes of the $\alpha$ chain are $\lambda=\pm 2t$ \cite{kitaev:physusp2001, Aguado}, see Eq. (\ref{equation: spectrum: mu=0, even N}). Recalling $a\,=\,-2i\,t$, leads to $N/2$ independent solutions of dimerised pairs $(x_l,\, y_l)$ with $y_l=\mp ix_l$ and the signs are with respect to the eigenvalues.   

The last cases belong to $\Delta=t$ ($a=0$), where we search for the solution of
\begin{align*}
	 \lambda\,x_1\,&=\,0,\\
	 \lambda\,y_{N/2}\,&=\,0,\\
 	 b\,x_{l+1}\,&=\,\lambda\,y_l,\\
 	-b\,y_l\,&=\,\lambda\,x_{l+1},
\end{align*}
where $l$ runs from $1$ to $(N-2)/2$. The first (second) line clearly states that either $\lambda$ is zero and/or $x_1$ ($y_{N/2}$).
The zero $\lambda$ means on the one hand that most entries vanish $x_2=x_3=\ldots=x_{N/2}=0$ and $y_1=y_2=\ldots=y_{(N-2)/2}=0$, since $b=2i t\neq 0$ to avoid a trivial Hamiltonian. On the other hand we have two independent solutions, first
\begin{align*}
	x_1\,&=\,1,\\
	y_{N/2}\,&=\,0,
\end{align*} 
and second 
\begin{align*}
	x_1\,&=\,0,\\
	y_{N/2}\,&=\,1,
\end{align*}
describing the isolated MZM's at opposite ends of the chain. In the case of $\lambda=\pm 2t$, we have $N-2$ independent solutions in form of pairs $(y_l,\,x_{l+1})$ with $y_l\,=\,\pm i\,x_{l+1}$. 

The non trivial solutions for $\vec{v}_\beta$ follow by replacing $x_l\rightarrow \xcal_l$, $y_l\rightarrow \ycal_l$ and $t\rightarrow -t$ everywhere.
\subsection{$N$ odd}
The eigenvectors have similar shape
\begin{align*}
	\vec{v}_\alpha\,&=\,\left(x_1,\,y_1,\,x_2,\,y_2,\,\ldots ,\,x_{\frac{N-1}{2}},\,y_{\frac{N-1}{2}},\,x_{\frac{N+1}{2}}\right)^\mathrm{T},\\
	\vec{v}_\beta\,&=\,\left(\xcal_1,\,\ycal_1,\,\xcal_2,\,\ycal_2,\,\ldots ,\,\xcal_{\frac{N-1}{2}},\,\ycal_{\frac{N-1}{2}},\,\xcal_{\frac{N+1}{2}}\right)^\mathrm{T},
\end{align*}
but the last entry is different compared to the even $N$ case. Although both subchains have the same spectrum, it is possible to consider a superposition of eigenstates of the full Hamiltonian which belongs to only one chain, for example $\alpha$. We consider $\vec{v}_\beta$ to be zero. 

The eigenvector system for $\vec{v}_\alpha$ reads
\begin{align*}
	a\,y_1\,&=\,\lambda\,x_1,\\
	-b\,y_{\frac{N-1}{2}}\,&=\,\lambda\,x_\frac{N+1}{2},
\end{align*}
and
\begin{align*}
	b\,x_{i+1}\,-\,a\,x_i\,&=\,\lambda\,y_i,\\
	a\,y_{l+1}\,-\,b\,y_l\,&=\,\lambda\,x_{l+1},
\end{align*}
with $l =1,\,\ldots,\,\frac{N-3}{2}$ and $i=1,\,\ldots,\,\frac{N-1}{2}$. 

If we consider $a,\,b$ and $\lambda$ all to be different from zero, we find again that the entries of $\vec{v}_\alpha$ are Fibonacci polynomials obeying the same recursion formula as in the even $N$ case and lead to the same solution
\begin{align*}
	y_i\,&=\,\frac{\lambda}{a}\,T_i\,x_1,\\
	x_l\,&=\,\left[T_l\,-\,\frac{b}{a}\,T_{l-1}\right]\,x_1,
\end{align*}
where $l=1,\,\ldots,\frac{N+1}{2}$ and $T_{l,\,(i)}$ is as before. The ansatz $f_1\,=\,e^{2i\,k d}$, $f_2\,=\,e^{-2i\,k d}$ for the extended states influences $T_l$ ($T_i$ analogously)
\begin{align*}
	T_{l}\,=\,\frac{\sin(2\,k d\,l)}{\sin(2\,k d)},
\end{align*}
and leads via
\begin{align*}
	y_{\frac{N+1}{2}}\,=\,0,
\end{align*}
to the equidistant quantization $k\,\equiv\,k_n\,=\,\frac{n\,\pi}{N+1}$ with $n\,=\,1,\,\ldots,\,(N-1)/2$, due to the number of eigenvectors of the single SSH-like chain. Both chains share the same spectrum for odd $N$ and thus we have in total $n\,=\,1,\,\ldots,\,N$, \mbox{$n\neq (N+1)/2$.}

We report here shortly on all other parameter situations.

i) If we consider $a$ and $b$ to be different from zero, but $\lambda=0$, we find only one state
\begin{align}\label{equation: appendix: zero energy mode mu = 0, N odd on the alpha chain}
	x_{l+1}\,=\,\left(\frac{\Delta-t}{\Delta+t}\right)^l\,x_1,
\end{align}
and $l$ runs from $1$ to $(N-1)/2$.

ii) If $\Delta = t$, i.e. $a=0$, but $\lambda\,=\,\pm 2t\,\neq 0$, we find $(N-1)/2$ solutions $(y_l,\,x_{l+1})$ with $y_l\,=\,\pm i\,x_{l+1}$, \makebox{$l=1,\,\ldots,\,(N-1)/2$} and $x_1=0$ for all.

The zero mode of this setting ($\Delta = t$) is a MZM localized on $x_1=1$, while all other components are zero. 
\newline{}\newline
iii) If $\Delta=-t$ ($b=0$) and $\lambda\neq0$ we find $(N-1)/2$ solutions of the form $(x_l,\,y_l)$ with $y_l\,=\,\pm i\,x_l$ \makebox{$l=1,\,\ldots,\,(N-1)/2$} and $x_{\frac{N+1}{2}}=0$ for all of them. The MZM is localised at $x_{\frac{N+1}{2}}=1$ for $b=0$.
\newline{}\newline
The results for the $\alpha$ chain follow again by replacing $x_l\rightarrow \xcal_l$, $y_l\rightarrow \ycal_l$ and $t\rightarrow -t$. \newline
\section{Spectrum for finite $\mu$}
\label{appendix: spectrum finite u}
The BdG Hamiltonian, expressed in the chiral basis $\hat{\Psi}_c=\left(\gamma_1^A,\,\gamma_2^A,\,\ldots,\,\gamma_N^A,\,\gamma_1^B,\,\gamma_2^B,\,\ldots,\,\gamma_N^B\right)^\mathrm{T}$ leads via $\hat{H}_\mathrm{kc}=\frac{1}{2} \hat{\Psi}_c^\dagger\,\mathcal{H}_c\, \hat{\Psi}_c$ to
\begin{align}\label{appendix euquation: Hamiltonian in chiral basis}
	\mathcal{H}_c\,=\,\left[\begin{matrix}
	0_{N\times N} & h\\
	h^\dagger & 0_{N\times N}
\end{matrix}			\right],
\end{align}
where the matrix $h$ is
\begin{align}\label{appendix equation: h}
	h=\left[\begin{matrix}
		-i\mu & a & \\
		-b & -i \mu & a\\
		& -b & -i \mu & a\\
		& & \ddots & \ddots & \ddots \\
		&&& -b & -i \mu & a\\
		&&&& -b & -i \mu 
	\end{matrix}\right]_{N \times N}.
\end{align}
As mentioned in Sec. \ref{section: influence of non zero chemical potential}, we look for a solution of
\begin{align}\label{appendix equation: eigenvector problem to hhdagger}
	hh^\dagger\, \vec{v} \,=\,\lambda^2\, \vec{v},
\end{align}
with $\vec{v} = \left(\xi_1,\,\xi_2\,\ldots,\,\xi_N\right)^\mathrm{T}$ to find the general quantization rule. The entries of the matrix $hh^\dagger$ are
\begin{align*}
	\left(hh^\dagger\right)_{n,m}\,&=\,\delta_{n,m}\,\left[\mu^2\,-\,a^2\,\left(1-\delta_{n,N}\right)\,-\,b^2\,\left(1-\delta_{n,1}\right) 
		\right]\\
		&\quad +\,i\mu\, \left(a-b\right)\,\left[\,\delta_{n,m+1}\,+\, \delta_{n+1,m}\right]\\
		&\quad +\,ab \,\left(\delta_{n,m+2}\,+ \,\delta_{n+2,m}\right),
\end{align*}
and Eq. (\ref{appendix equation: eigenvector problem to hhdagger}) becomes the Tetranacci sequence
\begin{align}\label{appendix equation: tetranacci recursion formula}
	\xi_{j+2}\,&=\,\frac{\lambda^2+a^2+b^2-\mu^2}{ab}\,\xi_j-\xi_{j-2}\notag\\
		&\quad -i\mu\left(\frac{a-b}{ab}\right)\,\left(\xi_{j+1} + \xi_{j-1}\right),
\end{align}
where $j=1,\,\ldots\,N-5$. The missing four boundary terms are
\begin{align*}
	\lambda^2\,\xi_1&=(\mu^2-a^2)\,\xi_1+i\mu (a-b)\,\xi_2+ab\,\xi_3,\\
	\lambda^2\,\xi_2&=(\mu^2-a^2-b^2)\,\xi_2+i\mu (a-b)\left(\xi_1+\xi_3\right)+ab\,\xi_4,\\
	\lambda^2\,\xi_{N-1}&=(\mu^2-a^2-b^2)\xi_{N-1}+i\mu (a-b)\left(\xi_N+\xi_{N-2}\right)\\
		&\qquad +\,ab\,\xi_{N-3},\\
	\lambda^2\,\xi_{N}&=(\mu^2-b^2)\,\xi_{N}+i\mu (a-b)\,\xi_{N-1}+ab\,\xi_{N-2}.
\end{align*}
We extend the Tetranacci sequence from $j = -\infty$ to $j = \infty$, i.e. the index limitations in Eq. (\ref{appendix equation: tetranacci recursion formula}) can be ignored, while $\vec{v}$ still contains only $\xi_1,\,\ldots,\,\xi_N$. Consequently, we can simplify the boundary conditions by using the recursion formula and further any restriction like $N>3$ does not exist. We find
\begin{align}\label{appendix equation: boundary condition 1}
	\xi_{N+1} \,= \,\xi_0 \,= \,0,\\
	b\,\xi_{N+2} \,= \,a\,\xi_N,\\
	\label{appendix equation: boundary condition 4}
	b\,\xi_{1} \,= \,a\,\xi_{-1}.
\end{align}
The procedure we followed in the context of Fibonacci polynomials was to obtain a closed form with the ansatz $\xi_j = r^j,\,r\neq 0$. So we do here on starting from Eq. (\ref{appendix equation: tetranacci recursion formula}). Thus, the characteristic equation for $r$ reads
\begin{align*}
	r^4- \frac{\lambda^2+a^2+b^2-\mu^2}{ab}\,r^2\,+\,1\,+\,i\mu \frac{a-b}{ab}\left(r^3+r\right) \,=\, 0,
\end{align*}
and we have to find all four zeros to determine $\xi_j$ in the end. We introduce two new variables
\begin{align}
	\zeta \,&=\, \frac{\lambda^2+a^2+b^2-\mu^2}{ab},\\
	\eta \,&=\,i\mu\frac{a-b}{ab},
\end{align}
to simplify the expressions in the following. The characteristic equation becomes
\begin{align*}
	r^4\,+\,\eta \,r^3\,-\,\zeta\,r^2\,+\,\eta\,r\,+\,1\,=\,0.
\end{align*}
Dividing by $r^2$ $(r\neq 0)$ and defining $S\defl r\,+\,r^{-1}$ leads to
\begin{align}\label{appendix equation: tetranacci: characteristic equation of S12}
	S^2\,+\,\eta\,S-\zeta-2\,=\,0,
\end{align}
where we can read out the solutions $S_{1,2}$
\begin{align}\label{appendix equation: tetranacci: S12 in explicit form}
	S_{1,2}\,=\,\frac{-\eta\pm \sqrt{\eta^2\,+\,4\,(\zeta+2)}}{2}.
\end{align}
The definition of $S$ amounts to an equation for $r$
\begin{align*}
	r^2\,-\,S\,r\,+\,1\,=\,0.
\end{align*}
Thus one can insert the solutions $S_{1,2}$ and solve for $r$. We find
\begin{align}\label{appendix equation: solution rpm12 in terms of S12}
	r_{\pm j}\,=\,\frac{S_j\pm\sqrt{S^2_j-4}}{2},\quad j = 1,2,
\end{align}
yielding directly $r_{j}\,r_{-j}\,=\, 1$. Here, we choose the ansatz
\begin{align}\label{appendix equation: tetranacci: ansatz for S12}
	S_{1,2}\,\defr\,2\,\cos(\kappa_{1,2}),\quad \kappa_{1,2}\in \mathbbm{C},
\end{align}
which is actually the definition of $\kappa_{1,2}$. Since the coefficients $S_{1,2}$ contain $\lambda$ through the variable $\zeta$, this is in the end an ansatz for $\lambda$. The expression for $\lambda$ follows easily from Eq. (\ref{appendix equation: tetranacci: characteristic equation of S12}) by inserting Eq. (\ref{appendix equation: tetranacci: ansatz for S12}). Using the definiton of $\eta$ and resolving for $\zeta$ first and in a second step for $\lambda$ we finally arrive at Kitaev's bulk formula
\begin{align*}
	\lambda\left(\kappa_{1,2}\right)\,=\,\pm\sqrt{\left[\mu\,+\,2t\cos\left(\kappa_{1,2}\right)\right]^2\,+\,4 \Delta^2\sin^2\left(\kappa_{1,2}\right)}.
\end{align*}
Notice that by construction we have $\lambda=\lambda\left(\kappa_1\right) = \lambda(\kappa_2)$. Alternatively the sum of $S_1$ and $S_2$ leads via Eqs. (\ref{appendix equation: tetranacci: S12 in explicit form}), (\ref{appendix equation: tetranacci: ansatz for S12}) to
\begin{align}\label{appendix equation: connection of k12}
	\cos\left(\kappa_1\right)\,+\,\cos\left(\kappa_2\right)\,=\,-\frac{\mu \,t}{t^2-\Delta^2}.
\end{align}
The use of Eq. (\ref{appendix equation: connection of k12}) on the dispersion relation will indeed yield
\begin{align}\label{appendix equation: degeneracy due to k12}
	\lambda\left(\kappa_1\right) = \lambda(\kappa_2).
\end{align}
Let us return to $\xi_j$. Since the recursion formula in Eq. (\ref{appendix equation: tetranacci recursion formula}) is linear, a superposition of all four solutions $r_{\pm 1}$, $r_{\pm 2}$
\begin{align}
	\xi_j\,&=\,c_1\,r_{+1}\,+\,c_2\,r_{-1}\,+\,c_3\,r_{+2}\,+\,c_4\,r_{-2},
\end{align}
is still a solution with some coefficients $c_{1,2,3,4}\in\mathbbm{C}$. From Eq.\eqref{appendix equation: solution rpm12 in terms of S12} it follows
\begin{align}\label{appendix equation: exponential form of fundamental solutions for tetranaccis}
	r_{\pm j} = e^{\pm i\kappa_{j}}
\end{align}
and thus
\begin{align}\label{appendix equation: tetranacci most general ansatz}
	\xi_j\,=\,c_1\,e^{i\kappa_1}\,+\,c_2\,e^{-i\kappa_1}\,+\,c_3\,e^{i\kappa_2}\,+\,c_4\,e^{-i\kappa_2}.
\end{align}
Further, Eq. (\ref{appendix equation: degeneracy due to k12}) implies that we consider a combination of states of the same energy. The usually following step would be to fix these constants, requiring four initial values. We can use e.g. $\xi_1$ as free parameter. Further setting $\xi_0 = \xi_{N+1}=0$, $\xi_{-1}= (b/a)\,\xi_1$ as the boundary conditions yield a sufficient number of constraints.

The remaining condition $a\,\xi_{N}= b\,\xi_{N+2}$ yields the quantization rule then. However, if one is not interested in the state $\vec{v}$ or in the general eigenstates of the Kitaev chain, but only in the quantization rule, one can use a much simpler approach. Using our ansatz for $\xi_j$ from Eq. (\ref{appendix equation: tetranacci most general ansatz}) and being aware of the fact that the boundary conditions yield a homogeneous system, we find
\begin{align*}
	B_{4\times 4}\left(\begin{matrix}
	c_1\\
	c_2\\
	c_3\\
	c_4
	\end{matrix}\right)\,=\,\vec{0},
\end{align*} 
where the boundary matrix $B$ is
\begin{widetext}
\begin{align*}
	B= \left(\begin{matrix}
	1 & 1& 1&1\\
	e^{i\kappa_1 (N+1)} & e^{-i\kappa_1 (N+1)} &e^{i\kappa_2 (N+1)} &e^{-i\kappa_2 (N+1)} \\
	b e^{i\kappa_1 (N+2)} -a e^{i\kappa_1 N}  & b e^{-i\kappa_1 (N+2)} -a e^{-i\kappa_1 N}& b e^{i\kappa_2 (N+2)} -a e^{i\kappa_2 N}
	& b e^{-i\kappa_2 (N+2)} -a e^{-i\kappa_2 N}\\
	 b e^{i\kappa_1 } -a e^{-i\kappa_1} &  b e^{-i\kappa_1 } -a e^{i\kappa_1} & b e^{i\kappa_2 } -a e^{-i\kappa_2} &  b e^{-i\kappa_2 } -a e^{i\kappa_2} 
\end{matrix}				\right).
\end{align*}
\end{widetext}
Demanding $\mathrm{det}\left(B\right)= 0$ avoids a trivial solution and leads to the quantization rule in Eqs. (\ref{equation: Kitaev chain full quantization rule}), (\ref{equation: definition of F for the quantization rule}).

\section{The closed formula of Tetranacci polynomials}
\label{appendix: section: The closed formula of Tetranacci polynomials}
The goal here is to obtain the general solutions of a polynomial sequence $\xi_j$, $j\in\mathbbm{Z}$, which obeys
\begin{align}\label{appendix equation: recursion formula tetranacci for closed formula}
	\xi_{j+2}\,&=\,\frac{\lambda^2+a^2+b^2-\mu^2}{ab}\,\xi_j-\xi_{j-2}\notag\\
		&\quad -i\mu\left(\frac{a-b}{ab}\right)\,\left(\xi_{j+1} + \xi_{j-1}\right),
\end{align}
with arbitrary initial values. We consider here $\xi_{-2},\ldots,\,\xi_{1}$, other choices are possible too, to be the initial values. We want to determine a closed form expression for all $\xi_j$'s. Similar to Eq. \eqref{appendix equation: tetranacci most general ansatz}, the general solution is given by a superposition of the four fundamental solutions $r_{\pm i}$ from Eq. \eqref{appendix equation: solution rpm12 in terms of S12}
\begin{align}\label{appendix equation: tetranacci ansatz}
	\xi_j\,=\,c_1\,r_{+1}^j\,+\,c_2\,r_{-1}^j\,+\,c_3\,r_{+2}^j\,+\,c_4\,r_{-2}^j,
\end{align}
with some constants $c_1,\ldots,\,c_4$ which follow from $\xi_{-2},\ldots,\,\xi_{1}$ via
\begin{align}\label{appendix equation: fixing the tetranacci constants}
\left[
	\begin{matrix}
	r_{+1}^{-2} &  r_{-1}^{-2} & r_{+2}^{-2} & r_{-2}^{-2}\\
	r_{+1}^{-1} &  r_{-1}^{-1} & r_{+2}^{-1} & r_{-2}^{-1}\\
	1 & 1 & 1 & 1\\
	r_{+1} &  r_{-1} & r_{+2} & r_{-2}
	\end{matrix}\right] \left(
	\begin{matrix}
	c_1\\
	c_2\\
	c_3\\
	c_4
\end{matrix}		
	\right)\,=\,\left(
	\begin{matrix}
	\xi_{-2}\\
	\xi_{-1}\\
	\xi_{0}\\
	\xi_{1}
\end{matrix}		
	\right).
\end{align}
Solving Eq. \eqref{appendix equation: fixing the tetranacci constants} and factorising $\xi_j$ into contributions of $\xi_{-2},\ldots,\,\xi_{1}$ yields
\begin{align}\label{appendix equation: tetranacci closed form}
	\xi_j\,=\,\sum\limits_{i=-2}^1 \xi_{i}\,X_i(j).
\end{align}
The functions $X_i(j)$ obey by construction (via Eq. \eqref{appendix equation: fixing the tetranacci constants})
\begin{align}\label{appendix equation: initial values for gamma}
	X_i(j)\,=\,\delta_{i,j}, \quad \mathrm{for~} i,j=\,-2,\,\ldots,\,1
\end{align}
for all values of $\lambda$, $\mu$, $a$, $b$ and further obey Eq. \eqref{appendix equation: recursion formula tetranacci for closed formula}.

Despite the short form of $\xi_j$ in Eq. \eqref{appendix equation: tetranacci closed form}, the formulas of $X_i(j)$ tend to be lengthy, such that we first introduce a short hand notation for their main pieces. We define
\begin{align}
	F_1(j)\defl\frac{r^j_{+1}-r^j_{-1}}{r_{+1}-r_{-1}}\,=\,\frac{r^j_{+1}-r^{-j}_{+1}}{r_{+1}-r^{-1}_{+1}},\\
	F_2(j)\defl\frac{r^j_{+2}-r^j_{-2}}{r_{+2}-r_{-2}}\,=\,\frac{r^j_{+2}-r^{-j}_{+2}}{r_{+2}-r^{-1}_{+2}},
\end{align}
where the r.h.s of both equalities arise due to $r_{i}\,r_{-i}=1$ for $i=1,2$. With $S_{1,2}$ from Eq. \eqref{appendix equation: tetranacci: S12 in explicit form} we find the $X_i(j)$ to be
\begin{align}
	\label{appendix equation: gamma -2}
	X_{-2}(j)\,&=\,\frac{F_2 (j)-F_1 (j)}{S_1-S_2},\\
	X_{-1}(j)\,&=\,\sum_{\sigma=1}^2\frac{F_\sigma (j+2)+ F_\sigma (j-1) F_{\bar{\sigma}}(2) -F_\sigma(3)F_{\bar{\sigma}}(j) }{\left(S_1-S_2\right)^2},\\
	X_{0}(j)&=\,\sum_{\sigma=1}^2\frac{F_\sigma(j+1) F_{\bar{\sigma}}(3)-F_\sigma(j+2) F_{\bar{\sigma}}(2)}{\left(S_1-S_2\right)^2}\notag\\
	&\quad -\sum_{\sigma=1}^2 \frac{F_\sigma(j-1)}{\left(S_1-S_2\right)^2},\\
	\label{appendix equation: gamma 1}
	X_{1}(j)\,&=\,\sum_{\sigma=1}^2\frac{F_\sigma(j+2)+F_\sigma(j)-F_\sigma(j+1)  F_{\bar{\sigma}}(2)}{\left(S_1-S_2\right)^2},
\end{align}
where $\bar{\sigma}$ is meant as "not $\sigma$", e.g. if $\sigma=1$ then we have $\bar{\sigma}=2$ and vice versa. The presence of $S_{1,2}$ in the form of $X_i(j)$ arises due to the definition of $F_{1,2}$, since they are Fibonacci polynomials with inital values $F_{1,2}(0)=0$, $F_{1,2}(1)=1$ and obey
\begin{align*}
	F_i(j+2)\,=\,S_i\,F_i(j+1)-F_i(j).
\end{align*} 
The proof is done by induction over $j$ and using the relation between $r_{\pm i}$ and $S_i$ according to the Eqs. \eqref{appendix equation: tetranacci: S12 in explicit form} \eqref{appendix equation: solution rpm12 in terms of S12}. 

The formulas for $X_{-2}(j)$ and $F_i$ are exact and hold for all values of $\mu$, $a$, $b$ ($t$, $\Delta$) and for all values of $\lambda$, regardless whether an eigenvector/ eigenvalue problem is considered or not. Notice that $\mu=0$ is a special situation, since 
\begin{align*}
	S_1\vert_{\mu=0}\,=\,-S_2\vert_{\mu=0}
\end{align*}
leads to
\begin{align*}
	r_{+1}\vert_{\mu=0}\,=\,-r_{-2}\vert_{\mu=0}
\end{align*}
for $\mu=0$. Thus, we find
\begin{align*}
	F_1(j)\vert_{\mu=0}\,=\,\left(-1\right)^{j-1}F_2(j)\vert_{\mu=0}
\end{align*}
yielding
\begin{align*}
	X_{-2}(2l+1)\vert_{\mu=0} \,&=\,0,\\
	X_{0}(2l+1)\vert_{\mu=0} \,&=\,0,\\
	X_{-1}(2l)\vert_{\mu=0} \,&=\,0,\\
	X_{1}(2l)\vert_{\mu=0} \,&=\,0,
\end{align*}
for all values of $l$. 

The closed formula of $\xi_j$ can be used in multiple ways. In the context of eigenvectors the exponential form of the fundamental solutions $r_{\pm i}$ according to Eq. \eqref{appendix equation: exponential form of fundamental solutions for tetranaccis} is the direct connection to the momenta $\kappa_{1,2}$, and their values follow from the quantisation rule in Eq. \eqref{equation: Kitaev chain full quantization rule}. The corresponding value of $\lambda = E\pm(\kappa_{1,2})$ follows then from Eq. \eqref{infinite chain: excitation spectrum}. The form of $F_{1,2}$ transforms into a ratio of $\sin(\kappa_{1,2}\,j)/\sin(\kappa_{1,2})$. However, once the energy $E_\pm(\kappa_{1,2})$ is known, the explicit use of $\kappa_{1,2}$ is not important, since $r_{\pm i}$ follow also directly from Eq. \eqref{appendix equation: solution rpm12 in terms of S12}.

\section{The zeros of the determinant}
\label{appendix: zeros of the determinant}
Our first step is to calculate the determinant of the Kitaev chain in closed form. We use the chiral basis where the BdG Hamiltonian is given by Eqs.
(\ref{appendix euquation: Hamiltonian in chiral basis}), (\ref{appendix equation: h}). The determinant is obviously
\begin{align}\label{equation: appendix: determinant of H_c in form of h}
	\mathrm{det}\left(\mathcal{H}_c\right)\,=\,\mathrm{det}\left(h\right)\,\mathrm{det}\left(h^\dagger\right)\,=\,\left\vert  \mathrm{det}\left(h\right)\right\vert^2,
\end{align}
and we need only the determinant of $h$. The calculation is performed with a sequence of polynomials\cite{Usmani} $h_0,\,\ldots,\,h_N$
\begin{align}\label{equation: appendix: generalized Fibonaccis for the zeros of the determinant}
	h_{j+1}\,=\,-i\mu\,h_j\,+\,ab\,h_{j-1},\quad j = 1,\ldots N-1
\end{align}
with the initial values $h_0 = 1$, $h_1 = -i\mu$ and the determinant of $h$ is 
\begin{align}\label{equation: appendix: determinant of h in form of h_N}
	\mathrm{det}\left(h\right)\,=\,h_N.
\end{align}
We notice the Fibonacci character\cite{Webb, Hoggatt, Oezvatan-2017} of the sequence in Eq. (\ref{equation: appendix: generalized Fibonaccis for the zeros of the determinant}) and continue with the calculation of the Binet form. The ansatz $h_j\propto R^j$ ($R\in\mathbbm{C}
\setminus\left\{0\right\}$) leads to
\begin{align*}
	R^2\,+\,i\mu\,R\,-\,ab\,=\,0,
\end{align*}
and the solutions $R_{1,2}$ are
\begin{align}\label{equation: appendix: definition of R_12}
	R_{1,2}\,=\,\frac{-i\mu\,\pm\sqrt{4\,ab-\mu^2}}{2}.
\end{align}
Our ansatz holds for all parameter choices of $\mu,\,\Delta$ and $t$ and $R_{1,2}$ obey
\begin{align}
	\label{equation: appendix: sum of R1 and R2}
	R_1\,+\,R_2\,&=\,-i\mu,\\
	\label{equation: appendix: product of R1 and R2}
	R_1\,R_2\, &=\, -ab.
\end{align} 
The general form of $h_j$ is given by a superposition of $R_1$ and $R_2$
\begin{align}\label{equation: appendix: general form of h_j}
	h_j\,=\,n_1\,R_1^{\,j}\,+\,n_2\,R_2^{\,j},
\end{align}
and $n_{1,2}$ are fixed by the initial values. The calculation can be simplified by extending the sequence $h_j$ backwards with Eq. (\ref{equation: appendix: generalized Fibonaccis for the zeros of the determinant}), because $h_{-1}=0$. The use of $h_{-1}$ and $h_0$ leads to 
\begin{align*}
	n_1 = \frac{R_1}{R_1-R_2},\quad n_2 = \frac{-R_2}{R_1-R_2},
\end{align*}
yielding the closed form of $h_j$
\begin{align*}
	h_j\,=\,\frac{R_1^{j+1}-R_2^{j+1}}{R_1-R_2}.
\end{align*}
We find the determinant of the Kitaev chain to be
\begin{align}\label{appendix: determinant of the Kitaev chain in closed form}
	\mathrm{det}\left(\mathcal{H}_c\right)\,=\,\left\vert \frac{R_1^{N+1}-R_2^{N+1}}{R_1-R_2} \right\vert^2,
\end{align}
for all values of $\mu,\,t,\,\Delta\in\mathbbm{R}$. The determinant does not vanish in general, due to Eq. (\ref{equation: appendix: definition of R_12}), but only for a specific combination of the parameters $\mu$, $t$, $\Delta$.

In the following we consider $t$ and $\Delta$ to be fixed values of our choice and we search for the values of $\mu$ such that the determinant vanishes. The Fibonacci character of $h_N$ enables us to factorize the determinant\cite{Webb, Hoggatt} and leads automatically to the zeros. The factorization follows from Eq. (\ref{equation: appendix: definition of R_12}) and the starting point is the square root:
\begin{align*}
	\sqrt{4\,ab-\mu^2}\,=\,\sqrt{4\,(t^2-\Delta^2)-\mu^2}.
\end{align*}
We have to consider in general three cases
\begin{enumerate}
	\item[a)] \hspace{1cm} $t^2\ge\Delta^2$ and $4\,(t^2-\Delta^2)\ge\mu^2$,
	\item[b)] \hspace{1cm} $t^2\ge\Delta^2$ and $4\,(t^2-\Delta^2)\le\mu^2$,
	\item[c)] \hspace{1cm} $t^2\le\Delta^2$ and $4\,(t^2-\Delta^2)\le\mu^2$,
\end{enumerate}
and we introduce the procedure in detail with the first scenario. 
\subsection{Case a)}
The most general form for $\mu$ is
\begin{align}\label{equation: appendix: case 1 general ansatz for mu}
	\mu\,=\,2\,\sqrt{t^2-\Delta^2}~f(\theta),
\end{align}
where the function $f(\theta)$ accounts for all possible ratios of $\mu$ and $\sqrt{t^2-\Delta^2}$. The case a) enforces the function $f(\theta)$ to be real valued, because both $\mu$ and $\sqrt{t^2-\Delta^2}$ are real. Further, we find that
\begin{align}\label{equation: appendix: f^2<1 for the case a)}
	f^2(\theta)\le 1,
\end{align}
since $4\,(t^2-\Delta^2)-\mu^2\ge 0$. Please note that Eq. (\ref{equation: appendix: f^2<1 for the case a)}) needs only to hold for $\theta$ on a finite set. From all possible functions $f(\theta)$, a convenient choice is $f= \cos(\theta)$. The reason behind our specific choice is the form of $R_{1,2}$, because $f$ leads in $\sqrt{4\,ab\,-\mu^2}$ to
\begin{align*}
	\sqrt{4\,ab\,-\mu^2}\,	&=\,\sqrt{4\,(t^2-\Delta^2)\,-\mu^2}\\
						  	&=\,\sqrt{4\,(t^2-\Delta^2)~\left[1-\cos^2(\theta)\right]}\\
						  	&=\,2 \sqrt{t^2-\Delta^2}\,\sin(\theta),
\end{align*} 
and $R_{1,2}(f)$ become
\begin{align*}
	R_{1,2}(f)\,	&=\,\frac{-i\mu \pm\sqrt{4\,ab\,-\mu^2} }{2}\\
				&=\,\frac{-2i \sqrt{t^2-\Delta^2} \cos(\theta)\pm 2 \sqrt{t^2-\Delta^2}\,\sin(\theta) }{2}.
\end{align*}  
Simplifications lead to
\begin{align*}
	R_{1,2}(f)\,=\,-i\,\sqrt{t^2-\Delta^2} \,\left\{\begin{matrix}
	e^{i\theta},\\
	e^{-i\theta}
\end{matrix}			 \right. .
\end{align*}
Let us focus on the determinant. We find $R_1^j-R_2^j$ to be
\begin{align*}
	R_1^j-R_2^j\,=\, \left[-i\,\sqrt{t^2-\Delta^2}\,\right]^j ~ 2i \sin(\theta\,j).
\end{align*}
Consequently the determinant reads
\begin{align}
	\mathrm{det}\left(\mathcal{H}_c\right)\,=\, (t^2-\Delta^2)^N ~ \left[\frac{\sin\left[\theta\,\left(N+1\right)\right]}{\sin\left(\theta\,\right)}\right]^2,
\end{align}
and vanishes for $\theta = n\,\pi /N+1$ ($n = 1,\,\ldots,\, N$) or \makebox{$t^2=\Delta^2$}. Since $\Delta$, $t$ and $\theta$ define together with $f_1 = \cos(\theta)$ the chemical potential, we find that the determinant of the Kitaev chain is zero if, and only if:
\begin{enumerate}
	\item[i)]  $\mu = 2\,\sqrt{t^2-\Delta^2}\,\cos\left(\frac{n\,\pi}{N+1}\right)$,
	\item[ii)] $\mu = 0$ and $t^2=\Delta^2$,
\end{enumerate}
for $n=1,\ldots,N$, $t^2\ge\Delta^2$ and for all $N$. A feature of odd $N$ is the value $n= N+1/2$ yielding $\theta = \pi/2$, i.e. $\mu = 0$ for all values of $\Delta$, $t$ for $t^2\ge\Delta^2$. In fact $\mu =0$ holds for odd $N$ everywhere, as we already know from previous discussion in appendix \ref{appendix: spectrum zero mu}. 

We found all zeros in case a) and we continue with b).
\subsection{Case b)}
We follow the same way of argumentation as above, but we have to keep in mind that $t^2-\Delta^2\ge 0 $, and \makebox{$4\,(t^2-\Delta^2)<\mu^2$}. The first step is to reshape the square root in $R_{1,2}$ 
\begin{align}\label{equation: appendix: manipulate the square root in case b)}
	\sqrt{4\,ab-\mu^2}\,=\,i\sqrt{\mu^2-4\,\left(t^2-\Delta^2\right)},
\end{align}
where we find a similar situation as in the previous scenario. Our ansatz is 
\begin{align}
\label{eq:mu-and-g}
	\mu = 2\,\sqrt{t^2-\Delta^2}~g(\theta),
\end{align}
where the function $g(\theta)$ is real and obeys
\begin{align}\label{equation: appendix: function g, large enough}
 	g^2(\theta)\ge 1,
\end{align}
since $\mu^2\ge 4\,(t^2-\Delta^2)$. The candidates of our choice are $g_{\pm}(\theta) = \pm\,\cosh(\theta)$, where $\theta$ is real. The square root becomes now
\begin{align*}
	\sqrt{\mu^2-4\,\left(t^2-\Delta^2\right)}\,=\,2\,\sqrt{\left(t^2-\Delta^2\right)}\, \sinh(\theta),
\end{align*}
and we find $R_{1,2}(g_+)$ to be
\begin{align*}
	R_{1,2}(g_+)\,	&=\,\frac{-i\mu \pm i \sqrt{\mu^2-4\,\left(t^2-\Delta^2\right)}}{2}\\
				&=\,-i\,\sqrt{t^2-\Delta^2}\,\left[\cosh(\theta)\mp \sinh(\theta)\right].
\end{align*}
Simplifications yield
\begin{align*}
	R_{1,2}(g_+)\,=\,-i\,\sqrt{t^2-\Delta^2}\,\left\{\begin{matrix}
	e^{-\theta}\\
	e^{\,\theta}
\end{matrix}				\right. ,
\end{align*}
and the determinant becomes:
\begin{align}\label{equation: appendix: determinant case b)}
	\mathrm{det}\left(\mathcal{H}_c\right)\,=\,\left(t^2-\Delta^2\right)^N~\left[\frac{\sinh \left(\theta\left[N+1\right)\right]}{\sinh(\theta)}\right]^2.
\end{align}
The determinant vanishes only if $t^2=\Delta^2$, which by virtue of Eq.~\eqref{eq:mu-and-g} implies $\mu = 0$, because the fraction of the hyperbolic sine functions is always positive. The use of $g_{1,-}=-\cosh(\theta)$ leads to Eq. (\ref{equation: appendix: determinant case b)}) again.
\subsection{Case c)}
We consider here $\Delta^2\ge t^2$ and $4\,(t^2-\Delta^2)-\mu^2\le 0$. We start by manipulating the square root in $R_{1,2}$
\begin{align}\label{equation: appendix: manipulate the square root in case c)}
	\sqrt{4\,ab-\mu^2}\,=\,\sqrt{4\,(t^2-\Delta^2)-\mu^2}\,=\,i\,\sqrt{\mu^2+ 4\,(\Delta^2-t^2)}.
\end{align}
Our ansatz is $\mu = 2\,\sqrt{\Delta^2-t^2}\,v(\theta)$ with a real valued function $v(\theta)$, without further restrictions, because
\begin{align*}
	\mu^2 = 4\,(\Delta^2-t^2)\,v^2(\theta)\ge -4\,(\Delta^2-t^2),
\end{align*}
in view of $\mu^2\ge -4 \,(\Delta^2-t^2)$. The square root in $R_{1,2}$ becomes in general
\begin{align*}
	i\,\sqrt{\mu^2+ 4\,(\Delta^2-t^2)}\,=\,i\,(\Delta^2-t^2)\,\sqrt{v^2(\theta)+1},
\end{align*}
and one sees immediately that $v(\theta)=\sinh(\theta)$, $\theta\in\mathbbm{R}$ is an appropriate choice. We find for $R_{1,2} $ the form
\begin{align*}
	R_{1,2}(v)\,=\,-i\,\sqrt{\Delta^2-t^2}\,\left\{\begin{matrix}
	-e^{-\theta}\\
	e^{\theta}
\end{matrix}		\right.,
\end{align*}
where the negative sign in front of the exponential forces us to distinguish between even and odd $N$. The determinant reads finally
\begin{align*}
	\mathrm{det}\left(\mathcal{H}_c\right)\,=\,(\Delta^2-t^2)^N\,\left\{\begin{matrix}
	\frac{\cosh\left[\theta\left(N+1\right)\right]}{\cosh\left(\theta\right)}, & N\,\text{even}\\
	\\
	\frac{\sinh\left[\theta\left(N+1\right)\right]}{\sinh\left(\theta\right)}, & N\,\text{odd}
\end{matrix}				\right.,
\end{align*}
and it is never zero, except for $\Delta^2=t^2$ at $\mu=0$. 
\subsection{Discussion of completeness of all scenarios}
In summary, for $\mu\neq 0$, we have only non trivial, zero determinants in case a). How can one be sure that no zero is missed especially in the settings b) and c)? This follows immediately from Eq. (\ref{appendix: determinant of the Kitaev chain in closed form}), because the determinant vanishes only if
\begin{align*}
	R_1^{N+1} = R_2^{N+1}.
\end{align*}
Consequently we need first of all $\vert R_1\vert = \vert R_2\vert$. The second part is to find the proper phase factors and all of them lie on a circle with radius $\vert R_1\vert$ in the complex plane. We have found non trivial solutions only for scenario a). 

In total, we found all conditions \makebox{$\mathrm{det}\left(\mathcal{H}_\mathrm{KC}\right) = 0$}. The general case is when the chemical potential is
\begin{align}\label{equation: appendix: cosin dependency for det = 0}
	\mu &= 2\,\sqrt{t^2-\Delta^2}\,\cos\left(\frac{n\,\pi}{N+1}\right),
\end{align}
with $t^2\ge \Delta^2$ and $n=1,\,\ldots,\,N$, i.e. the chemical potential corresponds to the energy levels of a linear chain with hopping $\sqrt{t^2-\Delta^2}$.
The case $\mu = 0$ and $t^2 = \Delta^2$ is included in Eq. (\ref{equation: appendix: cosin dependency for det = 0}). 

Further, the determinant of a Kitaev chain with odd number of sites is zero if $\mu=0$
%
%
%
%
for all values of $\Delta$ and $t$.
\section{The zero energy eigenstates}
\label{appendix: zero energy states}
The presence of zero energy modes is marked by $\mathrm{det}\left(\mathcal{H}_\mathrm{KC}\right) = 0$ and a natural question is to investigate their topological character, be it trivial or non-trivial. Hence, we have first to obtain these states. We use here again the SSH-like basis, e.g. the Hamiltonian from Eq. (\ref{equation: Kitaev Hamiltonian/ matrix in SSH basis}) for $\mu\neq 0$. We keep the notation for the eigenvector $\vec{\psi}=\left(\vec{v}_\alpha,\,\vec{v}_\beta\right)^\mathrm{T}$ with
\begin{align*}
	\vec{v}_\alpha\,&=\,\left(x_1,\,y_1,\,x_2,\,y_2,\,\ldots ,\,x_{N/2},\,y_{N/2}\right)^\mathrm{T},\\
	\vec{v}_\beta\,&=\,\left(\xcal_1,\,\ycal_1,\,\xcal_2,\,\ycal_2,\,\ldots ,\,\xcal_{N/2},\,\ycal_{N/2}\right)^\mathrm{T},
\end{align*}
for even $N$, but unlike in the previous calculation both SSH-like chains are coupled now. We consider first $N$ even, because the odd $N$ solutions have the same shape, as it turns out later. Further, we derive the general eigenvector problem including even non zero modes. Solving $\left(\lambda\,\mathbbm{1}_{2N}-\mathcal{H}_\mathrm{KC}^\mathrm{SSH}\right)\,\vec{\psi}\,=\,\vec{0}$ translates to
\begin{align*}
	\alpha\,\vec{v}_\alpha\,+\,\tau\,\vec{v}_\beta\,&=\,\lambda\,\vec{v}_\alpha,\\
	\beta\,\vec{v}_\beta\,+\,\tau^\dagger\,\vec{v}_\alpha\,&=\,\lambda\,\vec{v}_\beta.
\end{align*}
The reason to keep $\lambda$ first inside the calculation is the diagonal structure of $\tau$, $\tau^\dagger$ and $\mathbbm{1}_N$ as well as the entry structure of $\vec{v}_\alpha$ and $\vec{v}_\beta$, which enables us to identify easily the new contributions of $\tau\,\vec{v}_\beta$ and $\tau^\dagger\,\vec{v}_\alpha$ in comparison to the $\mu=0$, i.e.. $\tau = 0$, case from \makebox{appendix \ref{appendix: eigenvectors}}. The difficulty to write down $\left(\lambda\,\mathbbm{1}_{2N}-\mathcal{H}_\mathrm{KC}^\mathrm{SSH}\right)\,\vec{\psi}\,=\,\vec{0}$ reduces to take the correct signs of the $\mu$ terms. We have to solve ($l = 1,\,\ldots,\, N -1$)
\begin{align}\label{appendix equation: full eigenvectorsystem 1}
	b\,x_{l+1}\,-\,a\,x_l\,+\,i\mu\,\ycal_l\,&=\,\lambda\,y_l,\\
	a\,y_{l+1}\,-\,b\,y_l\,-i\mu\,\xcal_{l+1}\,&=\,\lambda\,x_{l+1},\\\notag\\
	a\,y_1\,-\,i\mu\,\xcal_1\,&=\,\lambda\,x_1,\\
	-a\,x_{\frac{N}{2}}\,+\,i\mu\,\ycal_{\frac{N}{2}}&=\,\lambda\,y_\frac{N}{2},
\end{align}
from $\alpha\,\vec{v}_\alpha\,+\,\tau\,\vec{v}_\beta\,=\,\lambda\,\vec{v}_\alpha$ and
\begin{align}
	a\,\xcal_{l+1}\,-\,b\,\xcal_l\,-i\mu\,y_l\,&=\,\lambda\,\ycal_l,\\
	b\,\ycal_{l+1}\,-\,a\,\ycal_l\,+i\mu\,x_{l+1}&=\,\lambda\,\xcal_{l+1},\\\notag\\
	b\,\ycal_1\,+\,i\mu\,x_1\,&=\,\lambda\,\xcal_1,\\
	\label{appendix equation: full eigenvectorsystem 8}
	-b\,\xcal_{\frac{N}{2}}\,-\,i\mu\,y_{\frac{N}{2}}\,&=\,\lambda\,\ycal_{\frac{N}{2}},
\end{align}
from $\beta\,\vec{v}_\beta\,+\,\tau^\dagger\,\vec{v}_\alpha\,=\,\lambda\,\vec{v}_\beta$. Extending the sequences $x_l$, $y_l$, $\xcal_l$ and $\ycal_l$ backwards leads to simplifications in the open boundary conditions
\begin{align*}
	y_0\,=\,x_{\frac{N}{2}+1}\,=\,\ycal_0\,=\,\xcal_{\frac{N}{2}+1}\,=\,0.
\end{align*} 
As we see from the particle-hole operator, see Eq. \eqref{equation: particle hole symmetry in real space in ssh basis}, a MZM requires either fully real or fully imaginary entries, which is not true for a generic solution of the eigenvector system Eqs. \eqref{appendix equation: full eigenvectorsystem 1}-\eqref{appendix equation: full eigenvectorsystem 8} of the Kitaev Hamiltonian for $\lambda \neq 0$. Thus, zero energy is essential for a MZM.

Zero energy has one advantage, because the chiral partner of a zero mode is itself a zero mode and superpositions of both will simplify the eigenvector problem even more. Acting with $\mathcal{C}$ from Eq. (\ref{equation: chiral symmetry even N}) on $\vec{\psi}$, all $y_l$ ($\xcal_l$) got into $-y_l$ ($-\xcal_l$), while all $x_l$ ($\ycal_l$) remain the same. Hence, $\vec{\psi}_A\defl \left(\vec{\psi}+\mathcal{C}\,\vec{\psi}\,\right)/2$ reads
\begin{align*}
	\vec{\psi}_A\,=\,\left(x_1,\,0,\,x_2,\,0,\ldots,x_{\frac{N}{2}},0\,\left\vert\,0,\ycal_1,\,0,\ycal_2,\ldots,0,\,\ycal_{\frac{N}{2}}\right.\right)^\mathrm{T},
\end{align*}
and "$\vert$" marks the boundary of both SSH-like chains. Similar  $\vec{\psi}_B\defl \left(\vec{\psi}-\mathcal{C}\,\vec{\psi}\,\right)/2$ is
\begin{align*}
	\vec{\psi}_B\,=\,\left(0,\,y_1,\,0,\,y_2,\ldots,0,y_{\frac{N}{2}}\left\vert\,\xcal_1,\,0,\xcal_2,\,0\ldots,\,\xcal_{\frac{N}{2}},\,0\right.\right)^\mathrm{T}.
\end{align*}
As we see, we decomposed $\vec{\psi}$ into $\vec{\psi}_{A,B}$. The decomposition is optional, but $\vec{\psi}_A$ ($\vec{\psi}_B$) has only non zero weight on $A$ type ($B$ type) Majorana positions $\gamma_j^A$ ($\gamma_j^B$) in the SSH-like basis, as depicted in Fig. (\ref{Figure: sawtooth pattern of  psi_A}). Thus, $\vec{\psi}_A$ obeys $(S+)$
\begin{align*}
	\left.\begin{matrix}
	b\,x_{l+1}\,-\,a\,x_l\,+\,i\mu\,\ycal_l\,=\,0\\
	b\,\ycal_{l+1}\,-\,a\,\ycal_l\,+i\mu\,x_{l+1}\,=\,0\\\\
	\ycal_0\,=\,x_{\frac{N}{2}+1} \,=\,0\\
	\end{matrix}\right\} \quad (S+),
\end{align*}
while $\vec{\psi}_B$ satisfies $(S-)$
\begin{align*}
	\left.\begin{matrix}
	a\,y_{l+1}\,-\,b\,y_l\,-i\mu\,\xcal_{l+1}\,=\,0\\
	a\,\xcal_{l+1}\,-\,b\,\xcal_l\,-i\mu\,y_{l}\,=\,0\\\\
	y_0\,=\,\xcal_{\frac{N}{2}+1}\,=\,0,\\
	\end{matrix}\right\} \quad (S-),
\end{align*}
and $l$ runs from $1$ to $N-1$. As we see, $(S+)$ turns into $(S-)$ by exchanging $a$'s and $b$'s, $\mu$ into $-\mu$ and the standard letters into the calligraphic ones. Thus, we need only to solve one set of equations and the solution of the second follows immediately.

We focus on $(S+)$ and we ignore the index limitations during the following calculation. Decoupling leads to
\begin{align}
	\label{equation: appendix: v_+, growing rule ycal}
	b^2\,\ycal_{l+1}\,=\,\left(2\,ab\,-\,\mu^2\right)\,\ycal_l\,-\,a^2\,\ycal_{l-1},\\
	\label{equation: appendix: v_+, growing rule x}
	b^2\,x_{l+1}\,=\,\left(2\,ab\,-\,\mu^2\right)\,x_l\,-\,a^2\,x_{l-1},
\end{align}
Fibonacci polynomials\cite{Hoggatt, Webb, Oezvatan-2017}. The Binet form needs initial values and we have to think about the number of free entries we have here. These degrees of freedom are given by the dimension of the zero energy subspace, i.e. the number of zero energy states. So far, the chiral symmetry implies their pairwise presence, but not their absolute quantity. Each zero of the determinant is twice degenerated, as we see from Eq. (\ref{appendix: determinant of the Kitaev chain in closed form}). Hence, we have in total only two zero energy modes and each has one unspecified entry. We choose $x_1$ as a fixed number. 

The naive choice would be to take $x_1$, $x_2$, $\ycal_1$ and $\ycal_2$ as initial values, where the last three are expressed in terms of $x_1$. Instead we use 
the $l=0,\,1$ expressions and introduce $x_0$ via (S+)
\begin{align*}
	b\,x_{1}\,-\,a\,x_0\,+\,i\mu\,\ycal_0\,=\,0,
\end{align*}
because $x_1$ is our choice and $\ycal_0 = 0$. We find $x_0 = x_1 \,b/a$. The term $y_1$ follows from (S+)
\begin{align*}
	b\,\ycal_{1}\,-\,a\,\ycal_0\,+i\mu\,x_{1}\,=\,0,
\end{align*}
which reduces to $y_1 = -i\mu\, x_1/b$. 

The Binet form follows again from a power ansatz \makebox{$x_l \propto z^l$}. The fundamental solutions for both sequences are
\begin{align*}
	z_{1,2}\,=\,\frac{2\,ab\,-\,\mu^2\pm \sqrt{(2\,ab\,-\,\mu^2)^2\,-\,4\, a^2\,b^2}}{2\,b^2}.
\end{align*}
We use Eq. (\ref{equation: appendix: cosin dependency for det = 0}) to get $2\,ab\,-\,\mu^2 = -2ab\,\cos\left(2\,\frac{n\pi}{N+1}\right)$ and we obtain 
\begin{align*}
	z_{1,2}\,=\,\frac{-2ab\,\cos\left(2\,\frac{n\pi}{N+1}\right)\pm 2i\,ab\,\sin\left(2\,\frac{n\pi}{N+1}\right) }{2\,b^2}.
\end{align*}
Finally, we have 
\begin{align*}
	z_{1,2}\,=\,-\frac{a}{b}\, \left\{\begin{matrix}
	e^{-2i\,\theta_n}\\
	e^{2i\,\theta_n}
	\end{matrix}\right. ,
\end{align*}
with $\theta_n \defl n\pi /(N+1)$. The general solution is given by the superposition of $z_1$ and $z_2$
\begin{align*}
	x_l\,=\,\alpha\,z_1^l\,+\,\beta\,z_2^l,
\end{align*}
and we find both coefficients with $x_1$ and $x_0$ to be
\begin{align*}
	\begin{pmatrix}
	\alpha\\
	\beta
	\end{pmatrix}\,=\,\frac{x_1}{z_2-z_1}\,\begin{pmatrix}
	\frac{b}{a}\,z_2-1\\
	1-z_1\,\frac{b}{a}
	\end{pmatrix}.
\end{align*}
With this $x_l$ becomes
\begin{align*}
	x_l\,=\,\frac{x_1}{z_1-z_2}\left[z_1^l-z_2^l-\frac{b}{a}\,z_1z_2\left(z_1^{l-1}-z_2^{l-1}\right)\right].
\end{align*}
Using the expressions for $z_{1,2}$, we find
\begin{align*}
	x_l\,=\,x_1\,\frac{\sin(2\,\theta_n l)+ \sin[2\,\theta_n (l-1)]}{\sin(2\,\theta_n)}\,\left(-\frac{a}{b}\right)^{l-1},
\end{align*}
or in the most compact form
\begin{align}\label{equation: appendix: vec{v}_+ entry x_l}
	x_l\,=\,x_1\,\frac{\sin[\theta_n (2l-1) ]}{\sin(\theta_n)}\,\left(-\frac{a}{b}\right)^{l-1},
\end{align}
Similar, we obtain $\ycal_l$
\begin{align}\label{equation: appendix: vec{v}_+ entry ycal_l}
	\ycal_l\,=\,\ycal_1\,\frac{z_2^l-z_1^l}{z_2-z_1}\,=\,x_1\,\left(\frac{-i\mu}{b}\right)\,\frac{\sin(2\,\theta_n\,l)}{\sin(2\,\theta_n)}\,\left(-			\frac{a}{b}\right)^{l-1},
\end{align}
which simplifies to
\begin{align*}
	\ycal_l= -x_1\,\mathrm{sign}(t+\Delta)\,\frac{\sin(2\,\theta_n\,l)}{\sin(\theta_n)}\,\left(-\frac{a}{b}\right)^{\frac{2l-1}{2}},
\end{align*}
where $-a/b$ is always positive since $t^2\ge \Delta^2$. The last step is to check if the open boundary conditions are satisfied. Obviously $\ycal_0 = 0$ holds and we get for $x_{\frac{N}{2}+1}$ the form
\begin{align*}
	x_{\frac{N}{2}+1}&\propto \, \sin\left\{\theta_n\left[2\left(\frac{N}{2}+1\right)-1 \right] \right\}=0.
\end{align*}
Hence, the vector $\vec{\psi}_A$ is an eigenvector of the Kitaev BdG Hamiltonian.
The vector $\vec{\psi}_B$ has the entries
\begin{align}\label{equation: appendix: vec{v}_- entry xcal_l}
	\xcal_l\,=\,\xcal_1\,\frac{\sin[\theta_n (2l-1) ]}{\sin(\theta_n)}\,\left(-\frac{b}{a}\right)^{l-1},
\end{align}
and
\begin{align}\label{equation: appendix: vec{v}_- entry y_l}
	y_l\,=\,-\xcal_1\,\mathrm{sign}(t-\Delta)\frac{\sin(2\,\theta_n\,l)}{\sin(\theta_n)}\,\left(-\frac{b}{a}\right)^{\frac{2l-1}{2}}
\end{align}
where $\xcal_1$ is free to choose. 
The case of odd $N$ is similar. We use
\begin{align*}
	\vec{v}_\alpha\,&=\,\left(x_1,\,y_1,\,x_2,\,y_2,\,\ldots ,\,x_{\frac{N-1}{2}},\,y_{\frac{N-1}{2}},\,x_{\frac{N+1}{2}}\right)^\mathrm{T},\\
	\vec{v}_\beta\,&=\,\left(\xcal_1,\,\ycal_1,\,\xcal_2,\,\ycal_2,\,\ldots,\,\xcal_{\frac{N-1}{2}},\,\ycal_{\frac{N-1}{2}},\,\xcal_{\frac{N+1}{2}}\right)^\mathrm{T},
\end{align*}
and $\vec{\psi}=\left(\vec{v}_\alpha,\,\vec{v}_\beta\right)^\mathrm{T}$. The vectors $\vec{\psi}_{A,B}\,=\,\left(\vec{\psi}\,\pm\,C\vec{\psi}\,\right)$ 
become now
\begin{align}\label{equation: appendix: vec_+ for odd N}
	\vec{\psi}_A&=\left(x_1,\,0,\,x_2,\,0,\ldots,x_{\frac{N+1}{2}}\left\vert\,0,\ycal_1,\,0,\ycal_2,\ldots,0,\,\ycal_{\frac{N-1}{2}}\right.\right)^\mathrm{T},\\
	\label{equation: appendix: vec_- for odd N}
	\vec{\psi}_B&=\,\left(0,\,y_1,\,0,\,y_2,\ldots,0,y_{\frac{N-1}{2}}\left\vert\,\xcal_1,\,0,\xcal_2,\,0\ldots,\,\xcal_{\frac{N+1}{2}}\right.\right)^\mathrm{T}.
\end{align}
As we see, we have to respect different index limitations for $x_j$ $(\xcal_j)$ and $\ycal_i$ ($y_i$), but apart from this small change everything else remains as in the even $N$ case. The vector $\vec{\psi}_A$ obeys now
\begin{align*}
	\left.\begin{matrix}
	b\,x_{j+1}\,-\,a\,x_j\,+\,i\mu\,\ycal_j\,=\,0\\
	b\,\ycal_{i+1}\,-\,a\,\ycal_i\,+i\mu\,x_{i+1}\,=\,0\\\\
	\ycal_0\,=\,\ycal_{\frac{N+1}{2}} \,=\,0\\
	\end{matrix}\right\} \quad (\tilde{S}+),
\end{align*}
with $j = 1,\,\ldots,\,(N-1)/2$,  $i = 1,\,\ldots,\,(N-3)/2$ and $\vec{\psi}_B$ satisfies
\begin{align*}
	\left.\begin{matrix}
	a\,\xcal_{j+1}\,-\,b\,\xcal_j\,-\,i\mu\,y_j\,=\,0\\
	b\,y_{i+1}\,-\,b\,y_i\,-i\mu\,\xcal_{i+1}\,=\,0\\\\
	y_0\,=\,y_{\frac{N+1}{2}} \,=\,0\\
	\end{matrix}\right\} \quad (\tilde{S}-).
\end{align*}
The only important change compared to the even $N$ case are the new open boundary conditions, while the Fibonacci character remains. Hence, we ignore the index limitation during the calculation of those entries as in the even $N$ case and we get the same results for $x_l$, $\xcal_l$, $y_l$ and $\ycal_l$, see Eqs. (\ref{equation: appendix: vec{v}_+ entry x_l}) - (\ref{equation: appendix: vec{v}_- entry y_l}). 

The boundary conditions are satisfied, since $y_0\,=\,\ycal_0=0$,
\begin{align*}
	\ycal_{\frac{N+1}{2}}\propto \sin\left(2\,\theta_n\frac{N+1}{2}\right)\,=\,\sin[\theta_n\,(N+1)]=0,
\end{align*}
and $y_\frac{N+1}{2}=0$. 
A last check for the odd $N$ case is done by choosing $n=N+1/2$, i.e. $\theta_n = \pi/2$, which leads back to the old $\mu=0$ limit. 
Applying $\theta_n\rightarrow \pi/2$ on $x_l$ leads to 
\begin{align*}
	x_l\,=\,x_1\,\left(\frac{a}{b}\right)^{l-1}\,=\,x_1\,\left(\frac{\Delta -t }{\Delta +t}\right)^{l-1},
\end{align*}
after some steps, while all $\ycal_l\propto \mu$ are zero. Similar we find $\xcal_l$ from $x_l$ upon changing $a$ with $b$, while $y_l=0$ for all $l$. Hence, we recover our result for the $\alpha$ ($\beta$) chain, see Eq. (\ref{equation: zero mode on alpha N odd})-(\ref{equation: zero mode on beta N odd}).

The remaining questions is whether these zero energy modes are Majorana zero modes or not. The use of the particle hole operator in the SSH-like basis from Eq. (\ref{equation: particle hole symmetry in real space in ssh basis}), i.e. complex conjugation, reveals that the expressions $x_l /x_1$, $\ycal_l/xl_1$, $\xcal_l /\xcal_1$ and $y_l/\xcal_1$ are always real quantities, for both even and odd $N$. Thus $\vec{\psi}_A$ ($\vec{\psi}_B$) is a MZM if $x_1$ ($\xcal_1$) is either real or pure imaginary. 

The MZM mode $\vec{\psi}_A$ ($\vec{\psi}_B$) has only zero weight on $\gamma_j^A$ ($\gamma_j^B$). Superpositions of both vectors can be MZM too if the coefficients are chosen properly. For example \makebox{$\vec{\psi}=\vec{\psi}_A\,+\,\vec{\psi}_B$} has no zero entry. Hence, it is a mixed type MZM (for the correct choice of $x_1$ and $\xcal_1$).  

\bibliography{literature}

\begin{thebibliography}{47}%
\makeatletter
\providecommand \@ifxundefined [1]{%
 \@ifx{#1\undefined}
}%
\providecommand \@ifnum [1]{%
 \ifnum #1\expandafter \@firstoftwo
 \else \expandafter \@secondoftwo
 \fi
}%
\providecommand \@ifx [1]{%
 \ifx #1\expandafter \@firstoftwo
 \else \expandafter \@secondoftwo
 \fi
}%
\providecommand \natexlab [1]{#1}%
\providecommand \enquote  [1]{``#1''}%
\providecommand \bibnamefont  [1]{#1}%
\providecommand \bibfnamefont [1]{#1}%
\providecommand \citenamefont [1]{#1}%
\providecommand \href@noop [0]{\@secondoftwo}%
\providecommand \href [0]{\begingroup \@sanitize@url \@href}%
\providecommand \@href[1]{\@@startlink{#1}\@@href}%
\providecommand \@@href[1]{\endgroup#1\@@endlink}%
\providecommand \@sanitize@url [0]{\catcode `\\12\catcode `\$12\catcode
  `\&12\catcode `\#12\catcode `\^12\catcode `\_12\catcode `\%12\relax}%
\providecommand \@@startlink[1]{}%
\providecommand \@@endlink[0]{}%
\providecommand \url  [0]{\begingroup\@sanitize@url \@url }%
\providecommand \@url [1]{\endgroup\@href {#1}{\urlprefix }}%
\providecommand \urlprefix  [0]{URL }%
\providecommand \Eprint [0]{\href }%
\providecommand \doibase [0]{http://dx.doi.org/}%
\providecommand \selectlanguage [0]{\@gobble}%
\providecommand \bibinfo  [0]{\@secondoftwo}%
\providecommand \bibfield  [0]{\@secondoftwo}%
\providecommand \translation [1]{[#1]}%
\providecommand \BibitemOpen [0]{}%
\providecommand \bibitemStop [0]{}%
\providecommand \bibitemNoStop [0]{.\EOS\space}%
\providecommand \EOS [0]{\spacefactor3000\relax}%
\providecommand \BibitemShut  [1]{\csname bibitem#1\endcsname}%
\let\auto@bib@innerbib\@empty
\bibitem [{\citenamefont {Aguado}(2017)}]{Aguado}%
  \BibitemOpen
  \bibfield  {author} {\bibinfo {author} {\bibfnamefont {R.}~\bibnamefont
  {Aguado}},\ }\href@noop {} {\bibfield  {journal} {\bibinfo  {journal} {La
  Rivista del Nuovo Cimento}\ }\textbf {\bibinfo {volume} {40}},\ \bibinfo
  {pages} {523} (\bibinfo {year} {2017})}\BibitemShut {NoStop}%
\bibitem [{\citenamefont {Kitaev}(2001)}]{kitaev:physusp2001}%
  \BibitemOpen
  \bibfield  {author} {\bibinfo {author} {\bibfnamefont {A.~Y.}\ \bibnamefont
  {Kitaev}},\ }\href {http://stacks.iop.org/1063-7869/44/i=10S/a=S29}
  {\bibfield  {journal} {\bibinfo  {journal} {Physics-Uspekhi}\ }\textbf
  {\bibinfo {volume} {44}},\ \bibinfo {pages} {131} (\bibinfo {year}
  {2001})}\BibitemShut {NoStop}%
\bibitem [{\citenamefont {Alicea}(2010)}]{Alicea-2010}%
  \BibitemOpen
  \bibfield  {author} {\bibinfo {author} {\bibfnamefont {J.}~\bibnamefont
  {Alicea}},\ }\href {\doibase 10.1103/PhysRevB.81.125318} {\bibfield
  {journal} {\bibinfo  {journal} {Phys. Rev. B}\ }\textbf {\bibinfo {volume}
  {81}},\ \bibinfo {pages} {125318} (\bibinfo {year} {2010})}\BibitemShut
  {NoStop}%
\bibitem [{\citenamefont {Lutchyn}\ \emph {et~al.}(2010)\citenamefont
  {Lutchyn}, \citenamefont {Sau},\ and\ \citenamefont
  {Das~Sarma}}]{Lutchyn-2010}%
  \BibitemOpen
  \bibfield  {author} {\bibinfo {author} {\bibfnamefont {R.~M.}\ \bibnamefont
  {Lutchyn}}, \bibinfo {author} {\bibfnamefont {J.~D.}\ \bibnamefont {Sau}}, \
  and\ \bibinfo {author} {\bibfnamefont {S.}~\bibnamefont {Das~Sarma}},\ }\href
  {\doibase 10.1103/PhysRevLett.105.077001} {\bibfield  {journal} {\bibinfo
  {journal} {Phys. Rev. Lett.}\ }\textbf {\bibinfo {volume} {105}},\ \bibinfo
  {pages} {077001} (\bibinfo {year} {2010})}\BibitemShut {NoStop}%
\bibitem [{\citenamefont {Oreg}\ \emph {et~al.}(2010)\citenamefont {Oreg},
  \citenamefont {Refael},\ and\ \citenamefont {von Oppen}}]{Oreg-2010}%
  \BibitemOpen
  \bibfield  {author} {\bibinfo {author} {\bibfnamefont {Y.}~\bibnamefont
  {Oreg}}, \bibinfo {author} {\bibfnamefont {G.}~\bibnamefont {Refael}}, \ and\
  \bibinfo {author} {\bibfnamefont {F.}~\bibnamefont {von Oppen}},\ }\href
  {\doibase 10.1103/PhysRevLett.105.177002} {\bibfield  {journal} {\bibinfo
  {journal} {Phys. Rev. Lett.}\ }\textbf {\bibinfo {volume} {105}},\ \bibinfo
  {pages} {177002} (\bibinfo {year} {2010})}\BibitemShut {NoStop}%
\bibitem [{\citenamefont {Mourik}\ \emph {et~al.}(2012)\citenamefont {Mourik},
  \citenamefont {Zuo}, \citenamefont {Frolov}, \citenamefont {Plissard},
  \citenamefont {Bakkers},\ and\ \citenamefont {Kouwenhoven}}]{Mourik-2012}%
  \BibitemOpen
  \bibfield  {author} {\bibinfo {author} {\bibfnamefont {V.}~\bibnamefont
  {Mourik}}, \bibinfo {author} {\bibfnamefont {K.}~\bibnamefont {Zuo}},
  \bibinfo {author} {\bibfnamefont {S.}~\bibnamefont {Frolov}}, \bibinfo
  {author} {\bibfnamefont {S.}~\bibnamefont {Plissard}}, \bibinfo {author}
  {\bibfnamefont {E.}~\bibnamefont {Bakkers}}, \ and\ \bibinfo {author}
  {\bibfnamefont {L.}~\bibnamefont {Kouwenhoven}},\ }\href {\doibase
  10.1126/science.1222360} {\bibfield  {journal} {\bibinfo  {journal}
  {Science}\ }\textbf {\bibinfo {volume} {336}},\ \bibinfo {pages} {1003}
  (\bibinfo {year} {2012})}\BibitemShut {NoStop}%
\bibitem [{\citenamefont {Klinovaja}\ and\ \citenamefont
  {Loss}(2012)}]{Jelena-2012}%
  \BibitemOpen
  \bibfield  {author} {\bibinfo {author} {\bibfnamefont {J.}~\bibnamefont
  {Klinovaja}}\ and\ \bibinfo {author} {\bibfnamefont {D.}~\bibnamefont
  {Loss}},\ }\href {\doibase 10.1103/PhysRevB.86.085408} {\bibfield  {journal}
  {\bibinfo  {journal} {Phys. Rev. B}\ }\textbf {\bibinfo {volume} {86}},\
  \bibinfo {pages} {085408} (\bibinfo {year} {2012})}\BibitemShut {NoStop}%
\bibitem [{\citenamefont {Deng}\ \emph {et~al.}(2016)\citenamefont {Deng},
  \citenamefont {S.}, \citenamefont {Hansen}, \citenamefont {Danon},
  \citenamefont {Leijnse}, \citenamefont {Flensberg}, \citenamefont
  {Nyg\r{a}rd}, \citenamefont {Krogstrup},\ and\ \citenamefont
  {Marcus}}]{Deng-2016}%
  \BibitemOpen
  \bibfield  {author} {\bibinfo {author} {\bibfnamefont {M.~T.}\ \bibnamefont
  {Deng}}, \bibinfo {author} {\bibfnamefont {V.}~\bibnamefont {S.}}, \bibinfo
  {author} {\bibfnamefont {E.}~\bibnamefont {Hansen}}, \bibinfo {author}
  {\bibfnamefont {J.}~\bibnamefont {Danon}}, \bibinfo {author} {\bibfnamefont
  {M.}~\bibnamefont {Leijnse}}, \bibinfo {author} {\bibfnamefont
  {K.}~\bibnamefont {Flensberg}}, \bibinfo {author} {\bibfnamefont
  {J.}~\bibnamefont {Nyg\r{a}rd}}, \bibinfo {author} {\bibfnamefont
  {P.}~\bibnamefont {Krogstrup}}, \ and\ \bibinfo {author} {\bibfnamefont
  {C.~M.}\ \bibnamefont {Marcus}},\ }\href {doi.org/10.1126/science.aaf3961}
  {\bibfield  {journal} {\bibinfo  {journal} {Science}\ }\textbf {\bibinfo
  {volume} {354}},\ \bibinfo {pages} {1557} (\bibinfo {year}
  {2016})}\BibitemShut {NoStop}%
\bibitem [{\citenamefont {Szumniak}\ \emph {et~al.}(2017)\citenamefont
  {Szumniak}, \citenamefont {Chevallier}, \citenamefont {Loss},\ and\
  \citenamefont {Klinovaja}}]{Jelena-2017}%
  \BibitemOpen
  \bibfield  {author} {\bibinfo {author} {\bibfnamefont {P.}~\bibnamefont
  {Szumniak}}, \bibinfo {author} {\bibfnamefont {D.}~\bibnamefont
  {Chevallier}}, \bibinfo {author} {\bibfnamefont {D.}~\bibnamefont {Loss}}, \
  and\ \bibinfo {author} {\bibfnamefont {J.}~\bibnamefont {Klinovaja}},\ }\href
  {\doibase 10.1103/PhysRevB.96.041401} {\bibfield  {journal} {\bibinfo
  {journal} {Phys. Rev. B}\ }\textbf {\bibinfo {volume} {96}},\ \bibinfo
  {pages} {041401(R)} (\bibinfo {year} {2017})}\BibitemShut {NoStop}%
\bibitem [{\citenamefont {Zhang}\ \emph {et~al.}(2018)\citenamefont {Zhang},
  \citenamefont {Liu}, \citenamefont {Gazibegovic}, \citenamefont {Xu},
  \citenamefont {Logan}, \citenamefont {Wang}, \citenamefont {van Loo},
  \citenamefont {Bommer}, \citenamefont {de~Moor}, \citenamefont {Car},
  \citenamefont {Op~het Veld}, \citenamefont {van Veldhoven}, \citenamefont
  {Koelling}, \citenamefont {Verheijen}, \citenamefont {Pendharkar},
  \citenamefont {Pennachio}, \citenamefont {Shojaei}, \citenamefont {Lee},
  \citenamefont {Palmstr{\o}m}, \citenamefont {Bakkers}, \citenamefont
  {Das~Sarma},\ and\ \citenamefont {Kouwenhoven}}]{Zhang-2018}%
  \BibitemOpen
  \bibfield  {author} {\bibinfo {author} {\bibfnamefont {H.}~\bibnamefont
  {Zhang}}, \bibinfo {author} {\bibfnamefont {C.-X.}\ \bibnamefont {Liu}},
  \bibinfo {author} {\bibfnamefont {S.}~\bibnamefont {Gazibegovic}}, \bibinfo
  {author} {\bibfnamefont {D.}~\bibnamefont {Xu}}, \bibinfo {author}
  {\bibfnamefont {J.~A.}\ \bibnamefont {Logan}}, \bibinfo {author}
  {\bibfnamefont {G.}~\bibnamefont {Wang}}, \bibinfo {author} {\bibfnamefont
  {N.}~\bibnamefont {van Loo}}, \bibinfo {author} {\bibfnamefont {J.~D.~S.}\
  \bibnamefont {Bommer}}, \bibinfo {author} {\bibfnamefont {M.~W.~A.}\
  \bibnamefont {de~Moor}}, \bibinfo {author} {\bibfnamefont {D.}~\bibnamefont
  {Car}}, \bibinfo {author} {\bibfnamefont {R.~L.~M.}\ \bibnamefont {Op~het
  Veld}}, \bibinfo {author} {\bibfnamefont {P.~J.}\ \bibnamefont {van
  Veldhoven}}, \bibinfo {author} {\bibfnamefont {S.}~\bibnamefont {Koelling}},
  \bibinfo {author} {\bibfnamefont {M.~A.}\ \bibnamefont {Verheijen}}, \bibinfo
  {author} {\bibfnamefont {M.}~\bibnamefont {Pendharkar}}, \bibinfo {author}
  {\bibfnamefont {D.~J.}\ \bibnamefont {Pennachio}}, \bibinfo {author}
  {\bibfnamefont {B.}~\bibnamefont {Shojaei}}, \bibinfo {author} {\bibfnamefont
  {J.~S.}\ \bibnamefont {Lee}}, \bibinfo {author} {\bibfnamefont {C.~J.}\
  \bibnamefont {Palmstr{\o}m}}, \bibinfo {author} {\bibfnamefont {E.~P. A.~M.}\
  \bibnamefont {Bakkers}}, \bibinfo {author} {\bibfnamefont {S.}~\bibnamefont
  {Das~Sarma}}, \ and\ \bibinfo {author} {\bibfnamefont {L.~P.}\ \bibnamefont
  {Kouwenhoven}},\ }\href {https://www.nature.com/articles/nature26142}
  {\bibfield  {journal} {\bibinfo  {journal} {Nature}\ }\textbf {\bibinfo
  {volume} {556}},\ \bibinfo {pages} {74} (\bibinfo {year} {2018})}\BibitemShut
  {NoStop}%
\bibitem [{\citenamefont {Prada}\ \emph {et~al.}()\citenamefont {Prada},
  \citenamefont {San-Jose}, \citenamefont {de~Moor}, \citenamefont {Geresdi},
  \citenamefont {Lee}, \citenamefont {Klinovaja}, \citenamefont {Loss},
  \citenamefont {Nyg\r{a}rd}, \citenamefont {Aguado},\ and\ \citenamefont
  {Kouwenhoven}}]{Prada2019}%
  \BibitemOpen
  \bibfield  {author} {\bibinfo {author} {\bibfnamefont {E.}~\bibnamefont
  {Prada}}, \bibinfo {author} {\bibfnamefont {P.}~\bibnamefont {San-Jose}},
  \bibinfo {author} {\bibfnamefont {M.~W.~A.}\ \bibnamefont {de~Moor}},
  \bibinfo {author} {\bibfnamefont {A.}~\bibnamefont {Geresdi}}, \bibinfo
  {author} {\bibfnamefont {E.~J.~H.}\ \bibnamefont {Lee}}, \bibinfo {author}
  {\bibfnamefont {J.}~\bibnamefont {Klinovaja}}, \bibinfo {author}
  {\bibfnamefont {D.}~\bibnamefont {Loss}}, \bibinfo {author} {\bibfnamefont
  {J.}~\bibnamefont {Nyg\r{a}rd}}, \bibinfo {author} {\bibfnamefont
  {R.}~\bibnamefont {Aguado}}, \ and\ \bibinfo {author} {\bibfnamefont {L.~P.}\
  \bibnamefont {Kouwenhoven}},\ }\href {https://arxiv.org/pdf/1911.04512}
  {\bibinfo  {journal} {arXiv:1911.04512}\ }\BibitemShut {NoStop}%
\bibitem [{\citenamefont {Nadj-Perge}\ \emph {et~al.}(2013)\citenamefont
  {Nadj-Perge}, \citenamefont {Drozdov}, \citenamefont {Bernevig},\ and\
  \citenamefont {Yazdani}}]{Nadj-Perge-2013}%
  \BibitemOpen
\bibfield  {journal} {  }\bibfield  {author} {\bibinfo {author} {\bibfnamefont
  {S.}~\bibnamefont {Nadj-Perge}}, \bibinfo {author} {\bibfnamefont {I.~K.}\
  \bibnamefont {Drozdov}}, \bibinfo {author} {\bibfnamefont {B.~A.}\
  \bibnamefont {Bernevig}}, \ and\ \bibinfo {author} {\bibfnamefont
  {A.}~\bibnamefont {Yazdani}},\ }\href {\doibase 10.1103/PhysRevB.88.020407}
  {\bibfield  {journal} {\bibinfo  {journal} {Phys. Rev. B}\ }\textbf {\bibinfo
  {volume} {88}},\ \bibinfo {pages} {020407(R)} (\bibinfo {year}
  {2013})}\BibitemShut {NoStop}%
\bibitem [{\citenamefont {Nadj-Perge}\ \emph {et~al.}(2014)\citenamefont
  {Nadj-Perge}, \citenamefont {Drozdov}, \citenamefont {Li}, \citenamefont
  {Chen}, \citenamefont {Jeon}, \citenamefont {Seo}, \citenamefont {MacDonald},
  \citenamefont {Bernevig},\ and\ \citenamefont {Yazdani}}]{Nadj-Perge-2014}%
  \BibitemOpen
  \bibfield  {author} {\bibinfo {author} {\bibfnamefont {S.}~\bibnamefont
  {Nadj-Perge}}, \bibinfo {author} {\bibfnamefont {I.~K.}\ \bibnamefont
  {Drozdov}}, \bibinfo {author} {\bibfnamefont {J.}~\bibnamefont {Li}},
  \bibinfo {author} {\bibfnamefont {H.}~\bibnamefont {Chen}}, \bibinfo {author}
  {\bibfnamefont {S.}~\bibnamefont {Jeon}}, \bibinfo {author} {\bibfnamefont
  {J.}~\bibnamefont {Seo}}, \bibinfo {author} {\bibfnamefont {A.~H.}\
  \bibnamefont {MacDonald}}, \bibinfo {author} {\bibfnamefont {B.~A.}\
  \bibnamefont {Bernevig}}, \ and\ \bibinfo {author} {\bibfnamefont
  {A.}~\bibnamefont {Yazdani}},\ }\href {\doibase 10.1126/science.1259327}
  {\bibfield  {journal} {\bibinfo  {journal} {Science}\ }\textbf {\bibinfo
  {volume} {346}},\ \bibinfo {pages} {602} (\bibinfo {year}
  {2014})}\BibitemShut {NoStop}%
\bibitem [{\citenamefont {Klinovaja}\ \emph {et~al.}(2013)\citenamefont
  {Klinovaja}, \citenamefont {Stano}, \citenamefont {Yazdani},\ and\
  \citenamefont {Loss}}]{Jelena-2013}%
  \BibitemOpen
  \bibfield  {author} {\bibinfo {author} {\bibfnamefont {J.}~\bibnamefont
  {Klinovaja}}, \bibinfo {author} {\bibfnamefont {P.}~\bibnamefont {Stano}},
  \bibinfo {author} {\bibfnamefont {A.}~\bibnamefont {Yazdani}}, \ and\
  \bibinfo {author} {\bibfnamefont {D.}~\bibnamefont {Loss}},\ }\href {\doibase
  10.1103/PhysRevLett.111.186805} {\bibfield  {journal} {\bibinfo  {journal}
  {Phys. Rev. Lett.}\ }\textbf {\bibinfo {volume} {111}},\ \bibinfo {pages}
  {186805} (\bibinfo {year} {2013})}\BibitemShut {NoStop}%
\bibitem [{\citenamefont {Zvyagin}(2013)}]{Zvyagin-2013}%
  \BibitemOpen
  \bibfield  {author} {\bibinfo {author} {\bibfnamefont {A.~A.}\ \bibnamefont
  {Zvyagin}},\ }\href {\doibase 10.1103/PhysRevLett.110.217207} {\bibfield
  {journal} {\bibinfo  {journal} {Phys. Rev. Lett.}\ }\textbf {\bibinfo
  {volume} {110}},\ \bibinfo {pages} {217207} (\bibinfo {year}
  {2013})}\BibitemShut {NoStop}%
\bibitem [{\citenamefont {Kim}\ \emph {et~al.}(2018)\citenamefont {Kim},
  \citenamefont {Palacio-Morales}, \citenamefont {Posske}, \citenamefont
  {R\'{o}zsa}, \citenamefont {Palot\'{a}s}, \citenamefont {Szunyogh},
  \citenamefont {Thorwart},\ and\ \citenamefont {Wiesendanger}}]{Kim-2018}%
  \BibitemOpen
  \bibfield  {author} {\bibinfo {author} {\bibfnamefont {H.}~\bibnamefont
  {Kim}}, \bibinfo {author} {\bibfnamefont {A.}~\bibnamefont
  {Palacio-Morales}}, \bibinfo {author} {\bibfnamefont {T.}~\bibnamefont
  {Posske}}, \bibinfo {author} {\bibfnamefont {L.}~\bibnamefont {R\'{o}zsa}},
  \bibinfo {author} {\bibfnamefont {K.}~\bibnamefont {Palot\'{a}s}}, \bibinfo
  {author} {\bibfnamefont {L.}~\bibnamefont {Szunyogh}}, \bibinfo {author}
  {\bibfnamefont {M.}~\bibnamefont {Thorwart}}, \ and\ \bibinfo {author}
  {\bibfnamefont {R.}~\bibnamefont {Wiesendanger}},\ }\href
  {https://advances.sciencemag.org/content/4/5/eaar5251} {\bibfield  {journal}
  {\bibinfo  {journal} {Science Advances}\ }\textbf {\bibinfo {volume} {4}}
  (\bibinfo {year} {2018})}\BibitemShut {NoStop}%
\bibitem [{\citenamefont {Sau}\ and\ \citenamefont {Tewari}(2013)}]{Sau-2013}%
  \BibitemOpen
  \bibfield  {author} {\bibinfo {author} {\bibfnamefont {J.~D.}\ \bibnamefont
  {Sau}}\ and\ \bibinfo {author} {\bibfnamefont {S.}~\bibnamefont {Tewari}},\
  }\href {\doibase 10.1103/PhysRevB.88.054503} {\bibfield  {journal} {\bibinfo
  {journal} {Phys. Rev. B}\ }\textbf {\bibinfo {volume} {88}},\ \bibinfo
  {pages} {054503} (\bibinfo {year} {2013})}\BibitemShut {NoStop}%
\bibitem [{\citenamefont {Marganska}\ \emph {et~al.}(2018)\citenamefont
  {Marganska}, \citenamefont {Milz}, \citenamefont {Izumida}, \citenamefont
  {Strunk},\ and\ \citenamefont {Grifoni}}]{Marganska-2018}%
  \BibitemOpen
  \bibfield  {author} {\bibinfo {author} {\bibfnamefont {M.}~\bibnamefont
  {Marganska}}, \bibinfo {author} {\bibfnamefont {L.}~\bibnamefont {Milz}},
  \bibinfo {author} {\bibfnamefont {W.}~\bibnamefont {Izumida}}, \bibinfo
  {author} {\bibfnamefont {C.}~\bibnamefont {Strunk}}, \ and\ \bibinfo {author}
  {\bibfnamefont {M.}~\bibnamefont {Grifoni}},\ }\href {\doibase
  10.1103/PhysRevB.97.075141} {\bibfield  {journal} {\bibinfo  {journal} {Phys.
  Rev. B}\ }\textbf {\bibinfo {volume} {97}},\ \bibinfo {pages} {075141}
  (\bibinfo {year} {2018})}\BibitemShut {NoStop}%
\bibitem [{\citenamefont {Milz}\ \emph {et~al.}(2019)\citenamefont {Milz},
  \citenamefont {Izumida}, \citenamefont {Grifoni},\ and\ \citenamefont
  {Marganska}}]{Milz-2018}%
  \BibitemOpen
  \bibfield  {author} {\bibinfo {author} {\bibfnamefont {L.}~\bibnamefont
  {Milz}}, \bibinfo {author} {\bibfnamefont {W.}~\bibnamefont {Izumida}},
  \bibinfo {author} {\bibfnamefont {M.}~\bibnamefont {Grifoni}}, \ and\
  \bibinfo {author} {\bibfnamefont {M.}~\bibnamefont {Marganska}},\ }\href
  {\doibase 10.1103/PhysRevB.100.155417} {\bibfield  {journal} {\bibinfo
  {journal} {Phys. Rev. B}\ }\textbf {\bibinfo {volume} {100}},\ \bibinfo
  {pages} {155417} (\bibinfo {year} {2019})}\BibitemShut {NoStop}%
\bibitem [{\citenamefont {Lee}(2014)}]{Lee-2014}%
  \BibitemOpen
  \bibfield  {author} {\bibinfo {author} {\bibfnamefont {P.}~\bibnamefont
  {Lee}},\ }\href {\doibase 10.1126/science.1260282} {\bibfield  {journal}
  {\bibinfo  {journal} {Science (New York, N.Y.)}\ }\textbf {\bibinfo {volume}
  {346}},\ \bibinfo {pages} {545} (\bibinfo {year} {2014})}\BibitemShut
  {NoStop}%
\bibitem [{\citenamefont {Kao}(2014)}]{Kao-2014}%
  \BibitemOpen
  \bibfield  {author} {\bibinfo {author} {\bibfnamefont {H.-c.}\ \bibnamefont
  {Kao}},\ }\href {\doibase 10.1103/PhysRevB.90.245435} {\bibfield  {journal}
  {\bibinfo  {journal} {Phys. Rev. B}\ }\textbf {\bibinfo {volume} {90}},\
  \bibinfo {pages} {245435} (\bibinfo {year} {2014})}\BibitemShut {NoStop}%
\bibitem [{\citenamefont {Hegde}\ \emph {et~al.}(2015)\citenamefont {Hegde},
  \citenamefont {Shivamoggi}, \citenamefont {Vishveshwara},\ and\ \citenamefont
  {Sen}}]{Hegde-2015}%
  \BibitemOpen
  \bibfield  {author} {\bibinfo {author} {\bibfnamefont {S.}~\bibnamefont
  {Hegde}}, \bibinfo {author} {\bibfnamefont {V.}~\bibnamefont {Shivamoggi}},
  \bibinfo {author} {\bibfnamefont {S.}~\bibnamefont {Vishveshwara}}, \ and\
  \bibinfo {author} {\bibfnamefont {D.}~\bibnamefont {Sen}},\ }\href {\doibase
  10.1088/1367-2630/17/5/053036} {\bibfield  {journal} {\bibinfo  {journal}
  {New Journal of Physics}\ }\textbf {\bibinfo {volume} {17}},\ \bibinfo
  {pages} {053036} (\bibinfo {year} {2015})}\BibitemShut {NoStop}%
\bibitem [{\citenamefont {Zvyagin}(2015)}]{Zvyagin-2015}%
  \BibitemOpen
  \bibfield  {author} {\bibinfo {author} {\bibfnamefont {A.~A.}\ \bibnamefont
  {Zvyagin}},\ }\href {\doibase 10.1063/1.4928919} {\bibfield  {journal}
  {\bibinfo  {journal} {Low Temperature Physics}\ }\textbf {\bibinfo {volume}
  {41}},\ \bibinfo {pages} {625} (\bibinfo {year} {2015})}\BibitemShut
  {NoStop}%
\bibitem [{\citenamefont {Zeng}\ \emph {et~al.}(2019)\citenamefont {Zeng},
  \citenamefont {Moore}, \citenamefont {Rao}, \citenamefont {Stanescu},\ and\
  \citenamefont {Tewari}}]{Zeng-2019}%
  \BibitemOpen
  \bibfield  {author} {\bibinfo {author} {\bibfnamefont {C.}~\bibnamefont
  {Zeng}}, \bibinfo {author} {\bibfnamefont {C.}~\bibnamefont {Moore}},
  \bibinfo {author} {\bibfnamefont {A.~M.}\ \bibnamefont {Rao}}, \bibinfo
  {author} {\bibfnamefont {T.~D.}\ \bibnamefont {Stanescu}}, \ and\ \bibinfo
  {author} {\bibfnamefont {S.}~\bibnamefont {Tewari}},\ }\href {\doibase
  10.1103/PhysRevB.99.094523} {\bibfield  {journal} {\bibinfo  {journal} {Phys.
  Rev. B}\ }\textbf {\bibinfo {volume} {99}},\ \bibinfo {pages} {094523}
  (\bibinfo {year} {2019})}\BibitemShut {NoStop}%
\bibitem [{\citenamefont {Lieb}\ \emph {et~al.}(1961)\citenamefont {Lieb},
  \citenamefont {Schultz},\ and\ \citenamefont {Mattis}}]{Lieb-1961}%
  \BibitemOpen
  \bibfield  {author} {\bibinfo {author} {\bibfnamefont {E.}~\bibnamefont
  {Lieb}}, \bibinfo {author} {\bibfnamefont {T.}~\bibnamefont {Schultz}}, \
  and\ \bibinfo {author} {\bibfnamefont {D.}~\bibnamefont {Mattis}},\ }\href
  {\doibase https://doi.org/10.1016/0003-4916(61)90115-4} {\bibfield  {journal}
  {\bibinfo  {journal} {Annals of Physics}\ }\textbf {\bibinfo {volume} {16}},\
  \bibinfo {pages} {407 } (\bibinfo {year} {1961})}\BibitemShut {NoStop}%
\bibitem [{\citenamefont {Loginov}\ and\ \citenamefont
  {Pereverzev}(1997)}]{Loginov-1997}%
  \BibitemOpen
  \bibfield  {author} {\bibinfo {author} {\bibfnamefont {A.~V.}\ \bibnamefont
  {Loginov}}\ and\ \bibinfo {author} {\bibfnamefont {Y.~V.}\ \bibnamefont
  {Pereverzev}},\ }\href {\doibase 10.1063/1.593419} {\bibfield  {journal}
  {\bibinfo  {journal} {Low Temperature Physics}\ }\textbf {\bibinfo {volume}
  {23}},\ \bibinfo {pages} {534} (\bibinfo {year} {1997})}\BibitemShut
  {NoStop}%
\bibitem [{\citenamefont {Kawabata}\ \emph {et~al.}(2017)\citenamefont
  {Kawabata}, \citenamefont {Kobayashi}, \citenamefont {Wu},\ and\
  \citenamefont {Katsura}}]{Kawabata-2017}%
  \BibitemOpen
  \bibfield  {author} {\bibinfo {author} {\bibfnamefont {K.}~\bibnamefont
  {Kawabata}}, \bibinfo {author} {\bibfnamefont {R.}~\bibnamefont {Kobayashi}},
  \bibinfo {author} {\bibfnamefont {N.}~\bibnamefont {Wu}}, \ and\ \bibinfo
  {author} {\bibfnamefont {H.}~\bibnamefont {Katsura}},\ }\href {\doibase
  10.1103/PhysRevB.95.195140} {\bibfield  {journal} {\bibinfo  {journal} {Phys.
  Rev. B}\ }\textbf {\bibinfo {volume} {95}},\ \bibinfo {pages} {195140}
  (\bibinfo {year} {2017})}\BibitemShut {NoStop}%
\bibitem [{\citenamefont {Chiu}\ \emph {et~al.}(2016)\citenamefont {Chiu},
  \citenamefont {Teo}, \citenamefont {Schnyder},\ and\ \citenamefont
  {Ryu}}]{Review-Chiu-2016}%
  \BibitemOpen
  \bibfield  {author} {\bibinfo {author} {\bibfnamefont {C.-K.}\ \bibnamefont
  {Chiu}}, \bibinfo {author} {\bibfnamefont {J.~C.~Y.}\ \bibnamefont {Teo}},
  \bibinfo {author} {\bibfnamefont {A.~P.}\ \bibnamefont {Schnyder}}, \ and\
  \bibinfo {author} {\bibfnamefont {S.}~\bibnamefont {Ryu}},\ }\href {\doibase
  10.1103/RevModPhys.88.035005} {\bibfield  {journal} {\bibinfo  {journal}
  {Rev. Mod. Phys.}\ }\textbf {\bibinfo {volume} {88}},\ \bibinfo {pages}
  {035005} (\bibinfo {year} {2016})}\BibitemShut {NoStop}%
\bibitem [{\citenamefont {Kempkes}\ \emph {et~al.}(2019)\citenamefont
  {Kempkes}, \citenamefont {Slot}, \citenamefont {van~den Broeke},
  \citenamefont {Capiod}, \citenamefont {Benalcazar}, \citenamefont
  {Vanmaekelbergh}, \citenamefont {Bercioux}, \citenamefont {Swart},\ and\
  \citenamefont {Morais~Smith}}]{Kempkes-2019}%
  \BibitemOpen
  \bibfield  {author} {\bibinfo {author} {\bibfnamefont {S.~N.}\ \bibnamefont
  {Kempkes}}, \bibinfo {author} {\bibfnamefont {M.~R.}\ \bibnamefont {Slot}},
  \bibinfo {author} {\bibfnamefont {J.~J.}\ \bibnamefont {van~den Broeke}},
  \bibinfo {author} {\bibfnamefont {P.}~\bibnamefont {Capiod}}, \bibinfo
  {author} {\bibfnamefont {W.~A.}\ \bibnamefont {Benalcazar}}, \bibinfo
  {author} {\bibfnamefont {D.}~\bibnamefont {Vanmaekelbergh}}, \bibinfo
  {author} {\bibfnamefont {D.}~\bibnamefont {Bercioux}}, \bibinfo {author}
  {\bibfnamefont {I.}~\bibnamefont {Swart}}, \ and\ \bibinfo {author}
  {\bibfnamefont {C.}~\bibnamefont {Morais~Smith}},\ }\href@noop {} {\bibfield
  {journal} {\bibinfo  {journal} {Nature Materials}\ }\textbf {\bibinfo
  {volume} {18}},\ \bibinfo {pages} {1292} (\bibinfo {year}
  {2019})}\BibitemShut {NoStop}%
\bibitem [{\citenamefont {Wakatsuki}\ \emph {et~al.}(2014)\citenamefont
  {Wakatsuki}, \citenamefont {Ezawa}, \citenamefont {Tanaka},\ and\
  \citenamefont {Nagaosa}}]{wakatsuki:prb2014}%
  \BibitemOpen
  \bibfield  {author} {\bibinfo {author} {\bibfnamefont {R.}~\bibnamefont
  {Wakatsuki}}, \bibinfo {author} {\bibfnamefont {M.}~\bibnamefont {Ezawa}},
  \bibinfo {author} {\bibfnamefont {Y.}~\bibnamefont {Tanaka}}, \ and\ \bibinfo
  {author} {\bibfnamefont {N.}~\bibnamefont {Nagaosa}},\ }\href {\doibase
  10.1103/PhysRevB.90.014505} {\bibfield  {journal} {\bibinfo  {journal} {Phys.
  Rev. B}\ }\textbf {\bibinfo {volume} {90}},\ \bibinfo {pages} {014505}
  (\bibinfo {year} {2014})}\BibitemShut {NoStop}%
\bibitem [{\citenamefont {Altland}\ and\ \citenamefont
  {Zirnbauer}(1997)}]{altland:prb1997}%
  \BibitemOpen
  \bibfield  {author} {\bibinfo {author} {\bibfnamefont {A.}~\bibnamefont
  {Altland}}\ and\ \bibinfo {author} {\bibfnamefont {M.~R.}\ \bibnamefont
  {Zirnbauer}},\ }\href {\doibase 10.1103/PhysRevB.55.1142} {\bibfield
  {journal} {\bibinfo  {journal} {Phys. Rev. B}\ }\textbf {\bibinfo {volume}
  {55}},\ \bibinfo {pages} {1142} (\bibinfo {year} {1997})}\BibitemShut
  {NoStop}%
\bibitem [{\citenamefont {Wen}\ and\ \citenamefont {Zee}(1989)}]{Wen-1989}%
  \BibitemOpen
  \bibfield  {author} {\bibinfo {author} {\bibfnamefont {X.}~\bibnamefont
  {Wen}}\ and\ \bibinfo {author} {\bibfnamefont {A.}~\bibnamefont {Zee}},\
  }\href {\doibase https://doi.org/10.1016/0550-3213(89)90062-X} {\bibfield
  {journal} {\bibinfo  {journal} {Nuclear Physics B}\ }\textbf {\bibinfo
  {volume} {316}},\ \bibinfo {pages} {641 } (\bibinfo {year}
  {1989})}\BibitemShut {NoStop}%
\bibitem [{\citenamefont {Kouachi}(2006)}]{Kouachi}%
  \BibitemOpen
  \bibfield  {author} {\bibinfo {author} {\bibfnamefont {S.}~\bibnamefont
  {Kouachi}},\ }\href {http://eudml.org/doc/126465} {\bibfield  {journal}
  {\bibinfo  {journal} {ELA. The Electronic Journal of Linear Algebra
  [electronic only]}\ }\textbf {\bibinfo {volume} {15}},\ \bibinfo {pages}
  {115} (\bibinfo {year} {2006})}\BibitemShut {NoStop}%
\bibitem [{Note1()}]{Note1}%
  \BibitemOpen
  \bibinfo {note} {Note that $\lambda $ can be zero and it will for odd $N$.
  Hence, the standard formula to calculate the determinant of a partitioned
  $2\times 2$ matrix can not be used here, because it requires the inverse of
  one diagonal block. We use instead Silvester's formula\cite {Silvester}:
  $\protect \text {det}\left [\begin {matrix} A & B\\ C & D\\ \end {matrix}
  \right ] =\protect \text {det} \left [AD\protect \tmspace +\thinmuskip
  {.1667em}-\protect \tmspace +\thinmuskip {.1667em}CB\right ]$, where
  $A,\protect \tmspace +\thinmuskip {.1667em}B,\protect \tmspace +\thinmuskip
  {.1667em}C,\protect \tmspace +\thinmuskip {.1667em}D$ are square matrices of
  the same size and the only requirement is $\left [C,\protect \tmspace
  +\thinmuskip {.1667em}D\right ]=0$.}\BibitemShut {Stop}%
\bibitem [{\citenamefont {Silvester}(2000)}]{Silvester}%
  \BibitemOpen
  \bibfield  {author} {\bibinfo {author} {\bibfnamefont {J.~R.}\ \bibnamefont
  {Silvester}},\ }\href {http://www.jstor.org/stable/3620776} {\bibfield
  {journal} {\bibinfo  {journal} {The Mathematical Gazette}\ }\textbf {\bibinfo
  {volume} {84}},\ \bibinfo {pages} {460} (\bibinfo {year} {2000})}\BibitemShut
  {NoStop}%
\bibitem [{\citenamefont {Li}\ \emph {et~al.}(2018)\citenamefont {Li},
  \citenamefont {Zhang}, \citenamefont {Zhang},\ and\ \citenamefont
  {Song}}]{li:prb2018}%
  \BibitemOpen
  \bibfield  {author} {\bibinfo {author} {\bibfnamefont {C.}~\bibnamefont
  {Li}}, \bibinfo {author} {\bibfnamefont {X.~Z.}\ \bibnamefont {Zhang}},
  \bibinfo {author} {\bibfnamefont {G.}~\bibnamefont {Zhang}}, \ and\ \bibinfo
  {author} {\bibfnamefont {Z.}~\bibnamefont {Song}},\ }\href {\doibase
  10.1103/PhysRevB.97.115436} {\bibfield  {journal} {\bibinfo  {journal} {Phys.
  Rev. B}\ }\textbf {\bibinfo {volume} {97}},\ \bibinfo {pages} {115436}
  (\bibinfo {year} {2018})}\BibitemShut {NoStop}%
\bibitem [{\citenamefont {Webb}\ and\ \citenamefont {Parberry}(1969)}]{Webb}%
  \BibitemOpen
  \bibfield  {author} {\bibinfo {author} {\bibfnamefont {W.}~\bibnamefont
  {Webb}}\ and\ \bibinfo {author} {\bibfnamefont {E.}~\bibnamefont
  {Parberry}},\ }\href {https://www.mathstat.dal.ca/FQ/Scanned/7-5/webb.pdf}
  {\bibfield  {journal} {\bibinfo  {journal} {The Fibonacci Quarterly}\
  }\textbf {\bibinfo {volume} {7}} (\bibinfo {year} {1969})}\BibitemShut
  {NoStop}%
\bibitem [{\citenamefont {E.~jun. Hoggatt}\ and\ \citenamefont
  {T.~Long}(1974)}]{Hoggatt}%
  \BibitemOpen
  \bibfield  {author} {\bibinfo {author} {\bibfnamefont {V.}~\bibnamefont
  {E.~jun. Hoggatt}}\ and\ \bibinfo {author} {\bibfnamefont {C.}~\bibnamefont
  {T.~Long}},\ }\href {https://www.fq.math.ca/Issues/12-2.pdf} {\bibfield
  {journal} {\bibinfo  {journal} {The Fibonacci Quarterly}\ }\textbf {\bibinfo
  {volume} {12}},\  (\bibinfo {year} {1974})}\BibitemShut {NoStop}%
\bibitem [{\citenamefont {{\"O}zvatan}\ and\ \citenamefont
  {Pashaev}(2017)}]{Oezvatan-2017}%
  \BibitemOpen
  \bibfield  {author} {\bibinfo {author} {\bibfnamefont {M.}~\bibnamefont
  {{\"O}zvatan}}\ and\ \bibinfo {author} {\bibfnamefont {O.}~\bibnamefont
  {Pashaev}},\ }\href {https://arxiv.org/abs/1707.09151} {\bibfield  {journal}
  {\bibinfo  {journal} {arXiv:1707.09151}\ } (\bibinfo {year}
  {2017})}\BibitemShut {NoStop}%
\bibitem [{\citenamefont {Sirker}\ \emph {et~al.}(2014)\citenamefont {Sirker},
  \citenamefont {Maiti}, \citenamefont {Konstantinidis},\ and\ \citenamefont
  {Sedlmayr}}]{Sirker}%
  \BibitemOpen
  \bibfield  {author} {\bibinfo {author} {\bibfnamefont {J.}~\bibnamefont
  {Sirker}}, \bibinfo {author} {\bibfnamefont {M.}~\bibnamefont {Maiti}},
  \bibinfo {author} {\bibfnamefont {N.~P.}\ \bibnamefont {Konstantinidis}}, \
  and\ \bibinfo {author} {\bibfnamefont {N.}~\bibnamefont {Sedlmayr}},\ }\href
  {https://iopscience.iop.org/article/10.1088/1742-5468/2014/10/P10032/meta}
  {\bibfield  {journal} {\bibinfo  {journal} {J. Stat. Mech.}\ ,\ \bibinfo
  {pages} {P10032}} (\bibinfo {year} {2014})}\BibitemShut {NoStop}%
\bibitem [{\citenamefont {Shin}(1997)}]{Shin-97}%
  \BibitemOpen
  \bibfield  {author} {\bibinfo {author} {\bibfnamefont {B.~C.}\ \bibnamefont
  {Shin}},\ }\href
  {https://www.cambridge.org/core/journals/bulletin-of-the-australian-mathematical-society/article/formula-for-eigenpairs-of-certain-symmetric-tridiagonal-matrices/AED4F57F7E4C9F3FF659FD378CD085D6}
  {\bibfield  {journal} {\bibinfo  {journal} {Bulletin of the Australian
  Mathematical Society}\ }\textbf {\bibinfo {volume} {55}},\ \bibinfo {pages}
  {249} (\bibinfo {year} {1997})}\BibitemShut {NoStop}%
\bibitem [{\citenamefont {McLaughlin}(1979)}]{McLaughlin-1979}%
  \BibitemOpen
  \bibfield  {author} {\bibinfo {author} {\bibfnamefont {W.~I.}\ \bibnamefont
  {McLaughlin}},\ }\href {https://www.fq.math.ca/Scanned/17-2/mclaughlin.pdf}
  {\bibfield  {journal} {\bibinfo  {journal} {The Fibonacci Quarterly}\
  }\textbf {\bibinfo {volume} {17.2}} (\bibinfo {year} {1979})}\BibitemShut
  {NoStop}%
\bibitem [{\citenamefont {Waddill}(1992)}]{Waddill-1992}%
  \BibitemOpen
  \bibfield  {author} {\bibinfo {author} {\bibfnamefont {M.}~\bibnamefont
  {Waddill}},\ }\href {https://www.fq.math.ca/Scanned/30-1/waddill.pdf}
  {\bibfield  {journal} {\bibinfo  {journal} {The Fibonacci Quarterly}\
  }\textbf {\bibinfo {volume} {30}} (\bibinfo {year} {1992})}\BibitemShut
  {NoStop}%
\bibitem [{Note2()}]{Note2}%
  \BibitemOpen
  \bibinfo {note} {In fact $\mu =0$ supports complex wavevectors too, but their
  real part has to be zero or $\pi /2$, i.e. one has to use $iq$ or $\protect
  \frac {\pi }{2}+iq$.}\BibitemShut {Stop}%
\bibitem [{\citenamefont {Salkuyeh}(2006)}]{Salkuyeh}%
  \BibitemOpen
  \bibfield  {author} {\bibinfo {author} {\bibfnamefont {D.~K.}\ \bibnamefont
  {Salkuyeh}},\ }\href
  {http://www.sciencedirect.com/science/article/pii/S0096300305008209}
  {\bibfield  {journal} {\bibinfo  {journal} {Applied Mathematics and
  Computation}\ }\textbf {\bibinfo {volume} {176}},\ \bibinfo {pages} {442}
  (\bibinfo {year} {2006})}\BibitemShut {NoStop}%
\bibitem [{\citenamefont {Molinari}(2008)}]{Molinari}%
  \BibitemOpen
  \bibfield  {author} {\bibinfo {author} {\bibfnamefont {L.~G.}\ \bibnamefont
  {Molinari}},\ }\href
  {http://www.sciencedirect.com/science/article/pii/S0024379508003200}
  {\bibfield  {journal} {\bibinfo  {journal} {Linear Algebra and its
  Applications}\ }\textbf {\bibinfo {volume} {429}},\ \bibinfo {pages} {2221}
  (\bibinfo {year} {2008})}\BibitemShut {NoStop}%
\bibitem [{\citenamefont {Usmani}(1994)}]{Usmani}%
  \BibitemOpen
  \bibfield  {author} {\bibinfo {author} {\bibfnamefont {R.~A.}\ \bibnamefont
  {Usmani}},\ }\href {\doibase https://doi.org/10.1016/0898-1221(94)90066-3}
  {\bibfield  {journal} {\bibinfo  {journal} {Computers \& Mathematics with
  Applications}\ }\textbf {\bibinfo {volume} {27}},\ \bibinfo {pages} {59}
  (\bibinfo {year} {1994})}\BibitemShut {NoStop}%
\end{thebibliography}%

\end{document}